\documentclass[aip,jmp,superscriptaddress,preprint,author-year]{revtex4-1}

\usepackage{settings}  
\usepackage{macros}    

\graphicspath{{figures/}}

\begin{document}

\title{Definition, detection, and tracking of persistent structures \\ in atmospheric flows}

\author{Johannes von Lindheim}
\affiliation{Institute for Mathematics, Technical University Berlin, 10623 Berlin, Germany}

\author{Abhishek Harikrishnan}
\author{Tom Dörffel}
\affiliation{Department of Mathematics and Computer Science, Freie Universität Berlin, 14195 Berlin, Germany}

\author{Rupert Klein}
\affiliation{Department of Mathematics and Computer Science, Freie Universität Berlin, 14195 Berlin, Germany}
\email{rupert.klein@math.fu-berlin.de}

\author{P\'eter Koltai}
\affiliation{Department of Mathematics and Computer Science, Freie Universität Berlin, 14195 Berlin, Germany}

\author{Natalia Mikula}
\affiliation{Zuse Institute Berlin, Visual and Data-centric Computing, 14195 Berlin, Germany}

\author{Annette M\"uller}
\author{Peter N\'evir}
\author{George Pacey}
\affiliation{Institute for Meteorology, Freie Universität Berlin, 12165 Berlin, Germany}

\author{Robert Polzin}
\affiliation{Department of Mathematics and Computer Science, Freie Universität Berlin, 14195 Berlin, Germany}

\author{Nikki Vercauteren}
\affiliation{Department of Geosciences, University of Oslo, Oslo, Norway}

\date{Received: date / Accepted: date}

\begin{abstract}
Long-lived flow patterns in the atmosphere such as weather fronts, mid-latitude blockings or tropical cyclones often induce extreme weather conditions. As a consequence, their description, detection, and tracking has received increasing attention in recent years. Similar objectives also arise in diverse fields such as turbulence and combustion research, image analysis, and medical diagnostics under the headlines of ``feature tracking'', ``coherent structure detection'' or ``image registration'' - to name just a few. A host of different approaches to addressing the underlying, often very similar, tasks have been developed and successfully used. Here, several typical examples of such approaches are summarized, further developed and applied to meteorological data sets. Common abstract operational steps form the basis for a unifying framework for the specification of ``persistent structures'' involving the definition of the physical state of a system, the features of interest, and means of measuring their persistence.

\end{abstract}

\maketitle


\section{Introduction}


\subparagraph{The role of persistent structures in atmospheric flows.} In observational data and atmospheric flow simulations localized flow structures with different degrees of dynamical persistence (or coherence) can often be clearly identified. Examples include Rossby- and internal waves, vortices, convective cells, precipitation fronts, or convectively coupled Kelvin waves in the tropics \citep{CattoandPfahl2013,FrittsAlexander2003,Houze2009,WheelerKiladis1999}. All these phenomena have a sizeable effect on both the local weather and the larger-scale circulation. The importance of these structures stems from their impact on mass, momentum, and energy transport. Large-scale vortices, for example, generally transport heat and moisture polewards and upwards and thereby also redistribute the influx of solar energy around the globe. More locally, vortices have a large impact on the weather and climate, and extremely intense vortices can induce severe ecological and economical damage~\citep{WeinkleEtAl2018}. Weather fronts, as a second example, are important in weather forecasting as they can imply severe weather with large temperature variations and extreme rainfall \citep{CattoandPfahl2013}. To investigate the role of different types of persistent dynamical structures for atmospheric dynamics and their impact on extreme weather, methods for their detection and tracking in observational and numerical simulation data have attracted increasing attention in recent years.


\subparagraph{Coherent structures in turbulence research.} The notion of ``coherent structures'' has played an important role in turbulence research since the 1970s (see, \eg, the review paper by \citet{Hussain1983}) when they were systematically identified in experiments as quasi-deterministic flow patterns in turbulent boundary layers and shear flows. \citet{Hussain1983} notes, however, that Ludwig Prandtl already recognized the importance of related structures, which he called ``Turbulenzballen'' in the context of his mixing length theory for turbulence closure, as early as the 1920s, see, \eg, (\citet{Rotta1972}, p.~171ff). In the wake of the discoveries of the 1970s, coherent structures were mathematically characterized as patches of ``coherent vorticity'', \ie, as (time dependent) subdomains of the flow field within which the vorticity field was to be statistically correlated strongly in space and time. In terms of active physical mechanisms, constant density incompressible flow amounts to no more and no less than the self-induced advection, stretching/tilting, and diffusion of vorticity. In line with this, researchers found descriptions of coherence in terms of the Lagrangian flow representation to have clear advantages over Eulerian concepts because, owing to the important role of the advection of vorticity, strong temporal correlation naturally arises in ``turbulent eddies'' that consist of the same fluid elements for long times \citep{Hussain1983,Haller:2015kj,froyland2013analytic,Fro15,SchneideEtAl2019}. 


\subparagraph{Non-Lagrangian persistent structures in atmospheric flows.} Lagrangian evolution, while important, is much less dominant in atmospheric flows. Gravity, Earth's rotation, and moist processes all interact with the transport of vorticity in ways not present in classical incompressible flow turbulence. As a consequence, the atmosphere exhibits clearly identifiable flow patterns that have a sizeable impact on the weather, propagate continuously in time, and are rather \emph{persistent} in comparison with local flow time scales, but are decidedly non-Lagrangian. The examples cited in the first paragraph above are all in this category. \citet{Ouellette:2012ch} offers an overview of Lagrangian concepts of coherence that have evolved from the abovementioned early developments in turbulence research, and of Eulerian concepts that are less promising in turbulence but play a more prominent role in the analysis of geophysical flows. 

More specifically, an Eulerian dimensionless vorticity number was defined and successfully applied to identify atmospheric vortices on multiple scales by \citet{Schielicke:2019cx}. This approach is efficient for the detection of vortical structures in the atmosphere and, for the same reasons as in the setting of incompressible turbulence, one might surmise that Lagrangian concepts might be even more effective. The material boundaries of atmospheric vortices are not clearly identifiable, however, and material flows in and out of a vortex, in particular during intensification and weakening. In fact, the exchange of mass between a tropical cyclone and its environment is crucial since it can be interpreted as a ``thermodynamic engine'' \citep{PauluisZhang2017} with air serving as the carrier fluid and moisture as its energetic component. 

Similarly, material coherence seems not to be the relevant notion for weather fronts. Indeed, a substantial relative motion between different constituents of the medium exists in weather fronts, so that mixing of material occurs across such fronts with water droplets on the one hand, and air and aerosols on the other partially changing sides. Yet a front exhibits persistency in its dynamics and can be identified clearly in observations and simulation data. The glossary of the American Meteorological Society (AMS) defines a front as the interface or transition zone between two air masses of different density.  This definition further stresses that a front almost invariably separates air masses of different temperature, but notes that many other features may distinguish a front. This includes, \eg, changes in wind direction or moisture discontinuities. To detect fronts through existing approaches, a criterion is to be defined to identify the boundary of the structure (\eg, a curve or surface based on the magnitude of the temperature gradient). Based on the identified frontal boundaries, tracking approaches have been used to study their time evolution~\citep{Kern:2019ep}. More generally, several observables of the measured flow field are of interest in a meteorological phenomenon (strong wind, temperature changes, etc.) and this gives rise to definitions of more complex multi-criteria ``features'' that characterize the structure in question.

There is also a prominent example, however, that does highlight the importance of material coherence in the context of incipient hurricanes. \citet{DunkertonEtAl2009} suggest that a precondition for the intensification of a tropical depression to hurricane strength is that at the center of the depression there is the formation of a ``pouch'' of closed streamlines of the horizontal flow seen in a frame of reference that moves with the associated baroclinic wave. Owing to vertical mass fluxes and the superposition of wave and vortex modes of motion prevailing in this situation, it is not clear that air masses within the pouch remain the same as time progresses. Nevertheless, this formation of a pocket of closed streamlines in a moving frame of reference is very similar to the travelling vortices identified as Lagrangian coherent sets in an analysis, \eg, of the idealized Bickley jet test case in~\citep{rypina2007,KoltaiRenger2018}.


\subparagraph{Feature tracking in (visual) data analysis.}  An identification and tracking approach for ``persistent'' or ``coherent'' structures should be based on a sufficiently general abstract definition, so as to allow for fair intercomparison of methods, transferability of approaches between different physical phenomena, and a general systematization in the sense of scientific data analysis. These principles have long since guided researchers in the area of visual data analysis, see, \eg, \citep{GuentherKuhnHegeetal2017,SakuraiHegeKuhnetal2017}, and \citet[figure~27]{HeinzeEtAl2017}. In line with the coherent structure concept in turbulence research, and in anticipation of the examples from meteorology discussed below, such a generalized definition should include notions of spatial boundedness, longevity (persistence) on a suitably chosen reference time scale, and of mathematical characterizations of what makes the structure stand out from its surroundings in physical terms and how its presence can be diagnosed. Section~\ref{sec:discussion} of the present work will suggest a formal definition along these lines taking into account the practical examples from meteorology to be discussed in sections~\ref{sec:meteo_data} to \ref{sec:Applications} in the meantime.

To provide a quick overview, let us consider the question of what would a sufficiently general definition of coherence / persistence  look like that were to encompass the different examples mentioned above? We divide the answer into two parts. The first part involves the identification of a, possibly complex multi-variable, feature that will characterize a meteorologically relevant flow pattern. This feature should be defined based on physical principles that allow the structure of interest to be distinguished from its surroundings. Straightforward examples of single-scalar criteria are associated with the level sets of the $Q$-function \citep{hunt1988eddies} of the vorticity field (see eq.~\eqref{eq:Q} below), of the pressure, or of the magnitude of its gradient. These criteria can be combined to provide a more selective filter that specifically focuses on concentrated vortices: From regions with a large value of $Q$ select only those with a sufficiently low pressure.  Another more complex example is given by N\'evir's  Dynamic State Index (DSI) \citep{Nevir2004} (see eq.~\eqref{eq:DSI_PE}), which combines gradients of the potential vorticity, the Bernoulli function, and the potential temperature (or entropy) in one scalar criterion.

The second part of the definition involves the persistence or coherence of the structures. In this context we will discuss the use of feature tracking approaches to directly estimate the lifetime of a structure \citep{GuentherKuhnHegeetal2017,SakuraiHegeKuhnetal2017,HeinzeEtAl2017}, Lagrangian coherent set analysis which provides estimates of the diffusive mixing time scale of a given set of Lagrangian trajectories \citep{froyland2013analytic,Fro15,BaKo17,KoltaiRenger2018,FrKo21}, and a new approach called ``dynamic spectral clustering'', which applies key ideas from the Lagrangian coherent set approach, but applies them to surrogate dynamics that are learned from the data and that may or may not approximate any Lagrangian paths~\citep{KLNS20}.

The rest of the paper is organized as follows. Section~\ref{sec:meteo_data} summarizes the meteorological data which this work relies upon in the concrete motivational examples of weather fronts, tropical cyclones, and a synoptic-scale storm. Based on these examples, several features are introduced in section~\ref{sec:FeatureDefinitions} that formalize their phenomenological descriptions. Section~\ref{sec:Persistence} describes several practical methods that are currently in use in the meteorological and fluid dynamics communities to generate insights into the overall lifetime or into the time scale of persistence of structures. Section~\ref{sec:Applications} gives examples of concrete applications of these methods which demonstrate their practical utility in ongoing research. Section~\ref{sec:discussion} suggests a unifying mathematical generalization of the presented methodologies, thereby formalizing the assessment of persistence/coherence of meteorological structures, and section~\ref{sec:conclusion} draws conclusions and provides an outlook to future work.


\section{Meteorological phenomena and data} 
\label{sec:meteo_data}

Various detection and tracking methods are applied to scale dependent atmospheric phenomena in the next sections based on different meteorological data sets: In a case study of the 6th and 7th of August 2013, fronts on different scales are observed and analyzed using the COSMO-REA6 reanalysis data set \citep{bollmeyer2015}. Several examples of tropical cyclones are evaluated: These include idealized simulations of a single and a pair of cyclones using the EULAG code, \citep{prusa2008eulag}, as well as Hurricane ``Florence'' (2018) as a concrete case study based on ERA5 reanalysis data \citep{hersbach_era5_2020}. A synoptic-scale example is the extratropical cyclone ``Lothar'' (1999).


\subsection{Fronts}\label{subsec:front_data}

A meteorological front marks the boundary between large-scale air masses characterized by different values of largely homogeneous properties, such as temperature or humidity. The spatial and temporal scale of fronts in the atmosphere varies. Forecasters are primarily interested in synoptic-scale fronts, of length $\sim \qty{1\,000}{\kilo\meter}$ and width $\sim \qty{100}{\kilo\meter}$, as these are the main drivers of precipitation in the mid-latitudes. Indeed, \citet{CattoandPfahl2013} found that 50\% (and up-to 90\% locally) of extreme precipitation events occurred at or in proximity to a front in the mid-latitudes. Forecasters often manually position such fronts based on numerical weather prediction (NWP) output and sometimes use more technical algorithms. Synoptic fronts usually appear as elongated structures extending from the center of extratropical cyclones. Because of their elongated structure and because air masses pass through fronts as part of their dynamics, synoptic fronts can rarely be identified using Lagrangian coherent sets methods.

Fronts with smaller spatio-temporal extent are also found in the atmosphere and these are referred to as ``local fronts'' from here on. Local fronts are not associated with a large-scale extratropical cyclone but rather arise from numerous other sources. For example, uneven solar heating over mountains can lead to temperature gradients at orographic boundaries, \citep{Weissman2005Alpine}. Convection can also modify the local temperature and humidity field, thus creating temperature and humidity gradients. Local fronts typically have a shorter duration (1-6 hours) as well as a smaller spatial scale (convective or mesoscale, \ie, $\sim 10-\qty{100}{\kilo\meter}$), and remain largely stationary during their lifetime.


\subparagraph{Case Study: Fronts and convection over Germany 6 August 2013} Synoptic and local fronts will be analyzed based on the COSMO-REA6 reanalysis data set from the German Weather Service. The dataset has a horizontal resolution of about $0.055^{\circ}$, \ie, $\qty{6}{\kilo\meter}$, and an hourly time resolution. This high-resolution reanalysis data set has been developed based on the NWP model COSMO, for more details see \citet{bollmeyer2015}. In the course of the analysis, the temperature gradient and the dynamic state index (DSI) are calculated in the domain \qty{47}{\degree}-\qty{58}{\degree}N, \qty{2}{\degree}-\qty{18}{\degree}E on August~6,~2013.


\subsection{Tropical Cyclones}
\label{subsec:two_cyclones}

A tropical cyclone is an intensely rotating storm system that typically forms over the tropical oceans, is characterized by strong winds and low pressures, and is often accompanied by heavy rain. In the following sections, three examples for such storms with different levels of idealization will be analyzed.


\subparagraph{A single idealized tropical cyclone.} A single tropical cyclone in three spatial dimensions is considered. The data is generated by numerical simulations with the general purpose atmospheric flow model EULAG \citep[for further details see][]{prusa2008eulag,DoerffelEtAl2021}. The horizontal grid is chosen to cover \qty{4000}{\kilo\meter} with 192 grid point in each direction. The grid points are distributed to focus towards the center of the simulation domain to increase the effective resolution near the storm center. The minimum, maximum and mean grid spacings are \qty{10500}{\meter}, \qty{61700}{\meter}, and \qty{20900}{\meter}, respectively. The vertical grid covers equidistantly \qty{10000}{\meter} with 96 grid points.
The simulation is initialized with a tilted vortex of \qty{10}{\meter\per\second} maximum wind speed at the radius of maximum wind of \qty{100}{\kilo\meter}. The radial velocity profile relative to the storm center is uniform with height; baroclinicity is induced by the tilt of the centerline of \qty{160}{\kilo\meter} between bottom and top of the simulation domain. Output data are the time-dependent three-dimensional fields of the velocity components, which are post-processed to obtain values for the vertical component of relative vorticity ($\zeta = \omega_z$), the potential vorticity (PV), and for the Dynamic State Index (DSI).
This setup serves as a trivial test case to assess the ability of the subsequently discussed feature definitions and persistency measures in detecting tropical cyclones.



\subparagraph{Two Tropical Cyclones.} 
Analogous to the case of a single tropical cyclone, two tropical cyclones are simulated within a domain of \qty{8000}{\kilo\meter} $\times$ \qty{8000}{\kilo\meter}. Due to the increased complexity of the motion of the storms, static mesh refinement is not beneficial, hence we set 384 equidistantly distributed grid points in each horizontal direction. The vertical grid again covers \qty{10000}{\meter} but now with 192 grid points.
The first vortex is centered at (-\qty{500}{\kilo\meter}, \qty{500}{\kilo\meter}) with \qty{10}{\meter\per\second} maximum velocity at the radius of maximum wind $r_{mw} = \qty{200}{\kilo\meter}$, while the second cyclone is centered at (\qty{500}{\kilo\meter}, -\qty{500}{\kilo\meter}) with maximum velocity of \qty{20}{\meter\per\second} at $r_{mw} = \qty{100}{\kilo\meter}$. Both vortices are untilted.
This setup aims at testing the ability of the presented tracking methods to identify each storm as an individual entity even if temporal resolution is not sufficient for tracking them by, \eg, feature overlap.



\subparagraph{Case study: Hurricane Florence.} This intense and long-lived Hurricane in September 2018 with a maximal wind speed of \qty{150}{mph} and with the lowest pressure of \qty{937}{\hecto\pascal} caused catastrophic damage, especially in Cape Verde.
The analysis in section~\ref{sssec:trop_cyclones} of this storm is based on ERA5 Reanalysis data on a regular horizontal grid with a resolution of $\ang{0.25} \times\ang{0.25}$, \ie, about \qty{30}{\kilo\meter} \citep{era5pres,era5single}. To this end, the pressure field, the relative vertical vorticity, and the velocity fields in hourly time steps within the region of the rapid intensification of the hurricane, \ie, $10^{\circ}\text{N -- }\ang{30}$N latitude $\ang{315}\text{E -- }\ang{335}$E longitude, are converted into values at fixed height levels with a vertical resolution of \qty{0.2}{\kilo\meter}.
This data set offers a variety of interesting scenarios. As mentioned earlier, the early stages should be characterized by a ``pouch''-like flow structure that is readily identifiable by Lagrangian coherent set detection methods. Later stages, however, when the storm is fully developed, exhibit a more complex flow organization in which fluid particles are drawn in near the sea surface, lifted within the eyewall of the storm, and ejected in an asymmetric outflow layer above the storm \citep{Houze2009}. Thus, this dataset serves as a real-world example of persistent meteorological structures that challenges established detection methods and calls for a clean definition of what exactly are (the) physical properties that characterize a \emph{tropical storm}.




\subsection{Extratropical cyclones}\label{subsec:lothar}

Like tropical cyclones, extratropical cyclones are rotating storm systems characterized by low pressure and intense winds and precipitation. One key difference is that tropical cyclones are purely warm core and are not associated with synoptic fronts. While extratropical cyclones are generally less intense, they can produce wind speeds and extreme precipitation of a similar intensity to those of tropical cyclones. 


\subparagraph{Case study: Storm ``Lothar''.} As an example we consider the winter storm Lothar which was generated on December 24, 1999, in the western Atlantic, then travelled as a shallow low-level cyclone of moderate intensity until it hit Europe on December~26. Its high wind velocities caused material damages worth tens of billions of Euros and more than 50 people were killed~\citep[][]{wernli2002}. To enable the application of detection and tracking methods to this synoptic scale vortex, the ERA-Interim reanalysis data set on the $\qty{600}{\hecto\pascal}$ level with a horizontal resolution of about $\qty{80}{\kilo\meter}$ (T255 spectral, see \citet{dee2011era}) and a temporal resolution of $\qty{6}{\hour}$ is used for applying Lagrangian (or material) coherent set detection techniques. We consider the time range 1999-12-25 00:00:00 to 1999-12-27 18:00:00 during which storm Lothar traveled towards Europe. Shortly after Lothar, storm Martin followed on about the same storm track. For the same time period and level we also use the ERA 5 data set on an hourly basis and on a horizontal grid of about 30 km to calculate and analyze the DSI~\citep[][]{era5pres}.


\section{Defining features to detect meteorological structures}
\label{sec:FeatureDefinitions}

In this section, the atmospheric and weather events presented in section~\ref{sec:meteo_data} are formally characterized through the definition of specific features. These features are based on physical principles that distinguish the structure of interest, regardless of the degree of persistence of the structure.


\subsection{Detecting vortical structures}
\label{subsec:QCriterion}

Although vortices or vortical structures can be readily identified in the atmosphere by viewing visualizations of tornadoes, Kelvin--Helmholtz rolls etc., a definition of such a structure remains elusive. \citet{lugt1979dilemma} points out that the classical way of using the magnitude of vorticity $|\omega|$, with
\begin{equation}
	\omega = \nabla \times u\,,
	\label{eq:relative_vorticity}
\end{equation}
is insufficient because it misinterprets parallel shear flow with no rotation as a vortex. Over the years, numerous definitions have been proposed which can be roughly categorized based on the observer location as Eulerian, when individual snapshots in time are considered, or Lagrangian, when the analysis relies on path lines of fluid particles. The reader is referred to \citet{gunther2018state} for a review of the various methods and their advantages and disadvantages.

Some of the most popular approaches to vortex identification based on the velocity gradient tensor ($\nabla u$) involve the $Q$ \citep{hunt1988eddies}, $\lambda_2$ \citep{jeong1995identification} and $\Delta$ \citep{chong1990general} criteria. \citet{chakraborty2005relationships} found that all of these criteria identify similar regions as vortices if suitable non-zero thresholds are adopted. This shows that a particular choice among these criteria does not essentially affect the results once some guideline for the equivalence of thresholds has been established. Therefore, paying particular attention to the thresholding issue, we choose the $Q$-criterion to identify vortical structures in what follows. This criterion, as introduced originally, classifies spatial regions as vortices, when
\begin{equation}\label{eq:Q}
Q = \frac{1}{2} (\|A\|^{2} - \|S\|^{2}) > 0\,,
\end{equation}
where $A=\frac{1}{2}\left( \nabla u - (\nabla u)^T \right)$ is the (antisymmetric) vorticity tensor, $S=\frac{1}{2}\left( \nabla u + (\nabla u)^T \right)$ is the (symmetric) strain rate tensor, $\Vert\cdot\Vert$ is the Euclidean norm. This definition overcomes the shortcomings of choosing the vorticity magnitude, $|\omega|$, for vortex detection as it identifies regions where $A$ dominates over $S$, \ie, it finds regions with dominant rotation. In meteorological contexts, other criteria based on $\nabla u$ may be useful to determine unique properties. For instance, the kinematic vorticity number $W_{k}$ (see \citep{schielicke2016kinematic}) is useful to determine the size and circulation of a vortex. 


\subsection{Features for front detection}
\label{george_method}

Since fronts are often associated with clouds and precipitation, locating them is useful for forecasting and research purposes. Forecasters often manually position synoptic-scale fronts based on numerical weather prediction (NWP) output. However, as noted by \citet{RenardClark1965}, different forecasters produce different analyses and thus the final front location is often subjective depending on the forecaster. An additional problem is that archives of such analyses are sparse in both space and time. The study by \citet{RenardClark1965} was one of the first to recognize the need for objective front detection methods, and since then numerous methods have been implemented to automatically detect synoptic fronts in reanalysis data (see, \eg, \citet{Hewson1998}, \citet{Jenkner2010}, \citet{Schemm2015} and \citet{Rudisuhli2020}).

Following a similar approach to the studies cited above, the primary methods used to detect fronts in reanalysis data proceed by the following steps:
\begin{enumerate}

\item Input the two-dimensional horizontal slice of scalar temperature $\theta\colon \Omega\to \R$ at some vertical level, where $\Omega$ denotes the horizontal domain of interest, usually discretized by a two-dimensional (approximately) rectangular grid.

\item Smooth the $\theta$-field by $n_f$ iterations of a spatial weighted averaging procedure. On an equidistant Cartesian mesh this may read, \eg, as
	\begin{equation}
	\theta_{i, j}^{n}=\frac{1}{2} \theta_{i, j}^{n-1}+\frac{1}{8}\left(\theta_{i+1, j}^{n-1}+\theta_{i-1, j}^{n-1}+\theta_{i, j+1}^{n-1}+\theta_{i, j-1}^{n-1}\right), \quad n=1, \ldots, n_{\mathrm{f}}\,.
	\end{equation}

\item A point $x \in \Omega$ is then said to belong to a front if, for a certain chosen gradient magnitude threshold $\tau$ and some numerical approximation of the horizontal gradient, $\gradhor$, we have
	\begin{equation}
    \Vert \gradhor \theta^{n_f}(x)\Vert > \tau.
    \end{equation}
\end{enumerate}
While this method seeks to remove the subjective positioning of fronts, the inputs (\ie, scalar field, height level, level of smoothing, gradient thresholds) must still be chosen by the user eventually. Recent work of \citet{ThomasandSchultz2019} investigated the sensitivity of selecting different inputs in detail by comparing automatic front detection output. They found that using a potential temperature (PT) field with a \qty{2}{\kelvin} per \qty{100}{\kilo\meter} threshold at \qty{850}{\hecto\pascal} most closely  represented  frontal features  commonly associated with  extratropical  cyclones. Here, we use a PT field with a \qty{3}{\kelvin} per \qty{100}{\kilo\meter} gradient at \qty{800}{\hecto\pascal}. The increased temperature gradient threshold is to account for the increased spatial resolution of the dataset being used (\qty{6}{\kilo\meter}). Much of southern Germany lies above sea-level, so a higher height level is used to limit the need for downward interpolations and thus approximations. The level of smoothing needed to eliminate irrelevant small-scale noise is highly dependent on resolution of the dataset and requires manual tuning. Following the tuning, the temperature field was smoothed 100 times. Here, we do not use velocity thresholds as we do not seek to investigate the type of front at this stage.

\begin{figure}
	\centering
	\includegraphics[width=0.75\textwidth]{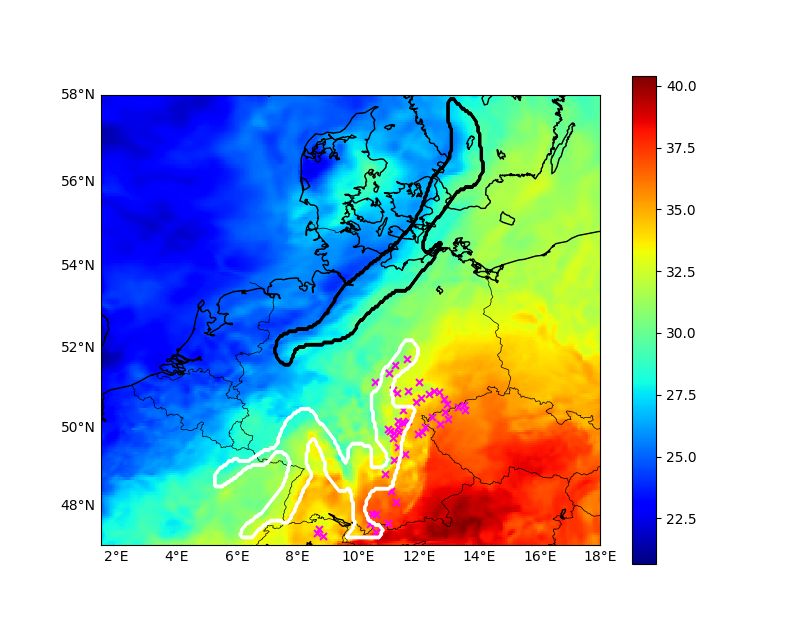}
	\caption{Synoptic front (black contour) and local front (white) detected in the potential temperature field (Celsius; shaded) over Germany on 6 August 2013 at 17UTC. Pink crosses represent locations of convective cells (see section~2.2 of \citet{WAPLER2015})}
	\label{synop_local_cells}
\end{figure}

Despite appropriate inputs, local fronts can still sometimes be detected along with the synoptic-scale front of interest. Local fronts are likely to be detected in high-resolution reanalysis datasets, which can resolve small scale features. As mentioned in section \ref{subsec:front_data}, local fronts typically have a shorter duration (\qty{1}{\hour} to \qty{6}{\hour}), a smaller spatial scale (convective or mesoscale) and remain more stationary. Thus, duration, size, and stationarity thresholds can be defined to separate local and synoptic-scale fronts \citep{Rudisuhli2020}. In regions of strong convection that self-organizes on larger scales in ``mesoscale convective systems'', however, the temperature field can experience rather chaotic and rapid changes  (cf.~figure~\ref{synop_local_cells}). Such features can last for several hours. In such cases, duration, size, and stationarity thresholds may not prove as effective, especially when considering a limited domain (\eg, Germany; \qty{47}{\degree}-\qty{58}{\degree}N, \qty{2}{\degree}-\qty{18}{\degree}E) as the duration of a synoptic front within this domain could be shorter than its total lifetime. An alternative would be to perform the front detection on a larger domain, but at a computational expense. In section~\ref{subsec:class_synop_loc} we introduce optimal transport as an alternative method for classifying synoptic and local fronts, and test it against the case of strong convection over Germany on August~6,~2013.


\subsection{The Dynamic State Index: A tool for diagnosing atmospheric processes}
\label{subsubsec:dsi}

The Dynamic State Index is a scalar field that can be used as an input parameter to identifying vortices or fronts by thresholding the field. The DSI, in its original version, indicates local deviations of the atmospheric flow field from a stationary, adiabatic, and inviscid solution of the non-hydrostatic compressible governing equations~\citep{Nevir2004}. It follows that the DSI can be used to diagnose whether a given atmospheric flow hosts non-steady, diabatic, or frictional processes, and it is applicable independent of scale. Thus, \eg, \citet{Weber2008storm} show how a characteristic dipole structure of the DSI indicates the presence of hurricanes on the meso-scale. On the convective scale, \citet{Claussnitzer2008,Gassmann2014,Weijenborg2015}  have shown that the DSI is strongly correlated with intensive precipitation processes. \citet{Mueller2018,Mueller2019} recently introduced DSI variants for the quasi-geostrophic model and for the Rossby model. Both studies confirm the high correlation of the DSI to precipitation patterns and its applicability to identify persistent weather situations, so-called blockings, on the synoptic scale.

In general, atmospheric processes can be described by energetic, thermodynamic, and vorticity-related quantities. The Dynamic State Index summarizes this information into one scalar. In this work, we consider the  Dynamic State Index $\mathrm{DSI}_{\rm PE}$ derived from the primitive equations. The DSI is given by the  Jacobian determinant: 
\begin{equation}
\mathrm{DSI}:= \frac{\partial(\Theta,B,PV)}{\partial(a,b,c)} = \frac{1}{\rho}\frac{\partial(\Theta,B,PV)}{\partial(x,y,z)}
\label{eq:DSI_PE}
\end{equation}
where $(x,y,z)$ are Cartesian space coordinates, and $(a,b,c)$ are Lagrangian mass coordinates related to the former by $dm = da \ db \ dc = \rho  dx \ dy \ dz$, so that the mass density $\rho$ takes the role of the Jacobian determinant. Moreover, $\Theta$ is the potential temperature, $PV$ denotes Ertel's potential vorticity given by
\begin{equation}
PV =  \rho^{-1}\boldsymbol{\omega}_a \cdot \nabla \Theta ,
\label{eq:EPV}
\end{equation}
where $\boldsymbol{\omega}_a = \nabla \times \mathbf{v} + 2{\bm\Omega}_E$ denotes the 3D absolute vorticity vector with 3D velocity $\mathbf{v}$ and angular velocity of the Earth~${\bm\Omega}_E$. Moreover, $B$ denotes the Bernoulli stream function defined by
\begin{equation}
B= \frac{1}{2}\mathbf{v}^2+\phi+h,
\end{equation}
where $h$ denotes the enthalpy expressed by $h= e+pv$ with internal energy $e$, pressure $p$ and specific volume~$v = 1/\rho$. 


\subparagraph{The DSI as an indicator of fronts and extratropical cyclones}
The Dynamic State Index \eqref{eq:DSI_PE} is calculated for the time period Dec.~23 to Dec.~30, 1999. Within this time period, storm Lothar formed in the Atlantic and crossed France and Germany on the 26th, followed by storm Martin which passed these regions one day later. The corresponding DSI field, based on the ERA~5 data set described in section~\ref{subsec:lothar}, is shown in figure~\ref{fig:Lothar_DSI}. For an animation of the DSI during the entire time period see the supplementary material. Increasing the resolution leads to finer DSI dipole structures as shown, \eg, in \citep{Mueller2018,Mueller2019}. 


\begin{figure}[h!]
	\centering
	\includegraphics[width=.7\linewidth]{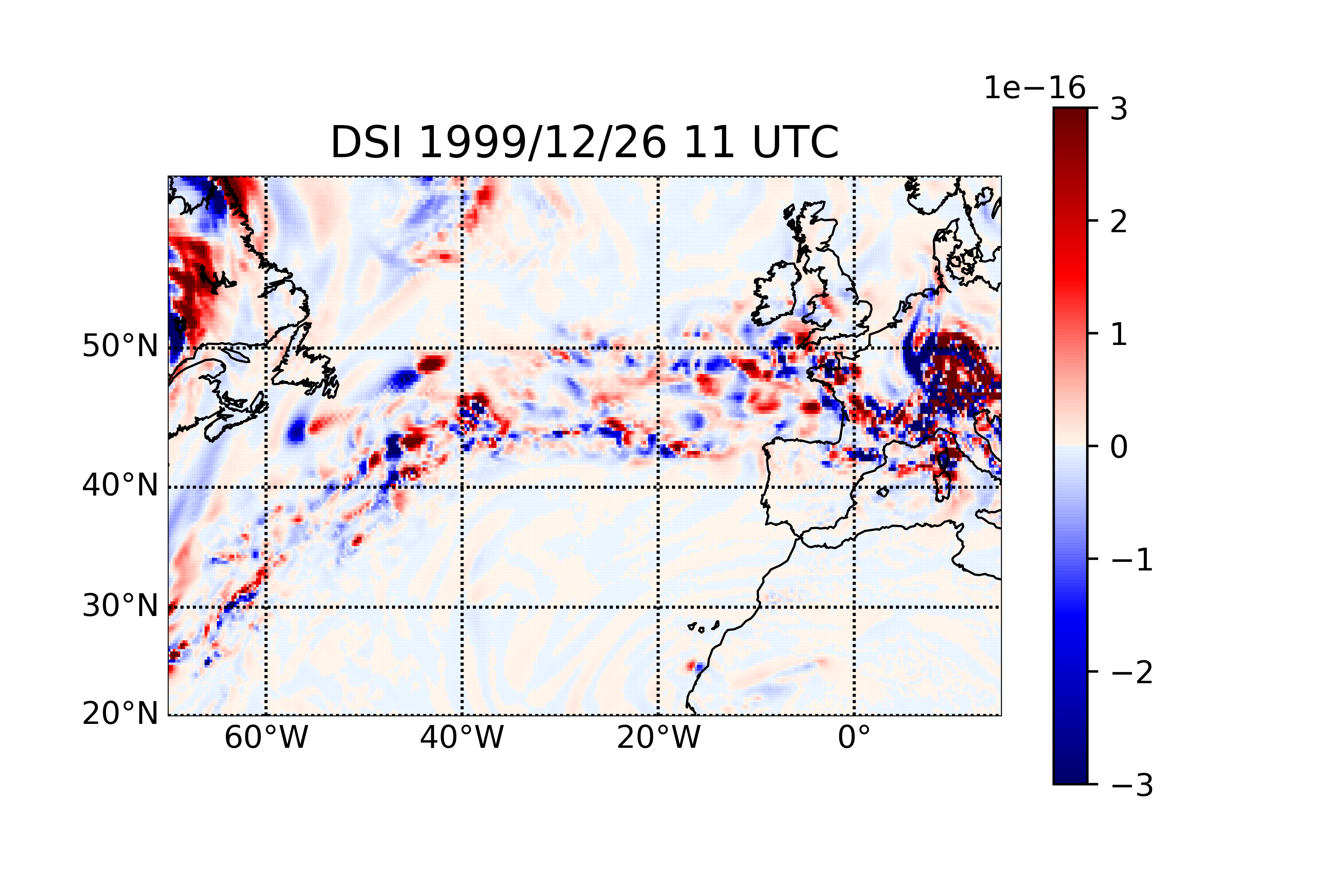}
	\caption{The DSI field indicates storm Lothar crossing southwest Germany. At this time storm Martin crossed the Atlantic as indicated by the distinct dipole structure eastwards of Canada.}
	\label{fig:Lothar_DSI}
\end{figure}


\subsection{Tropical cyclones: formalizing the feature}
\label{sssec:trop_cyclones}

Among the most powerful disturbances in the atmosphere are tropical cyclones (TCs).
A TC forms a cloud system of nearly axisymmetric or spiral structure around the vortex core that transforms sensible heat from the ocean surface and latent heat from cloud water condensation into kinetic energy of the dominant swirling wind fields, called the ``primary circulation''. Although there exists meteorological understanding of \emph{what a TC is}, \ie, there is a notion of a persisting structure evolving in space and time, a clear mathematical definition still is missing. This is partially due to the complex three-dimensional nature of such flows involving the overturning \emph{secondary circulation} which is directed radially inwards at low altitudes, up- and downwards at intermediate heights near and far from the vortex center, respectively, and predominantly outwards at high altitudes. And partially it is due to the fact that only lately researchers have gained access to high-resolutions three-dimensional simulation and reconnaissance data allowing for unprecedented in-depth exploration of more intricate structural properties.

There is a common understanding on the structure of a TC, however: Boundary layer air is transported inwards in spiral motions until it gets to be lifted in the convective ``eyewall'' region and subsequently pushed outwards on top of the TC. This, in particular, is the reason why a TC does not obviously fit into the framework of ``Lagrangian coherent structures'' \citep{Haller:2015kj,SchneideEtAl2019} or ``coherent sets'' \citep{Fro15,FrRoSa19} often used to characterize distinctive flow patterns in turbulent flows and other contexts.

In general, TCs are multiscale phenomena stretching from water phase transition processes and small scale turbulence over the lifecycle of individual clouds to the organized structure of the TC cloud system forming a cascade of scales in space and time. Due to the multi-scale nature of tropical storms, their prediction by means of numerical simulations remains a key challenge of modern weather forecasting. To remedy, simplifying the underlying equations, as \citet{PaeschkeEtAl2012} did, may lead to more insights on the physical mechanisms involved as well as open a route to predictions that need less computational resources.

\begin{figure}[htb]
 \centering
 \includegraphics[width=0.5\textwidth]{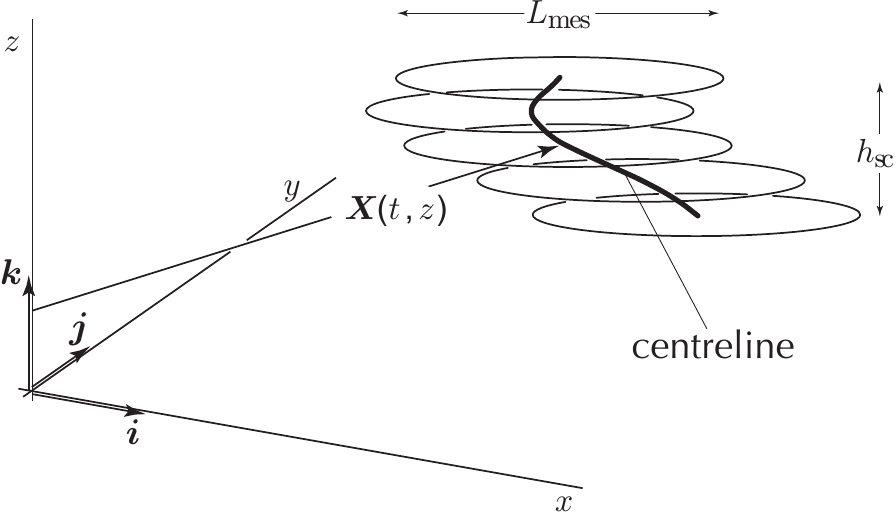}
 \caption{Idealized flow structure of a tilted TC (adopted from \citet{PaeschkeEtAl2012}).}
 \label{fig:tilted_vortex_schematic}
\end{figure}

The principle idea of \citet{PaeschkeEtAl2012} is to construct a TC as a nearly axisymmetric vortical flow organized along a \emph{tilted centerline} as schematically depicted in \autoref{fig:tilted_vortex_schematic}, and to construct an approximate set of governing equations by means of a matched asymptotic analysis.
Ultimately, they derived a closed set of equations describing the time evolution of the azimuthally averaged circumferential velocity $u_\theta(r,z,t)$ and the centerline coordinate $\vect X(t, z)$.
That is, there is a simplified model based on the structure of the storm. For the sake of assessing the predictive capability of the model against reanalysis data, a representation of the idealized structures of the storm within the non-idealized three-dimensional reanalysis data of \autoref{subsec:two_cyclones} is needed. In the spirit of the present paper, we therefore aim to describe the persistent structure of a tropical cyclone by the idealizations of \citet{PaeschkeEtAl2012} which, on their end, have been developed with empirical insights into the structure and symmetries of tropical cyclones in mind.

In a process of reverse-engineering we thus make use of the framework of \citet{PaeschkeEtAl2012} to identify tropical cyclones as nearly axisymmetric flow structures aligned along a tilted centerline and embedded into a more general large-scale atmospheric environment. The centerline is identified in this context by the centroid of some vorticity-based scalar field, $\omega$, in horizontal planes
\begin{equation} \label{eq:centerline}
 \vect X(t,z) =\frac{1}{|\Omega_p(z,t)|} \int\limits_{\Omega_p(z,t)} \begin{pmatrix} x \\ y \end{pmatrix} \omega(x,y,z,t) \, dx \, dy\,.
\end{equation}
The vertical component of vorticity defined in \eqref{eq:relative_vorticity} or the potential vorticity as defined in \eqref{eq:EPV} are reasonable choices for $\omega$, but so are the vortex criteria $\lambda_2$ or $Q$ as discussed in section~\ref{subsec:QCriterion}. In~\eq{eq:centerline}, $\Omega_p(z,t)$ denotes subsets of horizontal slices at height $z$ satisfying
\begin{subequations}
	\begin{align}
		i)\quad &\Omega_p(z,t) = \{ (x,y) \in \mathbb R^2: p^*(x,y,z,t) < p_{\rm thr} \} \\
		ii)\quad &\Omega_p \mbox{ is convex}\,, \label{eq:convexity}
	\end{align}
\end{subequations}
where $p_{\rm thr}>0$ is a threshold for the normalized pressure perturbation
\begin{equation} \label{eq:pres_perturbation}
p^*(x,y,z,t)=\dfrac{p(x,y,z,t)-p_{\rm min}(z,t)}{p_{\rm max}(z,t)-p_{\rm min}(z,t)}\,,
\end{equation}
with
\begin{equation}
		p_{\rm min}(z,t) = \min_{x,y}(p(x,y,z,t))\,,  \qquad 
		p_{\rm max}(z,t) = \max_{x,y}(p(x,y,z,t))\,.
\end{equation}
Hence, by appropriately choosing $p_{\rm thr}$, one ensures that $\Omega_p$ forms a domain of closed isobars.
The condition \eqref{eq:convexity} serves as an upper bound constraint on $p_{\rm thr}$.
With $p^*$ sufficiently smooth near a local minimum, a quadratic Taylor approximation yields convex level sets.
Higher-order deviations may impair this property as we depart from the minimum.
Therefore, there must exist a $p_{\rm thr}^c \in (0,1]$ for which level sets with $p^*< p_{\rm thr}^c$ are convex.
In the further analysis, we empirically ensure that $p_{\rm thr} < p^c_{\rm thr}$.
Typically, a value of $p_{\rm thr} = 0.1$ is chosen.
This definition of a tropical cyclone is in line with the notion of characterizing TCs by the pressure minimum near the center of the storm.

In the literature on tropical cyclones, they are commonly characterized in size by either the ``radius of gale force winds'' or the area enclosed by the ``outermost closed isobar'' of sea surface pressure \citep{merill1984}. The latter accounts for the depression that is associated to the force balance between pressure gradient on the one hand and the sum of Coriolis and radial forces on the other hand. The radial pressure gradient hence is necessary for maintaining the primary circulation of the storm. Restricting to sea surface pressure, however, neglects the vertical structure of the storm, may be distorted by superimposed large-scale pressure gradients, and typically overestimates the region that is dominated by the primary circulation. Therefore, sea-surface depression gives at most an estimate of the horizontal position of the storm, \ie, it is a necessary condition for the existence of a TC, but does not suffice to solely serve as a defining feature for the purpose of sections \ref{sec:Persistence} and \ref{sec:Applications}.

To overcome these limitations we analyze the pressure through the depth of the atmosphere and introduce an additional requirement based on a vortex criterion as discussed above. We define the feature of the \emph{vortex core} by the intersection
\begin{equation}
	\Omega(z,t) := \Omega_p(z,t) \cap \Omega_\omega(z,t)\,,
	\label{eq:feature_TC}
\end{equation}
where
\begin{equation}
	\Omega_\omega(z,t) := \{ (x,y) \in \mathbb R^2: \omega(x,y,z,t) > \omega_{thr} \}
	\label{eq:Omega_omega}
\end{equation}
with a threshold value $\omega_{thr}>0$. The feature definition in \eqref{eq:feature_TC} poses a multi-field criterion that is supposed to ensure a well-defined centerline by conditioning the integration of the vorticity variable on areas with sufficiently low pressure. This removes vorticity perturbations that are due to nearby fronts or other smaller scale events not belonging to the dominant vortex structure. In addition, this definition of $\Omega$ constrains  the integration domain also in the vertical direction because closed low pressure isobars disappear at higher altitudes.

The construction of the centerline above (cf.~\eqref{eq:centerline}) took $\Omega_p$ into account to limit the integration domain. If $\omega$, \ie, the integration weight in \eqref{eq:centerline}, encodes a criterion that ensures positivity in regions of vortical flow and is zero elsewhere, replacing $\Omega_p$ by $\Omega$ leads to the equivalent definition of the centerline $\smash{ \vec X }$ (since the integral is restricted to regions where $\omega>0$, cf.~\eqref{eq:Omega_omega}). Hence, a TC is identified by $\Omega$ as a vortical structure characterized by a well-defined centerline $\vect X(z,t)$ at a given instance in time. A notion of coherence, \ie, the persistence over a certain period in time, would allow one to characterize the lifecycle of a TC from the point where a \emph{coherent} flow forms until its dissipation or, possibly, its transition to a weather system with decidedly different characteristics occurs.


\section{Methods for the assessment of structural persistence}
\label{sec:Persistence}

Having defined different types of atmospheric structures with appropriate sets of formalized features, the next task is to quantitatively assess their persistence or coherence. Towards that goal, this section presents approaches to tracking level sets of scalar feature functions, hence enabling the user to not only follow the dynamical evolution of structures automatically, but to also quantify the temporal persistence of structures. In addition, approaches are presented to assessing the material coherence of structures, and to quantifying the degree of deformation of structure boundaries through their dynamical evolution.


\subsection{Tracking algorithms based on volume overlap}
\label{sssec:Persistence_overlap}

Volume overlap is a local tracking technique that extracts individual structures or features of a scalar field and seeks to match them with structures in subsequent timesteps based on their overlap. Tracking of a given structure progresses only when a certain overlap threshold, denoted as $\tau_{\rm overlap}$, is met or exceeded. This type of technique was applied to track cloud patterns in Meteosat infrared images \citep{arnaud1992automatic}, structures of vorticity magnitude in isotropic turbulence \citep{silver1995object, silver1996volume, silver1997tracking, silver1998tracking}, hairpin vortex packets in channel flow and compressible turbulent boundary layers \citep{o2009chasing}, and quadrant events in channel flow \citep{lozano2014time}. A major problem associated with this technique is that this method depends on a rather high temporal sampling rate to track a structure. As noted by \citet{lozano2014time}, some structures which move too fast or which are small enough will not be tracked. \citet{ji2006feature} introduced a technique based on the Earth Mover's Distance (EMD) or more commonly referred as Wasserstein distance which is closely related to the optimal transport problem, to resolve these issues and obtain a globally best matching result. 

In the following subsections, we describe two tracking approaches based on volume overlap. In section~\ref{subsubsec:simpleOverlap}, the standard volume overlap method with two different structure extraction techniques is explored. For cases where no overlapping regions can be identified in a future timestep, an alternative overlap method combined with image registration is shown in section~\ref{subsubsec: volumeOverlapIR}.


\subsubsection{Simple volume overlap}
\label{subsubsec:simpleOverlap}

Every tracking technique requires two principal steps, \ie, \textit{extraction} of the objects to be tracked and \textit{tracking} itself.

\begin{itemize}
\item [I.] \textit{Extraction: } The input is a time series of some $d$-dimensional (typically $d=2$ or $d=3$) scalar field $\alpha$ that is suitable for structure identification, for instance, vorticity magnitude, $\lambda_2$, $Q$-criterion, or $|DSI|$. We define
\begin{align}\label{eq:volOverlapGeneral}
	\Omega(t) &= \{ (x,y,z) \in \mathbb R^3: \alpha(x,y,z,t) > \tau \},
\end{align}
where $\tau$ represents an appropriate user-defined threshold.
The reader should note that, sometimes, a single condition is insufficient to identify the required structure. For instance, $Q > 0$ does not automatically correlate with low pressure and thus does not necessarily focus on the strongest vortices~\citep{jeong1995identification}. Therefore, an additional condition such as a normalized pressure perturbation as shown in \eqref{eq:pres_perturbation} can be enforced to ensure the existence of low-pressure regions within $Q$-criterion structures. Once the necessary conditions are introduced, individual structures are then identified as connected regions of $\Omega(t)$. 
	
\vspace{0.25cm}
	
In this work, two types of extraction techniques are used to separate the connected regions. The first is the \textit{contour-tree segmentation} which tracks contours of a level set \citep{Carr2003contour}. This has been successfully used to extract the Hurricane Florence as shown in section~\ref{subsec:Florence_cyclones}. The second is a \textit{Neighbor Scanning} approach based on the work of \citet{moisy2004geometry}. In this approach, every point on the scalar field is scanned successively. If one considers a point to be the center of a $3 \times 3$ cube, then its neighbors are the 26 points, directly adjacent across faces, edges, and vertices, that surround it. A limitation of this algorithm is that it is unable to distinguish between structures when they lie very close to each other. This is due to the fact that the Neighbor Scanning algorithm is unaware of how the final surface mesh based on the chosen physical criteria is going to be constructed for visualization. Therefore, the Neighbor Scanning algorithm is modified by incorporating the surface information from the popular \textit{Marching Cubes (MC)} algorithm. For further details on the implementation, please refer to \citep{harikrishnan2021geometry}. This extraction algorithm has been used to extract extratropical cyclones, weather fronts, and Tropical cyclones as shown in sections~\ref{subsec:results_lothar}, \ref{subsec:class_synop_loc}, and \ref{subsec:results_segm_cyclones}, respectively. For complex data, $\Omega$ identifies several 3-dimensional regions which could be represented as connected components
\begin{align}
	\Omega(t) &= \cup_{i\in I_{t}} \Omega_t^i
\end{align}
with $I_t = \{1,...,N_t\}$, where $N_t$ is the number of connected components in time frame $t$.

\item [II.] \textit{Tracking: } This part of the algorithm deals with the overlap of structures found at consecutive time steps. The latter can be assessed in at least two ways: (i) a specific structure (hereafter referred to as a main structure), which is user-defined, can be tracked, or (ii) all structures within the domain can be extracted and tracked. The former is beneficial when tracking specific features such as cyclones, whereas the latter is useful for understanding branching events which have been commonly observed in channel flow turbulence \citep{lozano2014time}. Following \citet{samtaney1994visualizing}, we reiterate the following structure interactions which may occur during its lifetime,
	
\begin{itemize}
\item \textbf{Creation:} A structure is created at $t_{n}$.
\item \textbf{Continuation:} The created structure overlaps with another at $t_{n + 1}$.
\item \textbf{Bifurcation:} It may also continue as group of structures at $t_{n + 1}$. In this case, the structure with the largest overlap is followed, if it also satisfies the overlap criterion.
\item \textbf{Amalgamation:} A group of structures at $t_{n}$ merge into a single structure at $t_{n + 1}$.
\item \textbf{Dissipation:} No structure can be matched at $t_{n + 1}$.
\end{itemize}
	
\vspace{0.25cm}
	
\textit{Individual structures:} Here, only the main structure is tracked. We manually select the area of interest (the connected component that refers to a desired structure) in the first time step $\Omega_1^s,$ $s\in I_1$ and exploit the fact that the regions in two consecutive time steps often overlap.  The overlap condition is formulated in terms of the ``percentage of overlap'' measured using the Dice Similarity Coefficient ($\DSC$) \citep{dice1945measures},
\begin{equation}
\label{eq:DSC}
\DSC(X, Y) = \frac{2|X \cap Y|}{|X| + |Y|}\,,
\end{equation}
where $|\Omega|$ denotes the volume of the domain $\Omega$. The value of $\DSC$ ranges from $0...1$, with a value of unity indicating perfect overlap. The overlap criterion then reads
\begin{equation}
\DSC(\Omega_{t}^s,\Omega_{t+1}^i)>\tau_{\rm overlap}\,.
\end{equation}
For two time steps $t, t+1$ the structure $\Omega^i_{t+1}$ for some $i\in I_{t+1}$ that satisfies the overlap condition with $\Omega^s_{t}$ is relabeled to $\Omega_{t+1}^s$. Then $\Omega^s = \{\Omega_1^s, ... ,\Omega_T^s\}$ defines the set of structures tracked over the time. \\

\textit{Branches of structures:} In this case, the main structure and all its interactions are tracked. The steps are similar to the previous case. However, when a structure bifurcates, all structures at $t_{n + 1}$ satisfying the overlap condition are tracked. When only one interaction occurs with the main structure, the event is termed as \textit{simple branching}. In a \textit{complex branching} scenario, the interactions may themselves have interactions of their own.
	
\end{itemize}

In general, this method is computationally inexpensive and has been successfully used to extract and track weather fronts, vortex cores for synthetical tropical cyclones, and hurricane Florence. In some specific cases, however, if extracted structures are small and/or moving fast and the data set does not have an appropriate time resolution, no overlapping areas can be determined. To overcome this issue, another overlap tracking methodology combined with \textit{image registration} is discussed in the subsequent subsection.


\subsubsection{Volume overlap with image registration}
\label{subsubsec: volumeOverlapIR}
\begin{figure}[h!]
	\includegraphics[width=\linewidth]{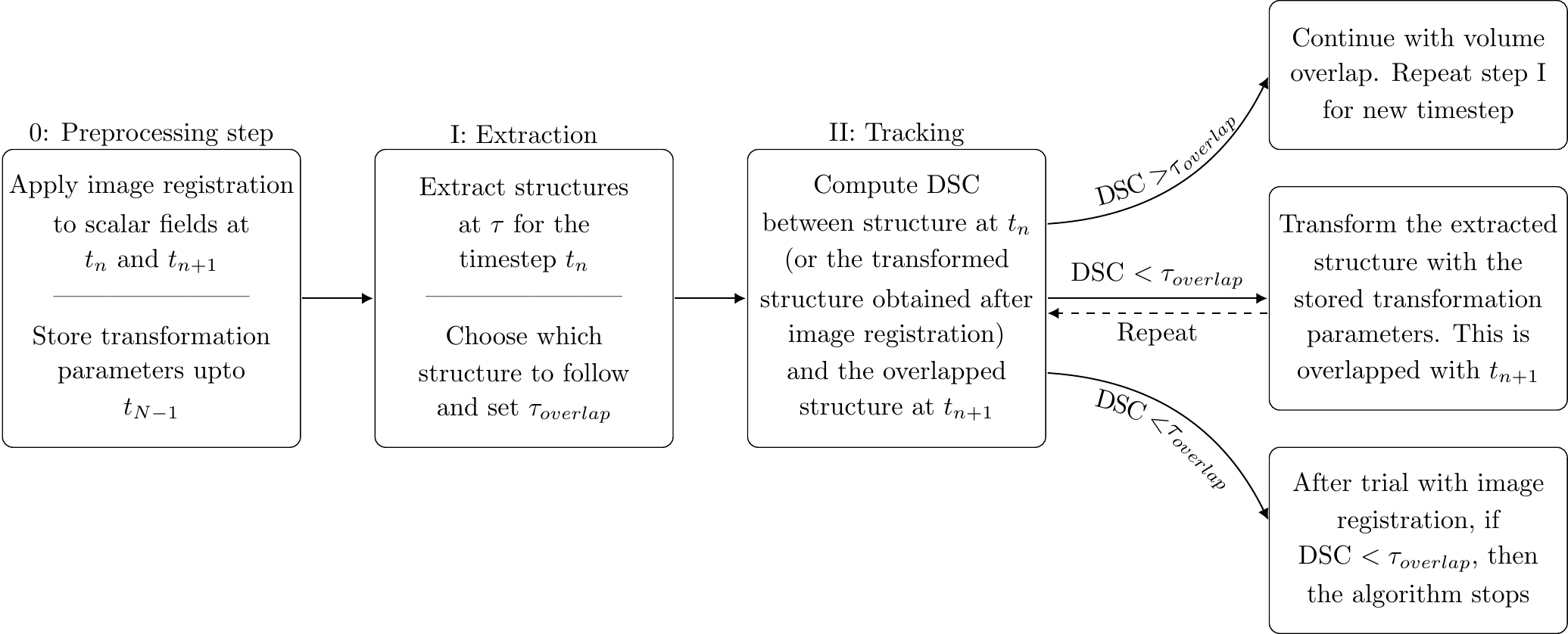}
	\caption{Steps involved in the tracking algorithm. Here, $t = \{n, n+1, .., N\}$ denotes the timesteps, $\tau$ is a user-defined threshold, $\tau_{\rm overlap}$ is the user-defined overlap threshold and DSC is the Dice Similarity Coefficient.}
	\label{fig:volumeOverlapSteps}
\end{figure}

Here we propose a technique similar to that of \citet{Badel2021} by combining volume overlap with Medical Image Registration (MIR). This combines the speed and simplicity that comes with the volume overlap and the ability to track small/fast objects independently of an overlap condition by computing the globally best matching results with MIR. The entire algorithm is written as a python module for the visualization software \emph{Amira ZIB edition} \citep{Stalling2005}. For MIR, we choose the open-source software \emph{elastix v4.8} \citep{KleinStaring:2010}. The approach involves the following three steps, which we also summarize in figure~\ref{fig:volumeOverlapSteps}:

\begin{itemize}

\item [0.] \textit{Preprocessing step:} Suppose we have a $d$-dimensional time-varying dataset with $N$ timesteps. MIR is initially applied between a timestep $t_{n}$ and a subsequent one $t_{n + 1}$. In MIR terms, the first timestep is called a \textit{fixed image} ($I_{F}$) and the second, a \textit{moving image} ($I_{M}$). This is followed by applying MIR to the next consecutive timesteps, \ie, $t_{n + 1}$ and $t_{n + 2}$ and repeated until $t_{N}$. By applying MIR, $I_{M}$ is transformed to fit $I_{F}$ for every instance, \ie, $\mathcal{T} : \Omega_{F} \subset \R^{d} \rightarrow \Omega_{M} \subset \R^{d}$, where $\mathcal{T}$ is the transformation, $\Omega_{F}$ and $\Omega_{M}$ correspond to the spatial domain of the fixed and moving image respectively. The goal in constructing $\mathcal{T}$ is to make $I_{M}(\mathcal{T}(x))$ as similar to $I_{F}(x)$ as possible. Although elastix allows numerous transformations (See \citep{KleinStaring:2010}), we found that nonrigid transformations, specifically B-spline, work best with the present meteorological datasets. The suitability of a transformation for a dataset depends on its complexity. One can check this objectively by computing the overlap between the deformed moving image $I_{M}\circ \mathcal{T}$, where $\circ$ denotes the composition of functions, and $I_{F}$ with a similarity measure such as DSC (see eq.~\eqref{eq:DSC}). For our dataset, DSC was computed for all available transformations. The reader should note that this does not include composite transformations, for instance translation $+$ affine. We choose B-spline as it has the highest DSC value. Since B-spline transformation is an ill-posed problem, $\mathcal{T}$ is penalized with a cost function $\mathcal{C}$ such that,
\begin{equation}
\label{eq:ImageRegistrationEq1}
	\hat{\mu} = arg \min_{\mu} \mathcal{C}(\mathcal{T}_{\mu}; I_{F}, I_{M})
\end{equation}
where the parameter vector $\mu$ contains values of the \textit{transformation parameters}. These are stored for every timestep up to~$t_{N-1}$.
	
\item [I.] \textit{Extraction:} Let us take the fixed image $I_{F}(x, t)$ at some time $t$. The connected regions or structures are extracted with the Neighbor Scanning + MC algorithm as described in section~\ref{subsubsec:simpleOverlap} at a user-defined threshold~$\tau$.
	
\vspace{0.25cm}
	
\item [II.] \textit{Tracking:} We manually select the structure $I_{F}(x, t)$ to follow. Then, the algorithm checks the same space of the structure at $t_{n + 1}$ and if a structure is found, it is extracted and the overlap is computed. If the overlap criterion is satisfied, the tracking proceeds as a \textit{simple volume overlap}. If not, the algorithm tries to find the appropriate structure with MIR. In this case, the extracted structure is copied onto an empty scalar field of the same dimension and size, denoted by~$I_{FM}$. The \emph{transformix} part of the \emph{elastix} software, which is responsible for identifying the transformations $\mathcal{T}$ between structures, is called with $\mu$ at $t_{n}$ and~$I_{FM}$. This generates another scalar field, $I_{MM}$, which has the transformed structure position. Finally, $I_{MM}$ is overlapped with $I_{M}$ to find the structure position at $t_{n + 1}$. This is repeated for all subsequent timesteps.
	
\end{itemize}

\noindent While the combination of volume overlap and image registration is successfully applied to track extratropical cyclones in section~\ref{subsec:results_lothar}, it is computationally expensive. The computational complexity lies in the additional preprocessing step.


\subsection{Material coherence - coherent sets}
\label{subsec:material_coherence}

As its name readily suggests, material coherence is a property of an object that is advected by a flow. Thus, it takes into consideration the transport of passive tracers without inertia by the atmospheric flow $\phi_t$ given by the time-dependent differential equation
\begin{equation}
    \label{eq:ODE}
    \dot{x}_t = v(t, x_t).
\end{equation}
Here, $v$ is the two- or three-dimensional wind field and $\phi_t$ is the flow from time $0$ to $t$ generated by \eqref{eq:ODE}, such that~$\phi_t x_0 = x_t$. We consider the flow on the interval $[0,T]$, \ie, on a finite \emph{time horizon}. The consideration of flow field is often restricted to a bounded and connected (full-dimensional) submanifold $\Omega:= \Omega_0 \subset \R^d$, and its evolution by the flow,~$\Omega_t := \phi_t\Omega$. We assume that $\phi_t$ preserves the $d$-dimensional volume measure $\mathrm{vol}_d$ for all times~$t$, although a generalization is possible by suitably adapting the volume measure~\citep{FrKw20}. The method from~\citep{BaKo17}, used here, incorporates this adaption.

\emph{Coherent sets}, as defined by~\citet{Fro15}, are subsets of $\Omega$ (and their time-evolved versions by the flow) that strongly resist filamentation by the dynamics. More precisely, $X\subset \Omega$ is called a coherent set if the ratio of time-averaged perimeter of the evolution of the set, $\frac{1}{T}\int_0^{T} \mathrm{vol}_{d-1}\left( \phi_t (\partial X) \right)\,dt$, and the minimal volume of the two-partition that it generates (\ie, the interior and exterior of the set), $\min\{\mathrm{vol}_d(X), \mathrm{vol}_d(\Omega\setminus X) \}$, is small:
\begin{equation}
	\label{eq:cohset}
	\frac{\frac{1}{T}\int_0^{T} \mathrm{vol}_{d-1}\left( \phi_t (\partial X) \right)\,\mathrm{d}t}{\min\{\mathrm{vol}_d(X), \mathrm{vol}_d(\Omega\setminus X) \}} \quad \text{is small.}
\end{equation}
Here, $\mathrm{vol}_{d-1}$ denotes the $(d-1)$-dimensional surface measure, and $\partial X$ denotes the boundary of the  set~$X$. Note that such a smallness requirement prevents a coherent set (or its complement $\Omega\setminus X$) to be too small in volume (\eg, point-like). Also, note that this ratio has the physical dimension of $1/\text{length}$, thus favoring larger sets over small ones. If additionally to advection by~\eqref{eq:ODE} the tracers would underlie some small diffusion as well, their diffusive leakage out of the material set~$\phi_t X$ would be proportional to the perimeter of the set~$X_t = \phi_t X$, cf.~\citep{schillling2021heat}. Thus, the intuition behind the definition \eqref{eq:cohset} is that the sets  $X_t$ should mix barely with their exterior even if the dynamics is perturbed by small (Gaussian) noise.

The natural question is how to find sets $X$ satisfying~\eqref{eq:cohset}. One could think of minimizing this ratio, however such a problem would not be computationally tractable. Hence, a relaxation of the \emph{set-based} optimization problem into a \emph{functional} optimization problem is used, cf.~\citep{Fro15}. The latter problem is solved by eigenfunctions of the partial differential eigenvalue equation
\begin{equation}
    \label{eq:cheeger}
    \Delta^{\mathrm{dyn}} f = \lambda f \text{ on }\Omega,
\end{equation}
where the operator $\Delta^{\mathrm{dyn}}$ is called the \emph{dynamic Laplacian}, and is given by
\begin{equation}
    \label{eq:DL}
    \Delta^{\mathrm{dyn}} f := \mathrm{div} ( M \nabla f), \quad\text{where } M(x) = \frac{1}{T}\int_0^{T} \big(W_x(t)^TW_x(t)\big)^{-1}
dt,
\end{equation}
with $W_x(t) = D_x (\phi_{t}x)$ being the spatial derivative of the flow map at~$x$. In continuum mechanics, the matrix $W_x(t)^TW_x(t)$ is known by the name \emph{Cauchy--Green deformation tensor}. The level sets~$X^0 = \{f < 0\} \subset \Omega$ and $X^1 = \{f > 0\} \subset \Omega$ are then coherent sets, their ``coherence'' decreasing with~$|\lambda|$. The boundary conditions for~\eqref{eq:cheeger} are homogeneous Dirichlet (\ie, $f=0$) if $\Omega$ is considered as an open domain in~$\R^d$, and homogeneous Neumann (\ie, normal derivative of $f$ equal to zero) otherwise. Simply speaking, in the former case coherent sets are not allowed to touch the boundary of~$\Omega$, cf.~\citep{FJ18}. Multiple eigenfunctions $(f_0,\ldots,f_{k-1})$ can be used to find multiple coherent sets $X^0,\ldots,X^{k-1}$ by assigning those points $x\in \Omega$ to the same coherent set for which the vectors $(f_0(x),\ldots,f_{k-1}(x))$ are close-by in $\R^k$; see~\citep{BaKo17,FrRoSa19}.

The method can be discretized consistently with the continuous formulation~\eqref{eq:cheeger}, and thus made to work on (potentially sparse and incomplete) \emph{finite trajectory data}
\begin{equation}
	\label{eq:trajdata}
	x_j^i := \phi_{t_j} \, x^i, \quad x^i \in \Omega, \quad t_1,\ldots,t_n \in [0,T],\ i = 1,\ldots,m.
\end{equation}
Both \citet{BaKo17} and \citet{FJ18} discretize the dynamic Laplace operator~\eqref{eq:DL} on a subspace of functions that is adapted to the data~\eqref{eq:trajdata}. The authors of the former study use a meshfree data analysis technique called \emph{diffusion maps} \citep{CoLa06}, while those of the latter use finite elements with adaptive re-meshing for the time steps~$j=1,\ldots,n$.


\subsection{Coherently evolving features: Dynamical Spectral Clustering}
\label{subsec:dyn_spec_cluster}

The method introduced in section~\ref{subsec:material_coherence} is well suited for identifying coherently evolving material sets in complex flows in the case where the underlying flow map of the dynamical system is known, or we have trajectory data of labeled tracer particles. In this section, we aim to treat the case where we are given the much more restricted information of a time series of non-negative feature-indicating functions derived from the given dynamical system, or sets of \emph{indistinguishable} tracer particles without labels as it is the case, \eg, in Particle Image Velocimetry (PIV). In these cases, we do not know, ``what went where''. Formally, we encode such given data as positive measures $\mu_1, \dots, \mu_n$; see \citep{tao11measure} for an introduction to measure theory. As for the coherent sets introduced in section~\ref{subsec:material_coherence}, if $\Omega$ is time-dependent, set $\Omega = \bigcup_t \Omega_t$, and extend the $\mu_t, t=1, \dots, n,$ by zero for brevity of notation. For non-negative feature-indicating functions, these measures are typically non-atomic or ``continuous''; examples include the absolute vorticity or the DSI, see section~\ref{subsubsec:dsi}. In contrast, in the case of tracer particle data, the measures are discrete, \ie, sums of Dirac delta functions~$\sum_i \delta_{x^i}$.

Similarly to section~\ref{subsec:material_coherence}, we would like to obtain well-separated \emph{clusters} of particles or regions with high feature indicator function values, which we will subsequently also call coherent sets, despite the lack of labels.
Contrary to standard clustering approaches, we would like to resolve the change of such coherent sets in time, \eg, by advection.
To this end, we introduce the notion of a \emph{transfer operator} or \emph{Frobenius--Perron operator} $L\colon L_2(\Omega, \mu)\to L_2(\Omega, \nu)$, encoding the dynamical information of a step forward in time.
Given a \emph{transition kernel} $K\in L_1(\Omega\times \Omega, \mu\times \nu)$, the associated transfer operator $L$ is defined by
\begin{equation}
(L\psi)(y) \coloneqq \int_\Omega K(x, y)\psi(x) \mathrm d\mu(x).
\end{equation}
Our goal is to obtain $n$ partitions $\Omega = X_1^0\dot\cup X_1^1 =\dots = X_n^0\dot\cup X_n^1$ from $n-1$ transfer operators
\begin{equation}
L_1\colon L_2(\Omega, \mu_1)\to L_2(\Omega, \mu_2),\dots, L_{n-1}\colon L_2(\Omega, \mu_{n-1})\to L_2(\Omega, \mu_n).
\end{equation}

Since the dynamical information $L_1, \dots, L_{n-1}$ is not given in the case of an unknown flow map or unlabeled particles, we first construct some ``surrogate dynamics'' using \emph{optimal transport} (OT).
For a non-negative, symmetric, and Lipschitz continuous ground cost function $c \in C(\X \times \Y)$ 
and given probability measures $\mu$ on $\X$ and $\nu$ on $\Y$, respectively, \emph{Kantorovich's relaxation} of OT aims to find a minimizer of 
\begin{equation}\label{Monge_Kantorovich_problem}
\OT(\mu,\nu) \coloneqq \inf_{\pi \in\Pi(\mu,\nu)} \int_{\X \times \Y} c(x,y) \dx \pi(x,y),
\end{equation}
where $\Pi(\mu,\nu)$ denotes the set of all joint probability measures $\pi$ on $\X \times \Y$ with marginals $\mu$ and $\nu$.
A frequent choice for $c$ is the squared Euclidean distance, in which case the $\OT$-functional is called (squared) \emph{Wasserstein-$2$-distance}, denoted by $\mathcal W_2^2$.
Here, we use \emph{unbalanced regularized OT}, which is better suited for practical applications.
Moreover, the regularization readily introduces a diffusion into the dynamics, which is an important ingredient to solve the associated partitioning problem outlined below.
For more details, we refer to \citep{KLNS20}.
In the case $n > 2$, we use \emph{multi-marginal unbalanced OT}, see \citep{BLNS21}.
The resulting optimal transport plans $\hat \pi_{t}$ yield corresponding transfer operators $L_{t}\colon L_2(\Omega)\to L_2(\Omega)$ via
\begin{equation}
L_{t} \coloneqq \mathrm{diag}(\mu_t^{-1/2}) \hat \pi_t^T \mathrm{diag}(\mu_{t+1}^{-1/2}), \quad t=1,\dots, n-1.
\end{equation}

Next, we compute a singular value decomposition of the concatenated transfer operator $L = L_1 L_2 \cdots L_{n-1}$ which encodes the dynamics from $t=1$ to~$t=n$. From the singular vector pair $f^2_\mathrm{left}, f^2_\mathrm{right}$ corresponding to the second largest singular value $\sigma_2$ of $L$, we obtain partition functions
\begin{equation}
f_1 \coloneqq f^2_\mathrm{left}, \quad f_2 \coloneqq L_1 f_1, \quad \dots, \quad f_n \coloneqq L_{n-1} f_{n-1} = f^2_\mathrm{right}.
\end{equation}
The desired partitions $\Omega = X_t^0\dot\cup X_t^1$ are obtained by thresholding these functions at zero, \ie, we set $X_t^0 = \{ f_t > 0 \}$, $t=1, \dots, n$.
If desired---\eg, for obtaining a clustering with more than two clusters using $k$-means, see, \eg, \citep{Luxburg2007, KLNS20}---,
further orthogonal partition functions can be obtained from singular vector pairs corresponding to smaller singular values $\sigma_j$,~$j>2$.

The name ``Dynamical Spectral Clustering'' is justified as follows: In the limiting case of a small regularization parameter $\epsilon$ of the OT, $L^*L$ is well approximated by the identity operator plus $\epsilon \Delta^{\mathrm{dyn}}$ introduced in the previous section, see \citep{Fro15, KLNS20}.
This operator is symmetric, such that the singular decomposition above now becomes an eigendecomposition.
Clearly, this yields the eigenvectors of the dynamic Laplacian $\Delta^{\mathrm{dyn}}$ itself, so that this is reminiscent of the usual eigendecomposition of a graph Laplacian in spectral clustering, see \citep{Luxburg2007}.

The obtained partitions are \emph{coherent} in the following way: Averaged over the whole time series, there is minimal diffusive leakage of ``mass'' (or ``particles'') between the respective parts $X_t^0, X_t^1$ of the partition under the extracted surrogate dynamics, relative to their ``masses'' $\mu_t(X_t^0)$, $\mu_t(X_t^1)$. Note that the quotation marks are placed here to indicate that ``mass'' is meant in the sense of an analogy, not literally in the sense of physical mass of the fluid medium. Since using the diffusive leakage is equivalent to using the boundary length in the low-diffusion-limit, this notion of coherence is similar to the previous section~\ref{subsec:material_coherence}, but with respect to the OT dynamics and given measures $\mu_1, \dots, \mu_n$ that are not necessarily equal to Lebesgue measures.

As a simple illustration, we applied the method to a data set of two synthetic tropical cyclones as described in section~\ref{subsec:two_cyclones} to detect and ``cut out'' different coherent features. The results are shown in section~\ref{subsec:results_segm_cyclones}.


\section{Applications}
\label{sec:Applications}


\subsection{Track of storms Lothar and Martin}
\label{subsec:results_lothar}


\subparagraph{Volume overlap}
Initially, the storms Lothar and Martin are tracked using the $Q$-criterion and the DSI as feature indicators and a tracking scheme relying on volume overlap. The efficacy of the tracking algorithm is also demonstrated (section~\ref{sssec:Persistence_overlap}) where we also test the image registration technique from medical imaging, which comes with a relaxed overlapping condition and thus allows for larger time intervals between individual time slices. The data described in section~\ref{subsec:lothar} has a temporal resolution of both \qty{1}{\hour} and \qty{6}{\hour}. The DSI is computed for the hourly data set and tracking solely utilizes the volume overlap technique. The $Q$-criterion is computed only every six-hours and, due to the poor time resolution for this data set, tracking is only possible when we combine volume overlap and image registration. For both cases, we set the values for three user-defined parameters namely, the threshold ($\tau$) of the $Q$-criterion or DSI scalar field, overlap threshold ($\tau_{\rm overlap}$) and the counter -- or label -- of the structure that is to be tracked. This ``structure number'' arises simply from counting all structures that have been detected in the initial data set. 
\begin{figure}[h!]
	\centering
	\includegraphics[width=\linewidth]{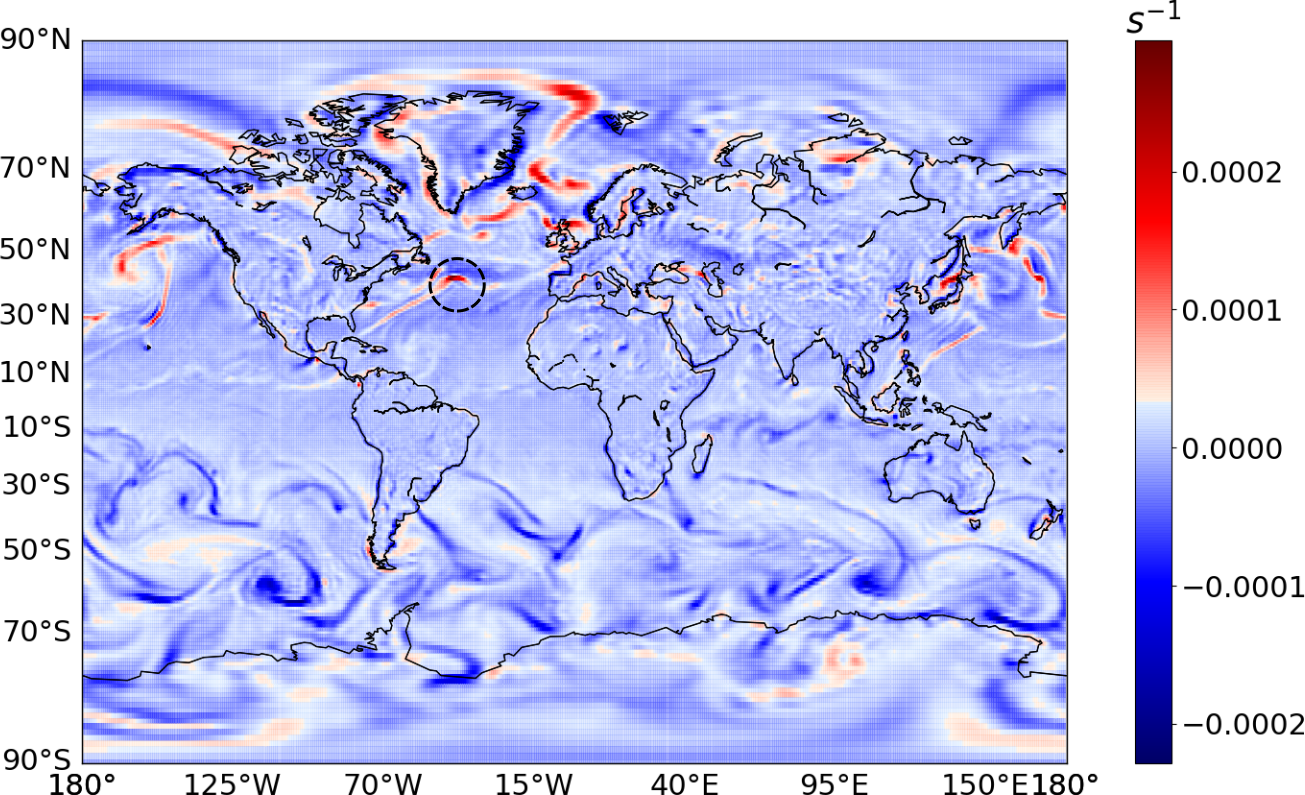}
	\caption{Vorticity magnitude plot of the dataset described in section~\ref{subsec:lothar} on 25 December 1999. The marked region is used as a start point for tracking the storm Lothar.}
	\label{fig:lotharVorticity}
\end{figure}

The value of $\tau$ is selected cautiously. If the chosen value is too low, an abundance of structures can be seen throughout the domain. This will result in the following problem: when a structure is selected for tracking at $t_{n}$, the same location may be occupied by a different structure at $t_{n + 1}$. The same location at $t_{n + 2}$ may be occupied by a yet another structure. Since overlap is the only parameter being checked between timesteps, this will result in the wrong structure being tracked. It should be noted that image registration is not active in this scenario as volume overlap succeeds in finding a structure at a future timestep. Therefore, a large value of $\tau$ is preferable. For $Q$-criterion and DSI, we choose the values of \num{15e-17} and \num{7.5e-17} respectively. Even though it seems logical to use an overlap threshold $\tau_{\rm overlap}$ greater than \num{0.9}, maintaining such a high overlap percentage is not feasible owing to the low time resolution (for both 1h and 6h time intervals). To ensure reliable tracking, we set $\tau_{\rm overlap} = 0.5$.

The final user-defined parameter is the starting position of the track. The Neighbor Scanning + MC algorithm described in section~\ref{sssec:Persistence_overlap} outputs a scalar field similar to the $Q$-criterion or DSI field but filled with numbers $\{0, 1, .. , n\}$, each of which corresponds to a structure. The number $0$ identifies regions where the overlap criterion was not satisfied and thus can be regarded as ``empty space'' as far as this analysis is concerned. To detect the point of departure for storm Lothar, we consider the vorticity magnitude as shown in figure~\ref{fig:lotharVorticity}. A spike is seen in the Atlantic Ocean off the East coast of the USA. The structure found in this region is used to initialize the tracking of storm Lothar. A similar procedure is followed to track storm Martin.

\begin{figure}[h!]
		\centering
		\includegraphics[width=\linewidth]{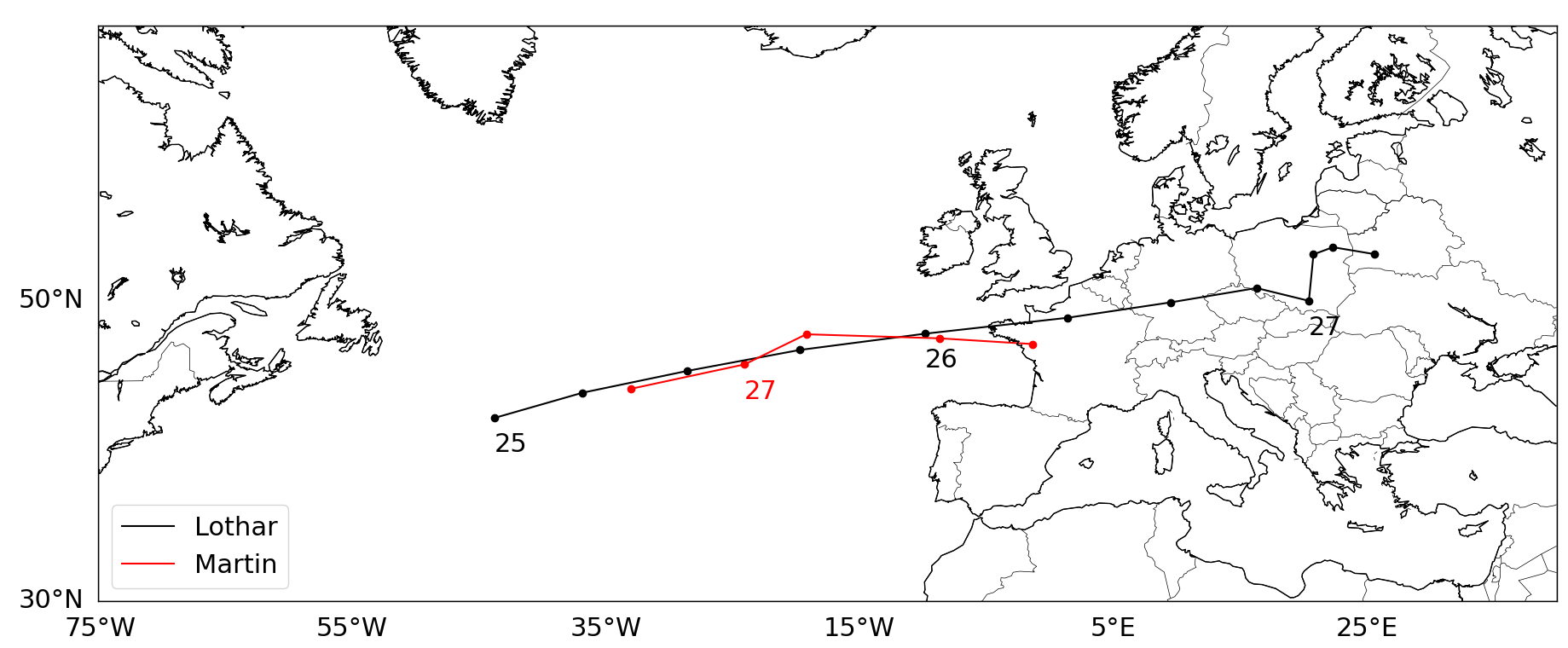}
		\caption{The black scatter points show Lothar's track from 25 December 1999 00:00:00 UTC to 27 December 1999 18:00:00 UTC. Red scatter points show Martin's track from 26th December 1999 18:00:00 UTC to 27th December 1999 12:00:00 UTC. Scatter points are mean values of the Q-criterion.}
		\label{fig:LotharMartinQ}
\end{figure}

\begin{figure}[h!]
		\centering
		\includegraphics[width=\linewidth]{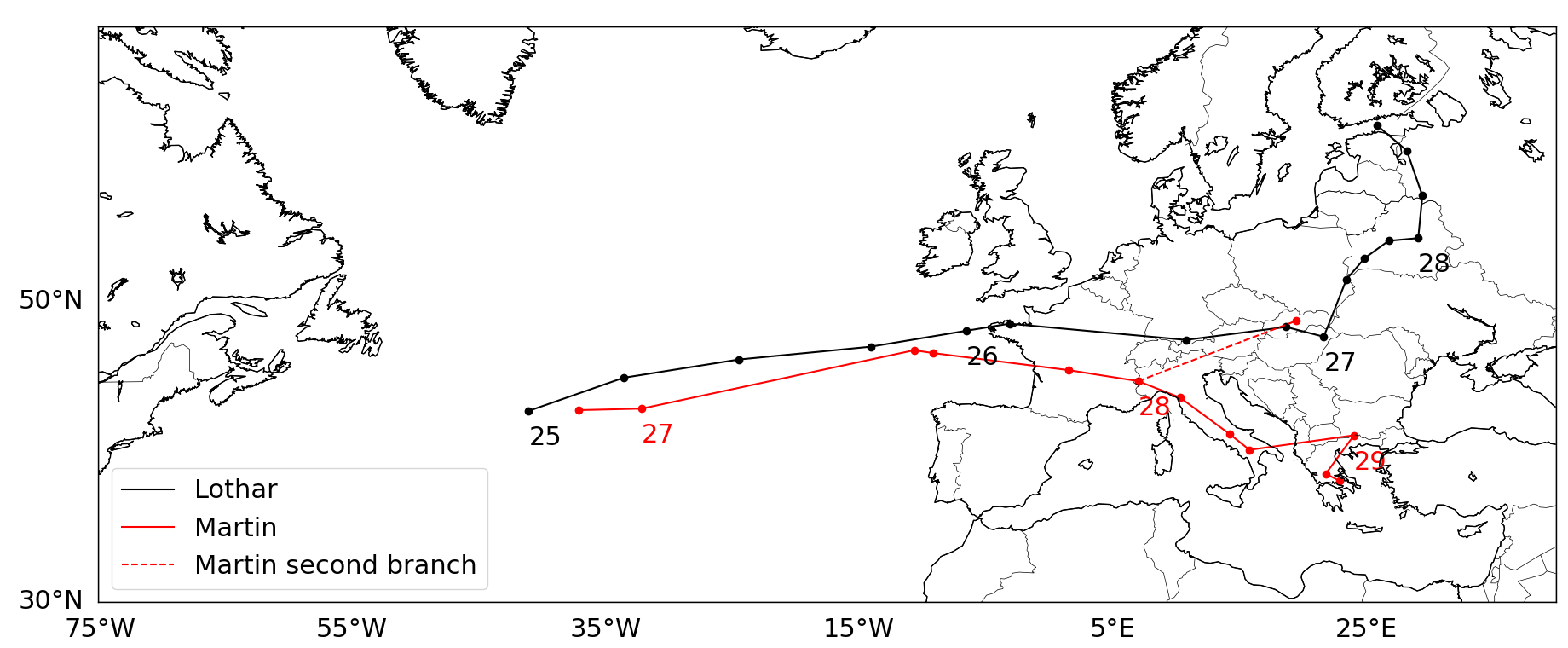}
		\caption{Similar to figure~\ref{fig:LotharMartinQ}, both Lothar and Martin's tracks are shown. Lothar's track is from 25 December 1999 00:00:00 UTC to 28 December 1999 18:00:00 UTC whereas Martin's track is from 26th December 1999 18:00:00 UTC to 29th December 1999 12:00:00 UTC. A second branch of Martin's track is shown until 28th December 1999 06:00:00 UTC. Scatter points are mean values of positive DSI.}
		\label{fig:LotharMartinDSI}
\end{figure}

The results for the trackings based on the $Q$-criterion and the DSI are shown in figure~\ref{fig:LotharMartinQ} and figure~\ref{fig:LotharMartinDSI}, respectively. Although a 2D plot is shown here, the structures are tracked in 3D. The scatter points are obtained by summing up all the planes in the wall-normal direction, \ie, along the height and then computing their means. The results for DSI are plotted at a 6-hour time interval to make them comparable to the $Q$-criterion results. Qualitatively, one observes that Lothar's track is hindcast similarly when the $Q$-criterion or the DSI are used. The track is also comparable to the \qty{6}{\hour} DSI results reported by \citet{Weber2008storm}. The reader is reminded that image registration is used for the \qty{6}{\hour} only. This shows that it is possible to track structures across poor time resolution data with image registration.

For storm Martin's track, several interesting observations suggest themselves. First, like Lothar's track, the approximate tracks of storm Martin based on the $Q$-criterion and the DSI are qualitatively similar until $27^{th}$ December 1999 at 18:0\qty{0}{\hour}. This is the maximum number of timesteps available for the \qty{6}{\hour} case. However, the storm is followed over a longer period with the DSI data. On $28^{th}$ December 1999, a branching event is detected, in which the storm splits into two structures of similar size. Naturally, the larger one, which has the larger overlap percentage, is followed subsequently, and this locates the storm as traversing Italy. If the other structure is tracked, one finds that it turns and moves North instead. 


\subparagraph{Lagrangian perspective}
Next, we analyze Lagrangian trajectories constructed based on the \emph{horizontal} velocity fields of storm Lothar using the Lagrangian coherent set approach. Thus, we aim to find Lagrangian coherent sets, as described in section~\ref{subsec:material_coherence}, by using space-time diffusion maps method~\citep{BaKo17}. We note that the $Q$-criterion, in contrast, yields an Eulerian local approach based on the velocity gradient tensor. We proceed as follows. First we construct trajectories $(x_{t}^{i})_{t}$, $t,i=1,...,N$ using numerical integration and interpolation of the 2D horizontal velocity fields~$v$. The starting points $x_{0}^{i}$ are uniformly generated. The Lagrangian coherent set method of \citet{BaKo17} then partitions the set of trajectories into a user-specified number of materially coherent subsets.
	
\begin{figure}
\begin{minipage}[t]{0.49\textwidth}
\includegraphics[width=\textwidth]{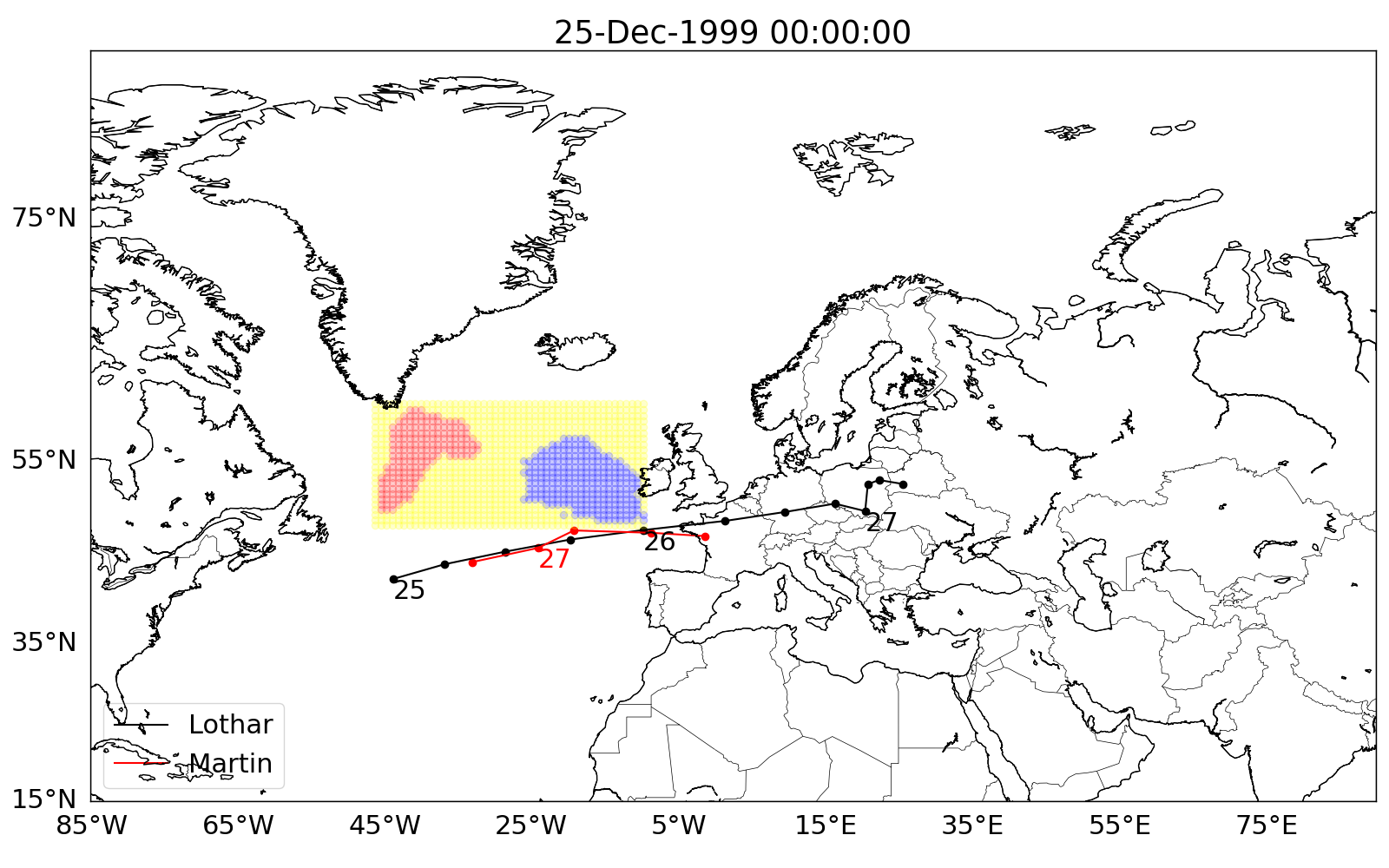}
\end{minipage}
\begin{minipage}[t]{0.49\textwidth}
\includegraphics[width=\textwidth]{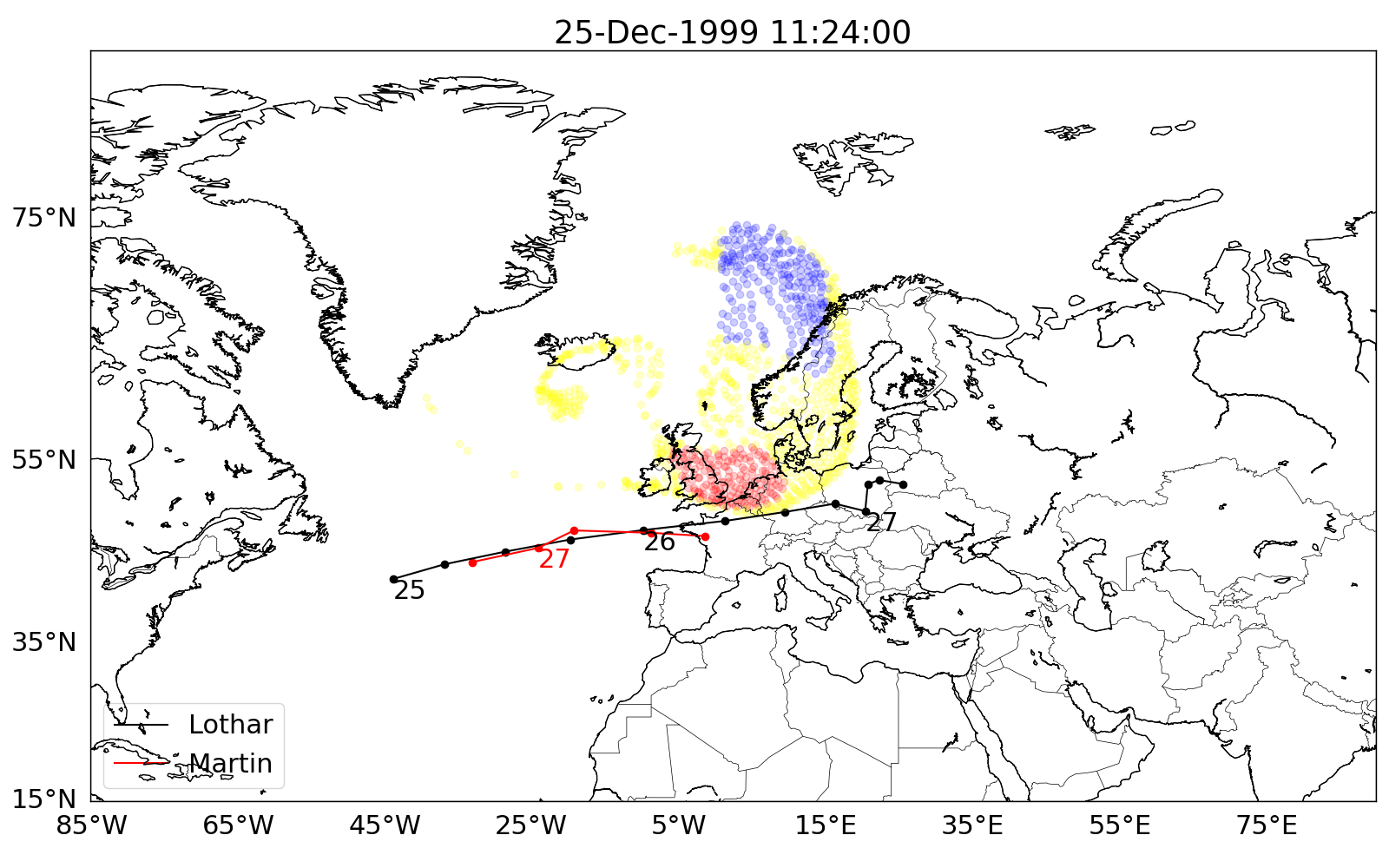}
\end{minipage}
\begin{minipage}[t]{0.49\textwidth}
\includegraphics[width=\textwidth]{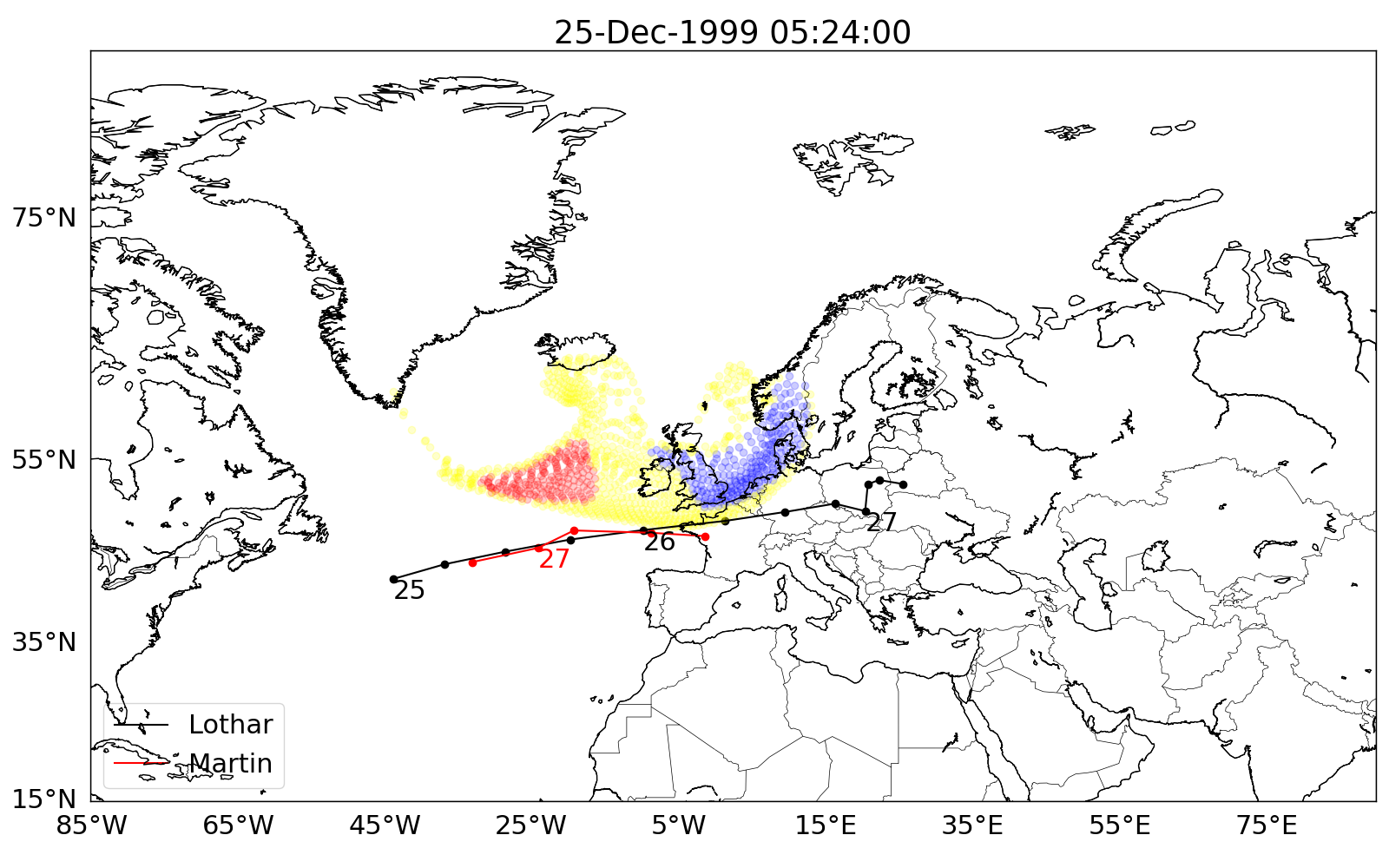}
\end{minipage}
\begin{minipage}[t]{0.49\textwidth}
\includegraphics[width=\textwidth]{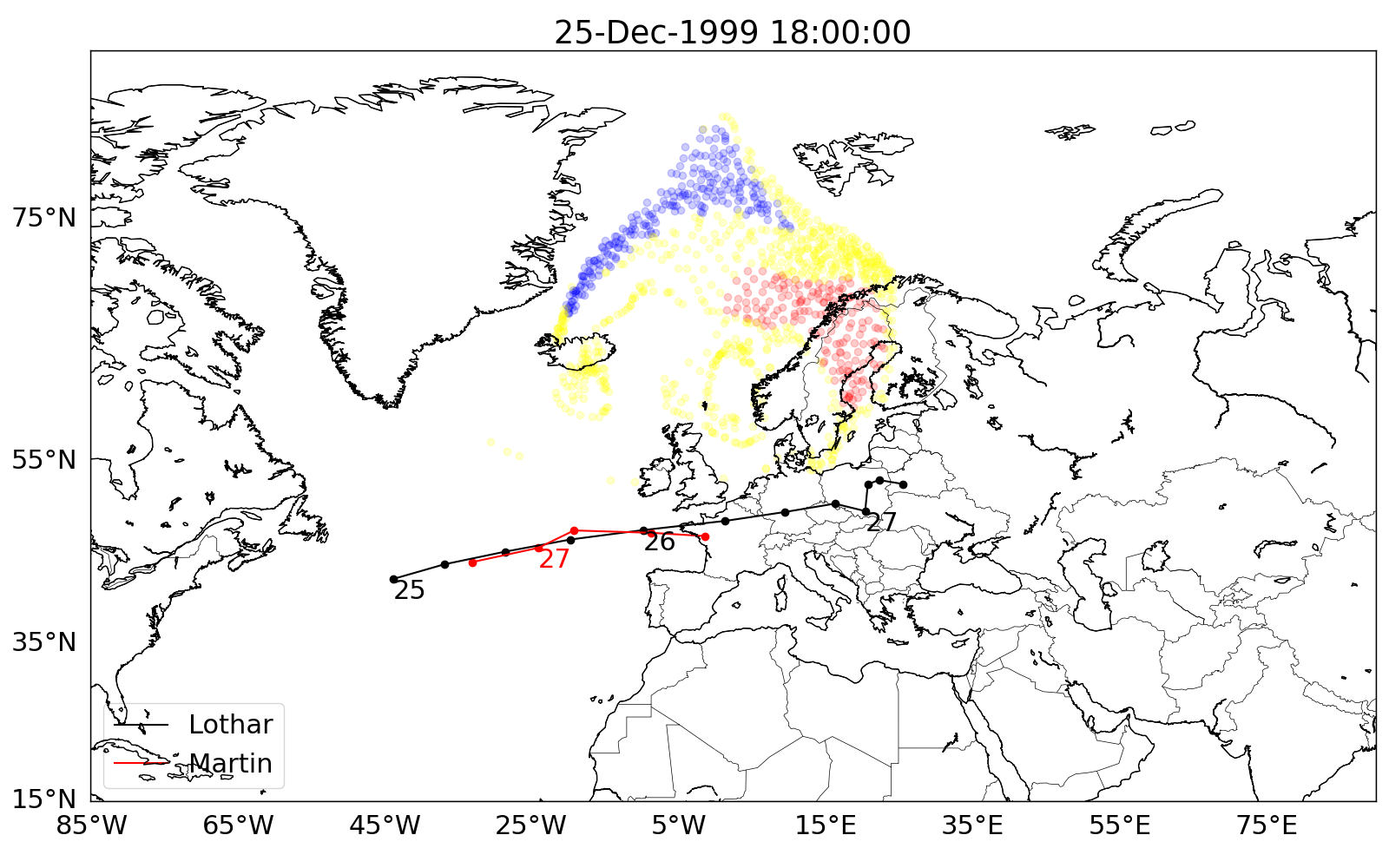}
\end{minipage}
\caption{Material coherent sets in the Lothar data set. Starting time is 25 December 1999 00:00:00 to final time 25 December 1999 18:00:00. The evolution under the flow is shown. Two material coherent sets are detected from 2500 trajectories.}  
\label{fig:LotharDFM1}
\end{figure}

\begin{figure}[h!]
\centering
\begin{minipage}[t]{0.49\textwidth}
\includegraphics[width=\textwidth]{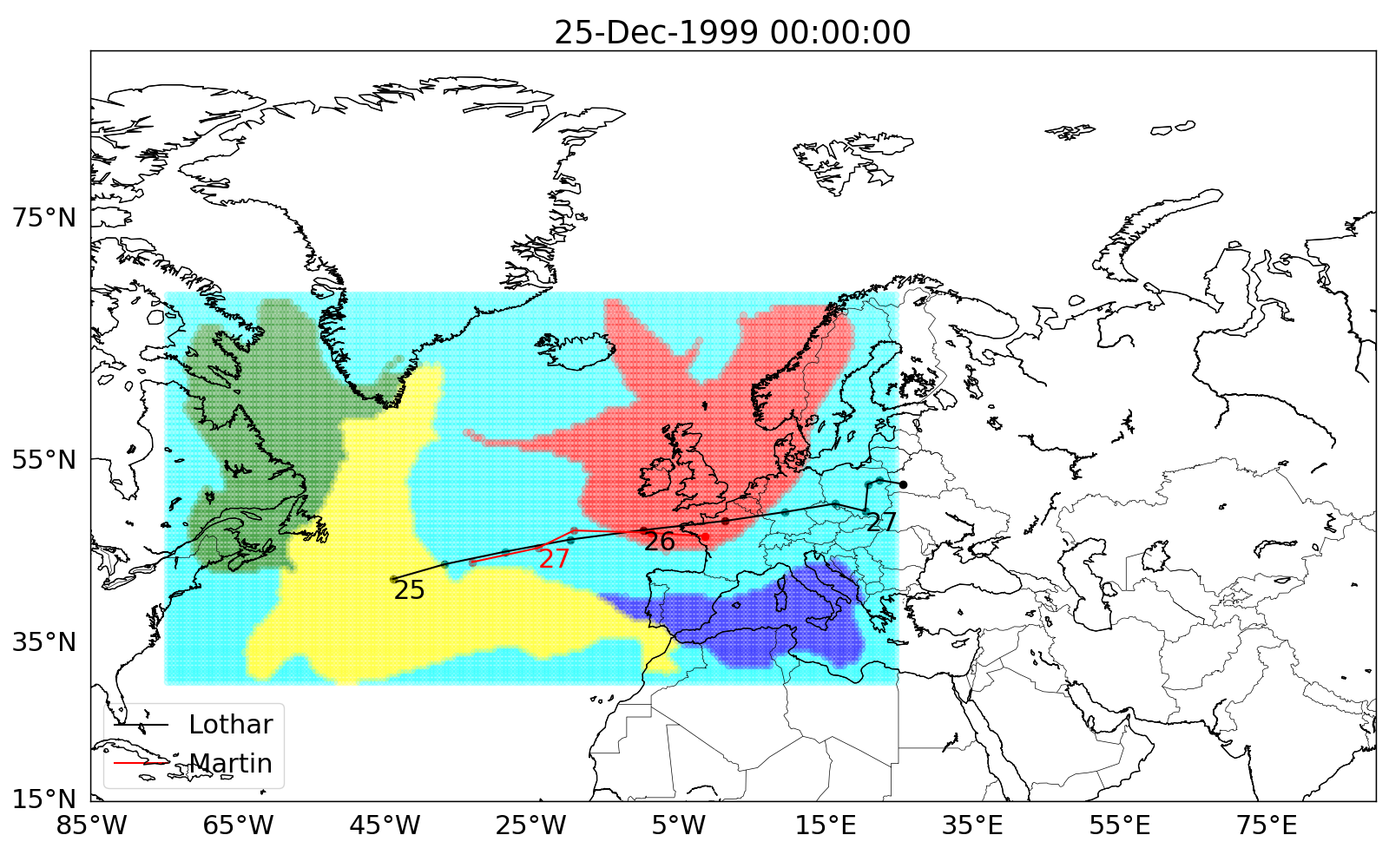}
\end{minipage}
\begin{minipage}[t]{0.49\textwidth}
\includegraphics[width=\textwidth]{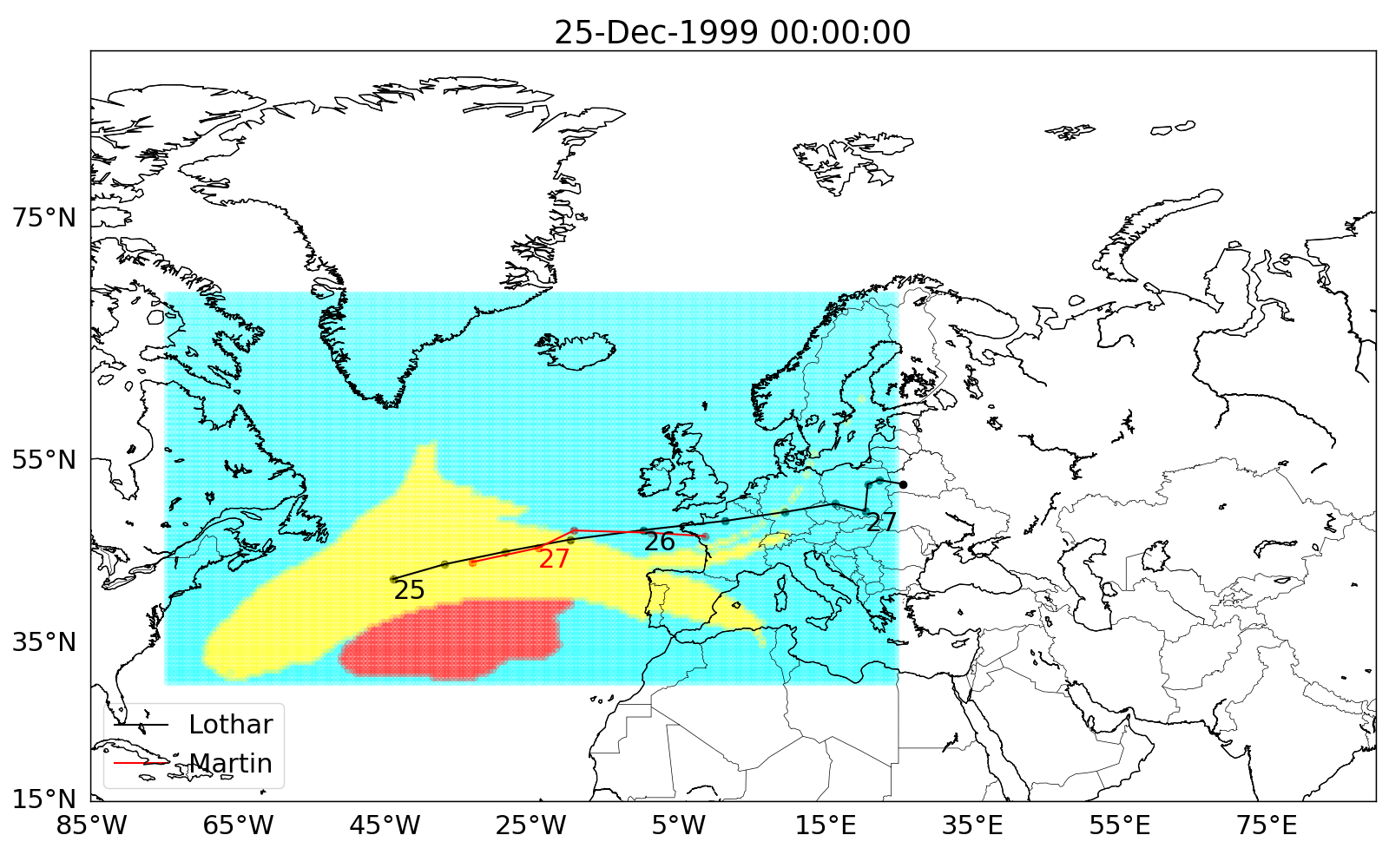}
\end{minipage}
\begin{minipage}[t]{0.49\textwidth}
\includegraphics[width=\textwidth]{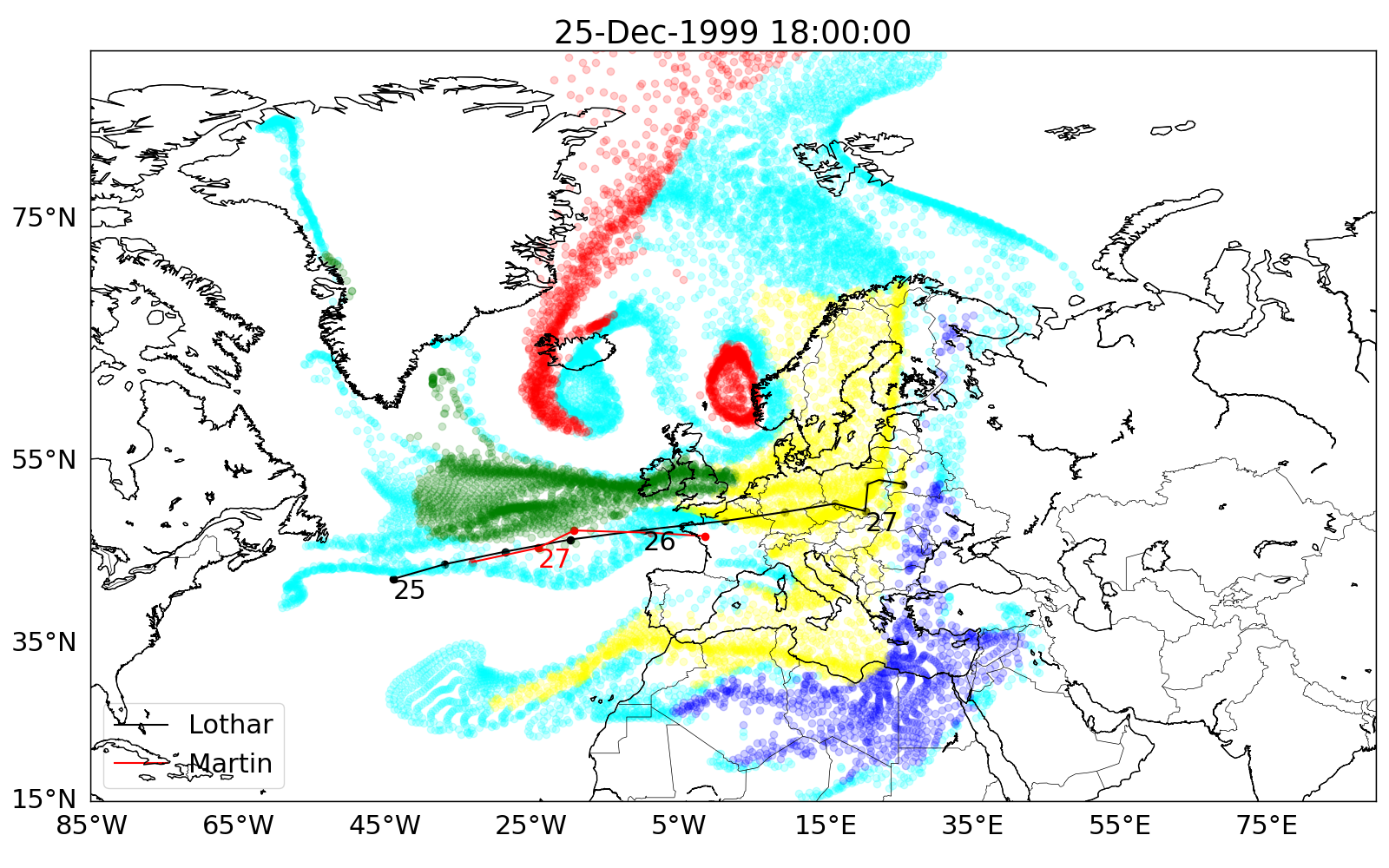}
\end{minipage}
\begin{minipage}[t]{0.49\textwidth}
\includegraphics[width=\textwidth]{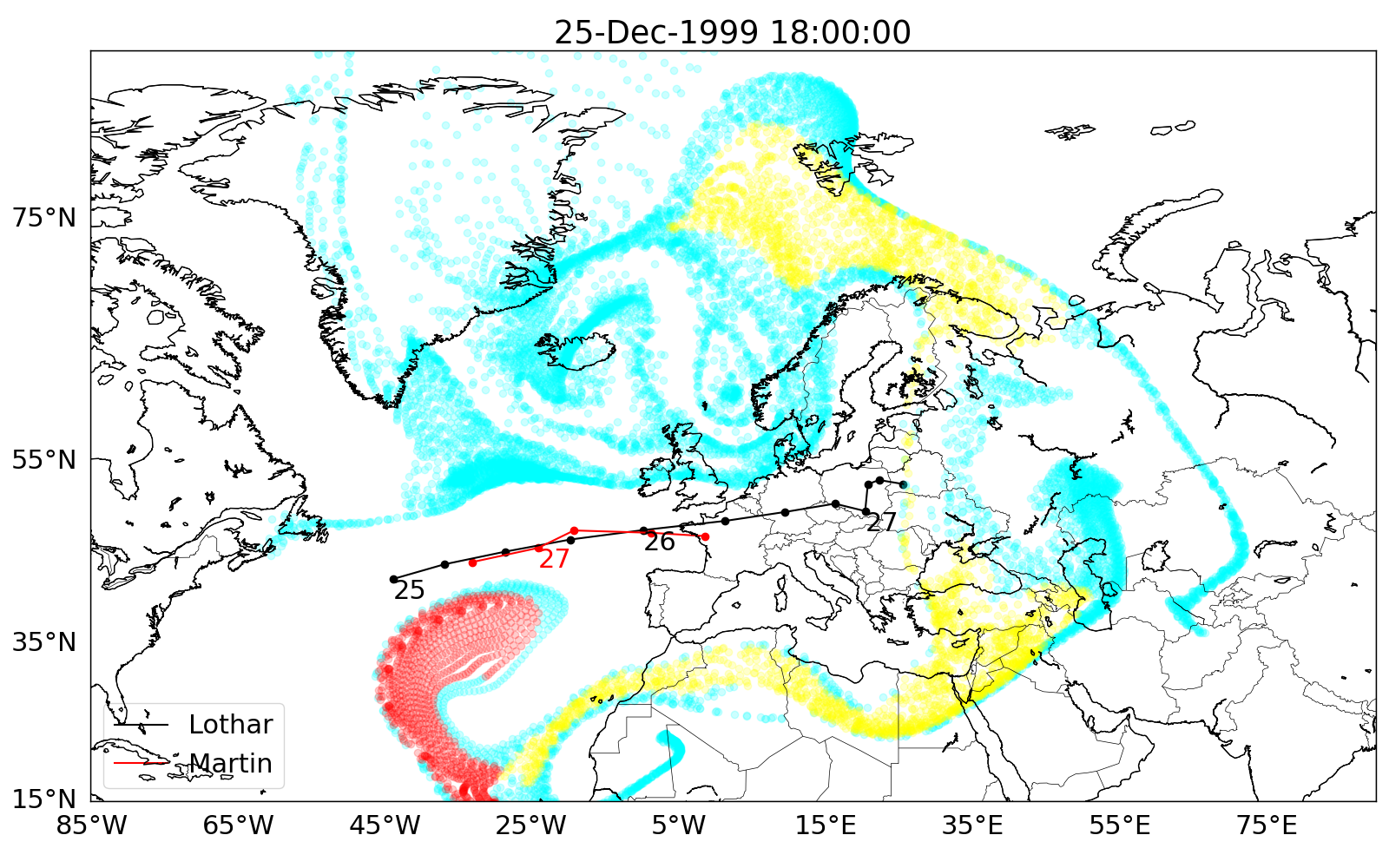}
\end{minipage}

\caption{The evolution under the flow, again for the area covering storm Lothar, is shown for height levels  825~hPa (left) and 550 hPa (right). The number of trajectories is~16000.}
\label{fig:LotharDFM2}

\end{figure}

The Lagrangian approach detects something else than the storm Lothar. In figure~\ref{fig:LotharDFM1} a domain of points on height level $\qty{775}{\hecto\pascal}$ is chosen which is bounded by \qty{48}{\degree}-\qty{61}{\degree}N and \qty{9}{\degree}-\qty{47}{\degree}W. The evolution under the flow is shown on Dec~25, 1999 for \qty{18}{\hour}. Two advective coherent sets emerge for the given initialization of points. The red and blue sets barely disperse, and their relative arrangement also stays rigid during the evolution. This suggests that they could be part of a jet stream. The velocity of particles is three times as fast compared to the velocity of the storm Lothar in figure~\ref{fig:LotharMartinQ}. There is a difference between the Lagrangian and Eulerian approach with respect to advection which yields a time lag.

Figure~\ref{fig:LotharDFM2} shows large-scale results on height levels $825$ and $\qty{550}{\hecto\pascal}$ for the same time horizon of \qty{18}{\hour}. The given initialization of points is bounded by \qty{30}{\degree}-\qty{70}{\degree}N and \qty{20}{\degree}-\qty{75}{\degree}W. The green set identified in the left column of figure~\ref{fig:LotharDFM2} can be associated to storm Lothar. It is least advected and stays close to the path identified with $Q-$criterion on 25th December at 18:00:00 UTC. The results suggest that the red and green set in the left column are coherent parts of a jet stream. The right column of figure~\ref{fig:LotharDFM2} shows a higher height level. The set presented in yellow follows Lothar's track and is divided into two sets at final time. The part of the set which moves up to the north could be again part of the jet stream. In the south of storm Lothar a set was detected which is barely advected and filamented. This set colored in red remains almost coherent.
 
The comparison of our results obtained by the Eulerian and the Lagrangian methods reveals that intense and rather localized storms in the middle latitudes are not detected by the Lagrangian coherent set approach applied to the horizontal flow. Instead, these structures propagate relative to the air masses indicating their wave-like nature. The Lagrangian information extracted by the material coherent set detection approach may reveal, though, where the air masses participating in the storm come from originally and where they go after having passed through the central region of the storm. In this sense, both the Lagrangian and Eulerian techniques yield valuable, yet complementary, information. 

\begin{figure}[htb]
	\centering
	\includegraphics[width=\textwidth]{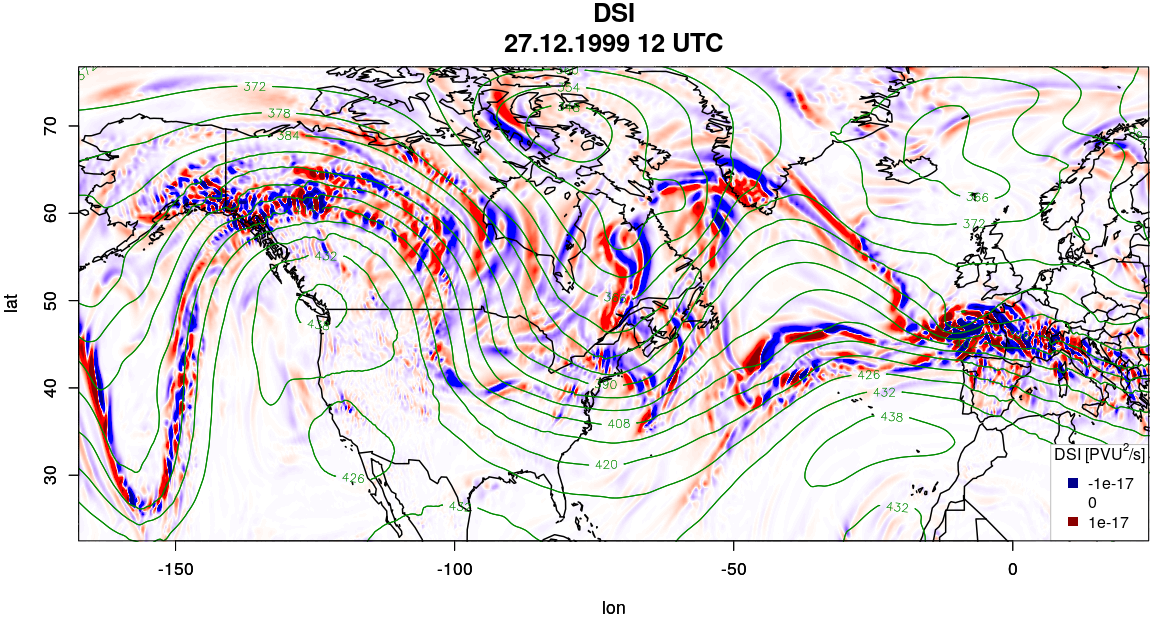}
	\caption{The DSI signalizes the flow surrounding the persistent weather pattern (high-over-low) at the west cost of Canada (high pressure area) and North America/Mexico (low pressure area). The green lines indicate the geopotential. At the regions of the persistent high and low pressure areas, the DSI is zero, because the flow is balanced. A strong DSI signal can be recognized over France and Germany, where storm Lothar hit Europe.}
	\label{fig:DSI_lothar_block}
\end{figure}

\begin{figure}[htb]
	\centering	
	\begin{minipage}[t]{0.98\textwidth}
		\includegraphics[width=\textwidth]{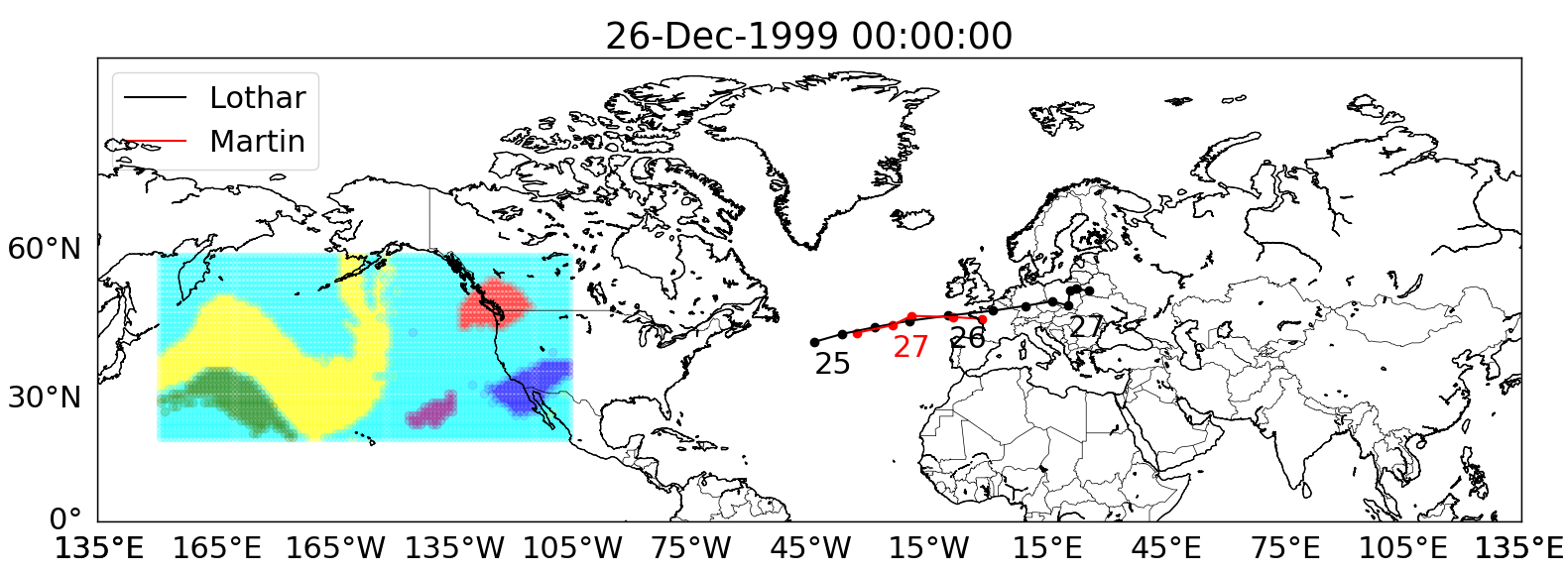}
	\caption{Applying material coherence: The rectangular box over the pacific indicates the initial configurations. 
	Tracks of storms Lothar and Martin are marked by the black and red lines, respectively.}
	\label{fig:DFM_lothar_inittialcond_block}
	\end{minipage}
	
	\begin{minipage}[t]{0.98\textwidth}
		\includegraphics[width=\textwidth]{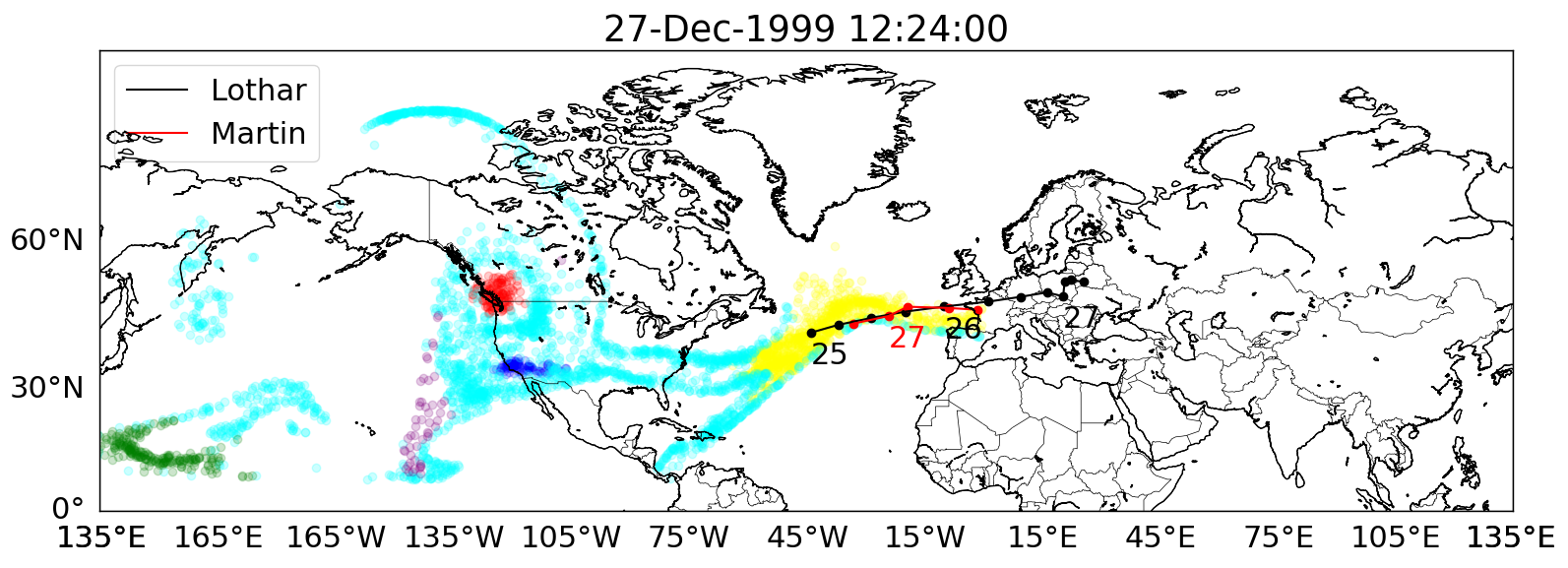}
	\end{minipage}	
	\caption{The persistent high-over-low can be identified as coherent sets (red marked domain: high pressure area, 
	dark blue marked domain: low pressure area). Storm Martin is embedded in a coherent set (yellow set) that represents the 
	meandering jet stream.}
	\label{fig:DFM_lothar_block}
\end{figure}


\subparagraph{Material coherence and a stationary weather pattern during the storms Lothar and Martin}
Once again, we apply the method to find material coherent sets to the case study of storms Lothar and Martin, but seed the initial tracers elsewhere. We find that the ability of identifying storms as coherent sets via material coherence depends very much on these initial configurations, \ie, on the initial domain and the considered time period. Figure~\ref{fig:DFM_lothar_inittialcond_block}, and figure~\ref{fig:DFM_lothar_block} show one example of an initial configuration and of how it has evolved 36:2\qty{4}{\hour} later, where Martin, embedded in a narrow elongated jet stream, can now also be identified as a coherent set (yellow set).

In general, we find that this method can diagnose stationary vortices such as persistent high and low pressure areas that form blocked weather states. At the same time as Lothar and Martin crossed Europe, a so-called high-over low block was diagnosed over Canada/Northern America/Mexico. The green lines in figure~\ref{fig:DSI_lothar_block} show the geopotential in a height of 600 hPa indicating these persistent high and low pressure areas. Regarding this blocked weather situation, we compare the coherent set analysis with the DSI field. Figure~\ref{fig:DFM_lothar_block} clearly shows the high-over-low as coherent sets, where the set marked in red indicates the steady high pressure area and the dark blue marked set identifies the low pressure area. The animation in the supplementary material shows the anticyclonal and cyclonal motion of the high and low pressure areas. In the regions of the persistent vortices the DSI is nearly zero, see figure~\ref{fig:DSI_lothar_block}. This can be explained as follows: The DSI signalizes imbalances caused by non-steady, diabatic and viscous conditions of atmospheric flows. In the opposite, DSI values equal to zero indicate adiabatic, and inviscid conditions of a steady basic state, see the definition in section~\ref{subsubsec:dsi}. Therefore, the DSI is (nearly) zero for a persistent vortex pattern, but indicates -- with non-zero values -- the flow surrounding the high pressure area and especially indicating strong storms such as Lothar. For the animations of the whole time period see the supplementary material. In conclusion, we note that the DSI and the coherent set method diagnose complementary flow conditions such that combining both methods might yield an interesting tool for the diagnosis of important weather situations. 


\subsection{Classification and Tracking of weather fronts}
\label{subsec:class_synop_loc}

\begin{figure}[htb]
	\centering
	\includegraphics[width=.98\textwidth]{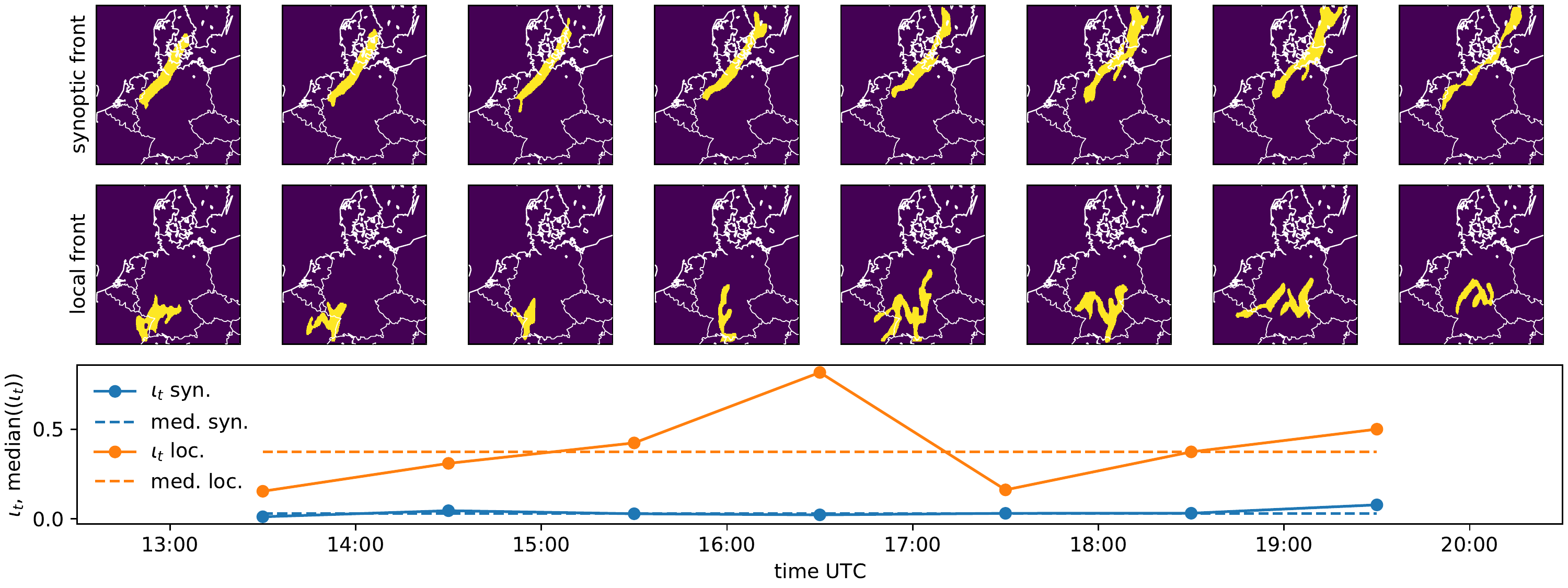}
	\caption{Wasserstein-indicators,  $\iota_t$, for a time series of local and synoptic fronts on the 6th of August 2013. The terms ``synoptic'' and ``local'' refer to the normalized approximate Wasserstein distance, $\iota_t$, from \eq{eq:WassersteinCoherenceFronts}, and ``synoptic median'' and ``local median'' refer to the medians of~$\iota_t$.}
	\label{fig:syn_loc_explanation}
\end{figure}

In this section, we aim for a classification of two different types of weather fronts based on their coherence properties. Weather fronts can be divided into \emph{local} (meso-scaled) and \emph{synoptic} (macro-scaled) fronts. Certain classes of local weather fronts have boundaries that change quite erratically, whereas synoptic fronts usually show a more persistent behavior of their boundaries, see section~\ref{george_method} for further details and detection of fronts. This motivates a classification algorithm based on comparing quantitative measures of persistence of the fronts at hand. We measure persistence of these front boundaries using the Wasserstein distance $\mathcal W_2$ introduced in section~\ref{subsec:dyn_spec_cluster}: If appropriately standardized, for given uniform probability measures $\mu^l_t, \mu^l_{t+1}$ from two timesteps $t$ and $t+1$ with support on the boundaries of local fronts and uniform probability measures $\mu^s_t, \mu^s_{l+1}$ supported on the boundaries of synoptic fronts, we expect in general that
\begin{equation}
\mathcal W_2(\mu^l_t, \mu^l_{t+1}) > \mathcal W_2(\mu^s_{t}, \mu^s_{t+1}). 
\end{equation}
To achieve comparable Wasserstein-distances for structures of different sizes and to avoid counting translations as changes of shape, we perform a standardization as follows. Take two measures $\mu_t, \mu_{t+1}$ representing two consecutive boundaries of a front and two random variables $X_t\sim \mu_t, X_{t+1}\sim  \mu_{t+1}$ with these laws, which represent equidistributed random positions on the front. Let their expectations be denoted by~$\E X_t, \E X_{t+1}.$ Set $s\coloneqq \frac 1 2 (\E\Vert X_t - \E X_t\Vert_2 + \E\Vert X_{t+1} - \E X_{t+1}\Vert_2)$ as the arithmetic mean of the expected Euclidean distance from their means. Then we normalize $X_t$ and $X_{t+1}$ by $\tilde X_t \coloneqq (X_t-\E X_t) / s$, $\tilde X_{t+1} \coloneqq (X_{t+1}-\E X_{t+1}) / s$ and set $\tilde\mu_t$ and $\tilde \mu_{t+1}$ to be their corresponding normalized laws. Finally, we compute our indicator value $\iota_t$ as
\begin{equation}\label{eq:WassersteinCoherenceFronts}
\iota_t \coloneqq \OT_\varepsilon(\tilde \mu_t, \tilde \mu_{t+1}).
\end{equation}
Here, we approximate the Wasserstein-distance $\mathcal W_2$ by regularized optimal transport $\OT_\varepsilon$, such that these distances can be computed using the fast Sinkhorn algorithm, see \citep{cuturi2013sinkhorn}. For any given front, the input to this method is a time series of compactly supported indicator functions $(1_t)_{t\in \{0, \dots, n\}}$ indicating the support of the corresponding front. It will then produce a series of real indicator values $(\iota_t)_{t\in \{0, \dots, n-1\}}$. A schematic explanation of the method using two examples from the mentioned data is given in figure~\ref{fig:syn_loc_explanation}.

\begin{figure}[h]
	\centering
	\includegraphics[width=.9\textwidth]{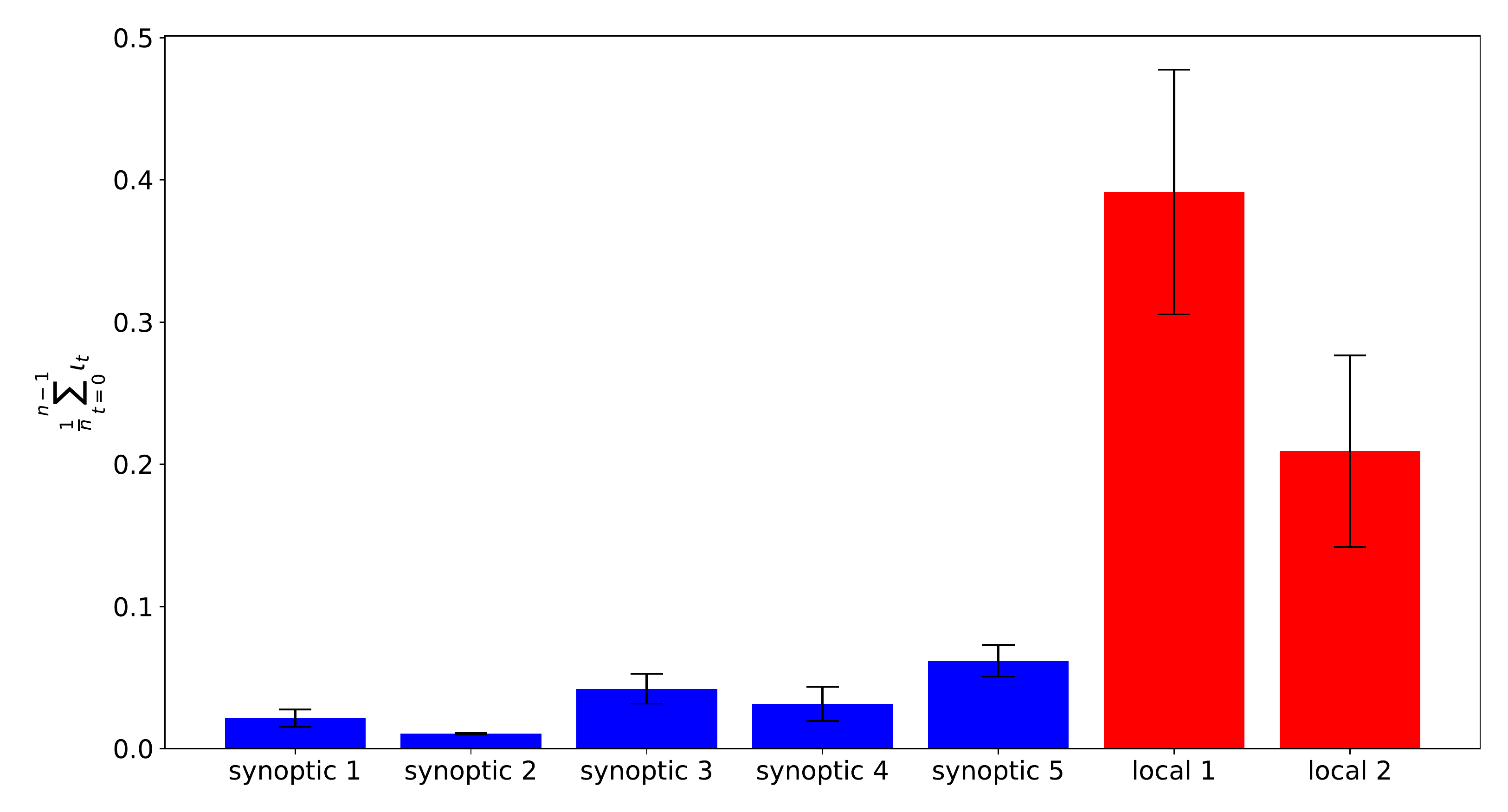}
	\caption{Means and standard deviations of the change indicator $\iota_t$ from \eq{eq:WassersteinCoherenceFronts} over the entire front lifetime for synoptic fronts (blue) and local fronts (red). See also figure~\ref{fig:frontOverlap} and \ref{fig:localFront2} for further information on the data used here.}
	\label{fig:indicators}
\end{figure}

We applied this method to a set of local and synoptic weather fronts that have been detected using the method from section~\ref{george_method} from data described in section~\ref{subsec:front_data}. The results for the mean indicator values of the whole corresponding time series of the lifetime of the front are shown in figure~\ref{fig:indicators}.
As expected, the indicator values for the local fronts are higher than those for the synoptic fronts. Overall, these first experiments using this small sample of case studies yield promising results for discriminating between these two types of fronts.

Next, we subject the dataset to the overlap method introduced in section~\ref{sssec:Persistence_overlap}. In this scenario, we only set two user-defined parameters: $\tau$, $\tau_{\rm overlap}$. The structure number is not set, and all structures are followed instead to study branching behavior. After the fronts are detected with the method described in section~\ref{george_method}, the result is a data set containing values $0$ and $1$ only, where $1$ indicates the presence of a front. This makes it trivial to select the threshold $\tau$ as any value less than $1$ is suitable. Since the data is less complicated than others, \ie, only one structure exists in most timesteps, any amount of overlap with a future timestep will positively identify the correct structure. In what follows, $\tau_{\rm overlap}$ has been set to~$0.1$.

\begin{figure}[h!]
	\centering
	\includegraphics[width=\linewidth]{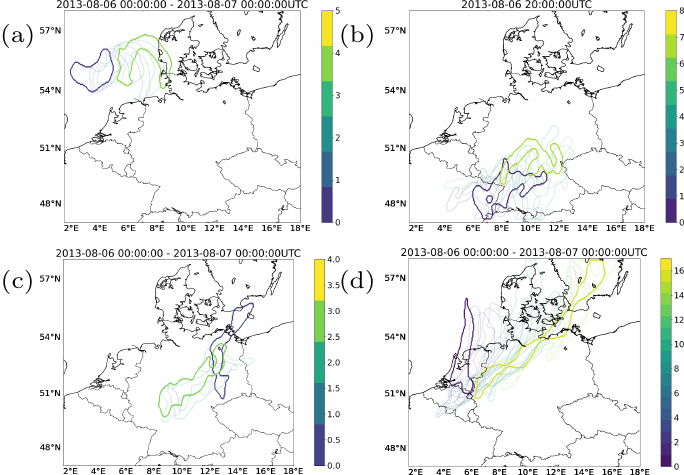}
	\caption{Three local fronts (a, b, c) and a synoptic front (d) are shown. The colormap indicates the timesteps. The first and last timestep in each case is shown with thicker lines.}
	\label{fig:frontOverlap}
\end{figure}

\begin{figure}[h!]
	\centering
	\begin{minipage}{\textwidth}
		\centering
		\includegraphics[width=\linewidth]{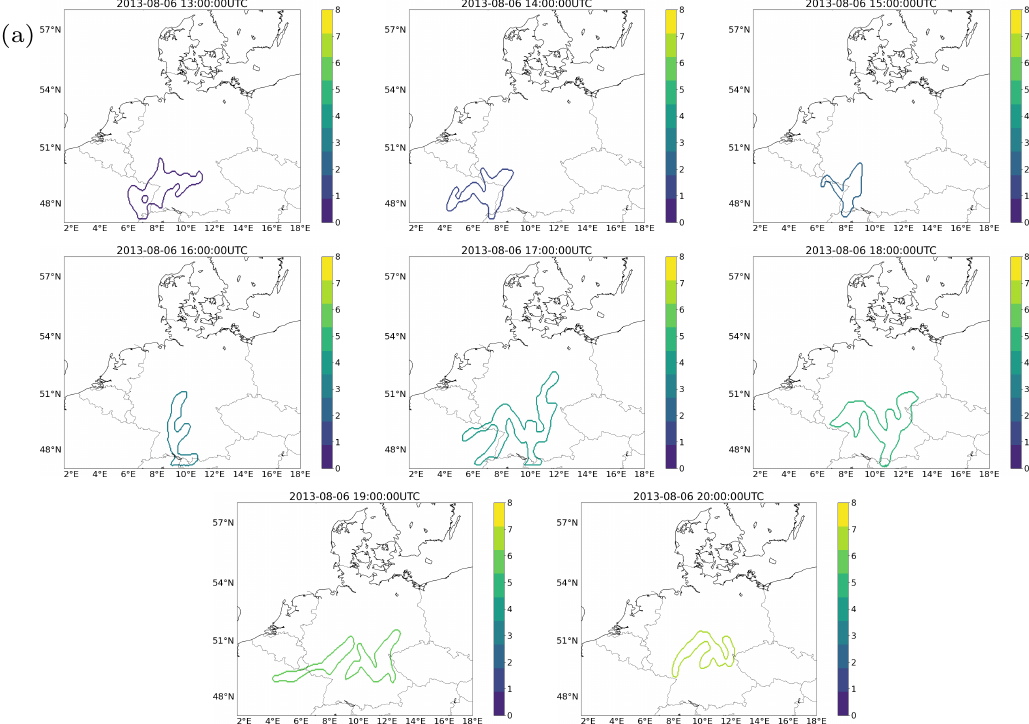}
	\end{minipage}
	\begin{minipage}{\textwidth}
		\centering
		\bigskip
		\includegraphics[width=0.3\linewidth]{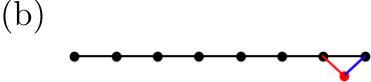}
	\end{minipage}
	\caption{(a) highlights the individual timesteps of Local front 2 (from figure~\ref{fig:frontOverlap}b). (b) shows a time graph for the same.}
	\label{fig:localFront2}
\end{figure}

The results of the tracking are shown in figures~\ref{fig:frontOverlap} and \ref{fig:localFront2}. The data set reveals three meso-scale local and one synoptic front as shown in figures~\ref{fig:frontOverlap} (a, b, c) and figure~\ref{fig:frontOverlap} (d), respectively. The local front from figure~\ref{fig:frontOverlap} (b) exhibits a simple branching event. This is documented by a sequence of the front boundary geometries in figure~\ref{fig:localFront2} (a) and by a time graph recording the branching event in figure~\ref{fig:localFront2}~(b). 

\subsection{Tracking Synthetical Tropical Cyclones}\label{subsec:results_segm_cyclones}

\begin{figure}
	\centering
	\includegraphics[width=.98\textwidth]{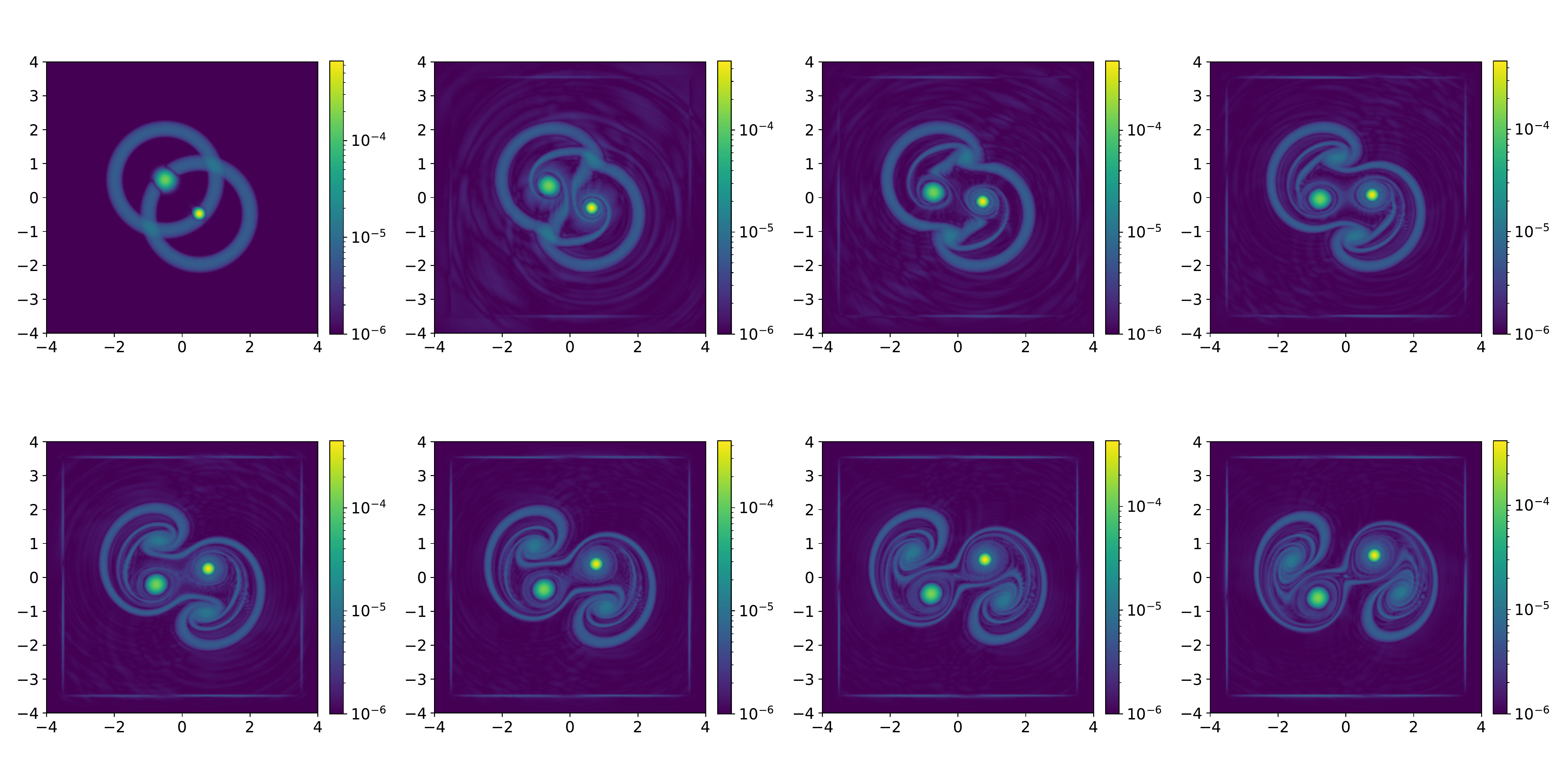}
	\caption{Time series of absolute vorticity data of two synthetical cyclones.}
	\label{fig:synth_cyclones}
	\vspace*{1cm}
	\includegraphics[width=.98\textwidth]{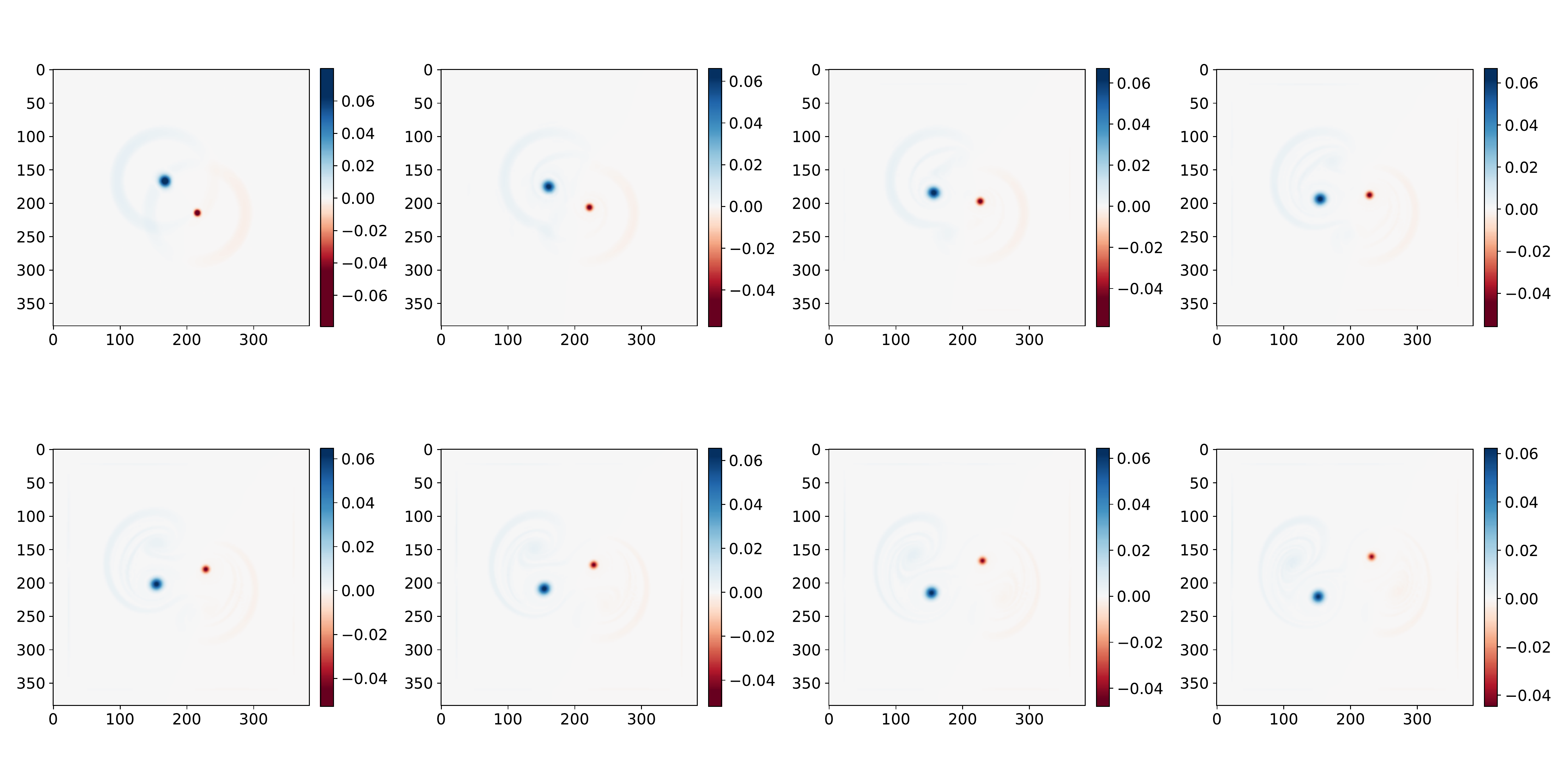}
	\caption{Resulting partition functions of dynamical spectral clustering.
	}
	\label{fig:partition_cyclones}
\end{figure}

As a simple proof-of-concept-example and illustration of dynamical spectral clustering, we applied the method from section~\ref{subsec:dyn_spec_cluster} to the synthetical data set of a simulation of two tropical cyclones described in section~\ref{subsec:two_cyclones}.
More specifically, we used $8$ snapshots of the absolute values of the vorticity variable from the simulation, which are displayed in figure~\ref{fig:synth_cyclones}.
The two moving centers of the cyclones are visible as two ``blobs'' of vorticity surrounded by rings of (negative) small-amplitude vorticity being the result of a mollifier applied to the velocity field ensuring that it goes to zero within a finite distance from the center.

For the multi-marginal optimal transport, we set the regularization parameter as $\varepsilon=(2\cdot 10^{-2}\cdot a)^2$, where $a$ is the side length of the square domain, and used $1.0\cdot \KL(\cdot, \mu_t)$ as the marginal penalization terms for all time steps $t$. Here, $\KL$ denotes the Kullback--Leibler-divergence, see \citep{BLNS21} for more details.
The corresponding partition function is shown in figure~\ref{fig:partition_cyclones}, where positive or negative values correspond to one structure or the other, respectively.
The two sets of the partition can be obtained by thresholding these functions at zero.
Although there is no ``overlap'' of the two bumps moving through the domain, optimal transport serves as a good proxy to interpolate between the snapshots, such that the clustering method detects the two cyclones as two separate persistent structures moving through the domain.

Next, both idealized tropical cyclone datasets (1TC and 2TC, see sections~\ref{subsec:two_cyclones}) are analyzed by means of the overlap technique from section~\ref{sssec:Persistence_overlap}. The first case, denoted 1TC, involves the three-dimensional flow of an idealized tilted tropical cyclone which precesses about a vertical axis. The second case, denoted 2TC, features two nearby cyclones of different intensity, again in three space dimensions. These cyclones can be highlighted by both the $Q$-criterion and the DSI. Due to the relative simplicity of the datasets, with at most two structures of interest, we set the detection threshold, $\tau$, for $Q$ or $\DSI$ more or less arbitrarily. As explained in the previous section, the choice of a particular threshold $\tau_{\rm overlap}$ for the spatial overlap of structures between successive output times is not of much consequence either as any amount of overlap with a structure at a future timestep will correspond to a correct continuation in this simple setup. Therefore, we set $\tau_{\rm overlap}$ to $0.1$.

\begin{figure}[h]
	\centering
	\includegraphics[width=.8\textwidth]{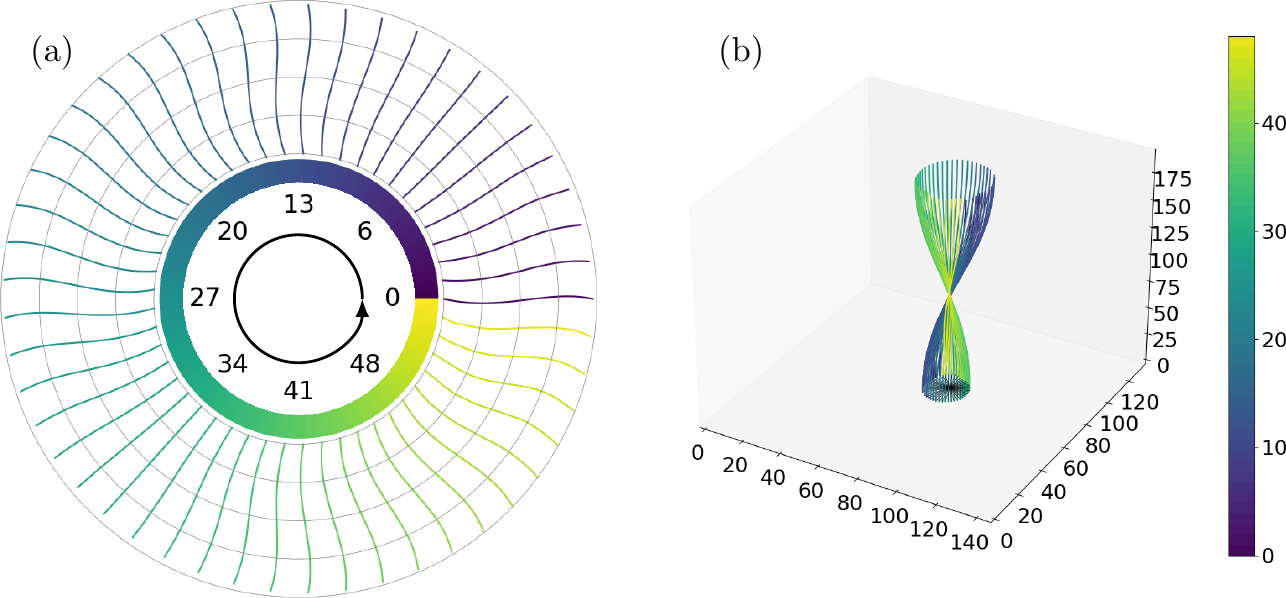}
	\caption{First 49 timesteps of tracking an idealized cyclone with volume overlap. Both (a) and (b) show the centerline of the Q-criterion structure. While (b) shows the centerline curves in a three-dimensional perspective, the shape of their projections onto the horizontal plane are shown in (a), with their centers shifted radially so as to untangle the individual plots. The curves in (b) are parameterized by a normalized height coordinate as indicated by the five concentric rings.
	}
	\label{fig:results_overlap_oneTC_Q}
\end{figure}

\begin{figure}[h]
	\centering
	\includegraphics[width=.8\textwidth]{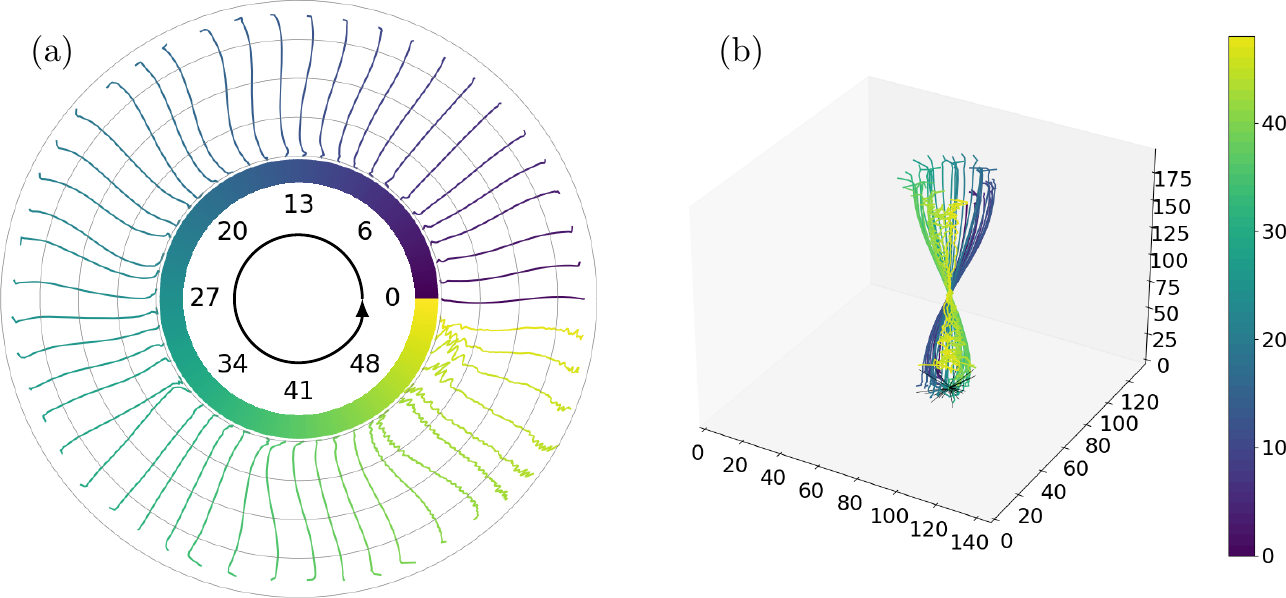}
	\caption{Similar to figure~\ref{fig:results_overlap_oneTC_Q}, (a) and (b) show the centerline of the mean DSI structure. Here, the `mean' refers to averaging the positive and negative DSI structures.}
	\label{fig:results_overlap_oneTC_DSI}
\end{figure}

\begin{figure}[h]
	\centering
	\includegraphics[width=.8\textwidth]{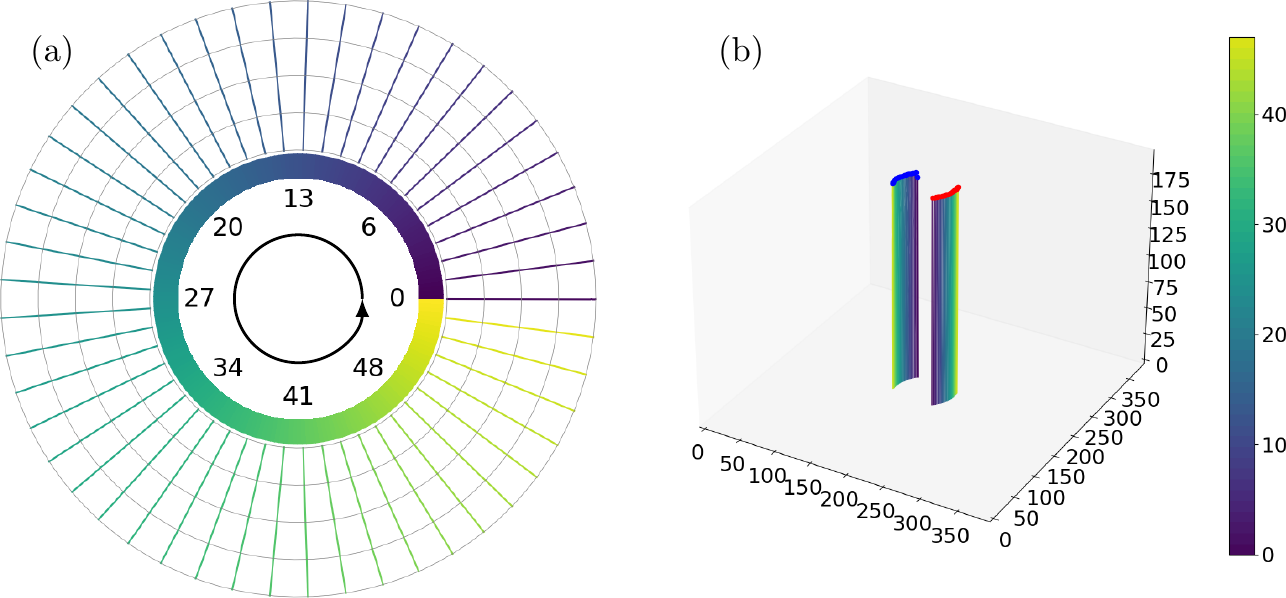}
	\caption{First 49 timesteps of tracking two idealized cyclones with volume overlap. Both (a) and (b) show the centerline of the Q-criterion structures. Red and blue scatter points indicate structure 1 and 2 respectively. DSI of structure 2 is shown in figure~\ref{fig:results_overlap_twoTC_DSI}.}
	\label{fig:results_overlap_twoTC_Q}
\end{figure}

\begin{figure}[h]
	\centering
	\includegraphics[width=.8\textwidth]{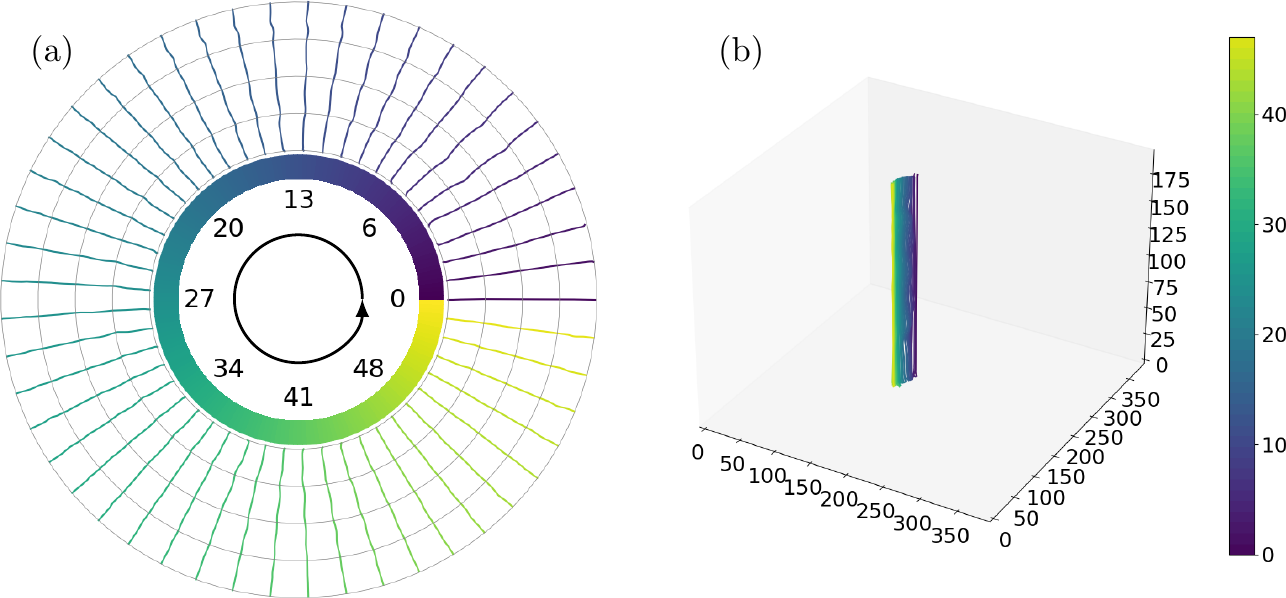}
	\caption{Similar to figure~\ref{fig:results_overlap_twoTC_Q}, (a) and (b) show the centerline of the DSI structure. The behavior of structure 1 is identical to the $Q-$criterion structure 1 shown in figure~\ref{fig:results_overlap_twoTC_Q} and is not shown here.}
	\label{fig:results_overlap_twoTC_DSI}
\end{figure}

The motion of the centerline for case 1TC as computed from \eqref{eq:centerline} is shown in figures~\ref{fig:results_overlap_oneTC_Q} and \ref{fig:results_overlap_oneTC_DSI} based on the $Q$-criterion and the $\DSI$, respectively. Within the first 49 timesteps one notices that the storm's centerline performs a complete precession about the domain center. In comparison to the results obtained from the $Q$-criterion, the DSI structure appears to distort severely close to the bottom and top of the domain after timestep 40, whereas this behavior is not observed using the $Q$-criterion. The reason for this difference is rooted in the nature of imbalances caused by the tilt of the storm.

The dominant contribution of the DSI is the tilt asymmetry (also present in the potential temperature, cf. \citet{DoerffelEtAl2021}). Being present at the initial time of the simulation in the form of small scale fluctuations of the centerline, the tilt near the top and bottom wall decreases as the vortex is balancing itself (through radiation of gravity waves) towards a configuration in which the tilt, and thus the dominant DSI mode, vanishes near the bottom and top boundaries. With a vanishing $\DSI$ signal, however, the averaging procedure of \eqref{eq:centerline} becomes unstable ultimately causing the oscillations observed in figure~\ref{fig:results_overlap_oneTC_DSI}. Hence, we conclude that (the absolute value of) the DSI may not be well-suited to indicate the presence of a tropical cyclone since small-scale imbalances have an impact orders of magnitude larger than the tilt or the center depression of the vortex. Similar results are shown in figures~\ref{fig:results_overlap_twoTC_Q}, and \ref{fig:results_overlap_twoTC_DSI} for the 2TC case. Here, we do not observe much difference in shape for either of the tracked structures. The shape of structure~2 for $Q$-criterion and $\DSI$ is exemplified in figure~\ref{fig:results_overlap_twoTC_Q} and \ref{fig:results_overlap_twoTC_DSI} respectively.

\begin{figure}[h!]
	\centering
	
	\begin{minipage}[t]{0.22\textwidth}
		\includegraphics[width=\textwidth]{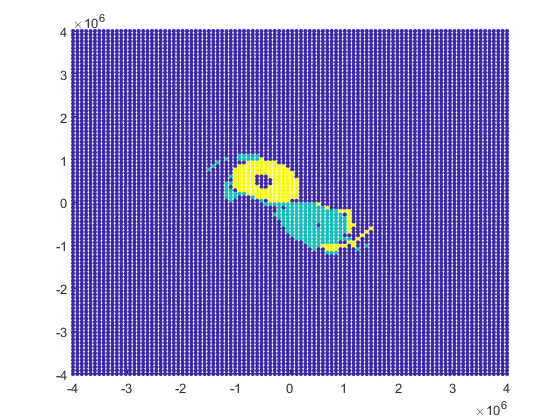}
	\end{minipage}
	\begin{minipage}[t]{0.22\textwidth}
		\includegraphics[width=\textwidth]{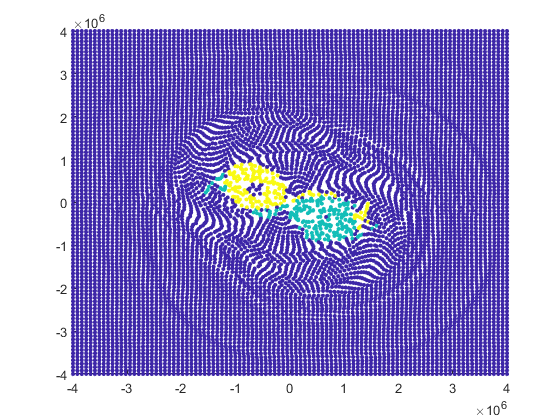}
	\end{minipage}
	\begin{minipage}[t]{0.22\textwidth}
		\includegraphics[width=\textwidth]{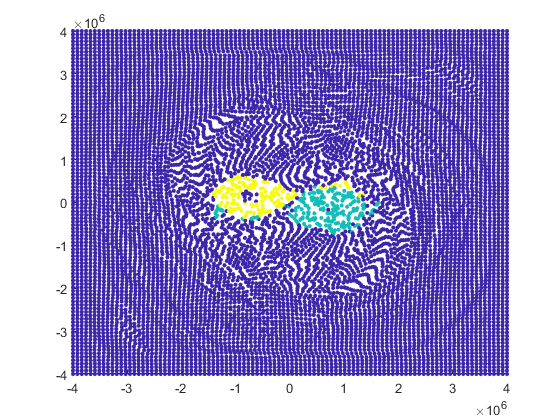}
	\end{minipage}
	\begin{minipage}[t]{0.22\textwidth}
		\includegraphics[width=\textwidth]{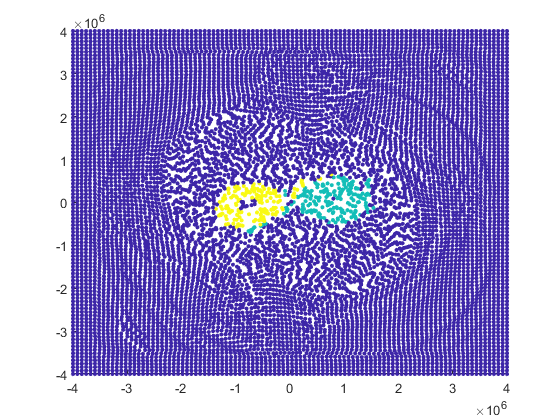}
	\end{minipage}

	\begin{minipage}[t]{0.22\textwidth}
		\includegraphics[width=\textwidth]{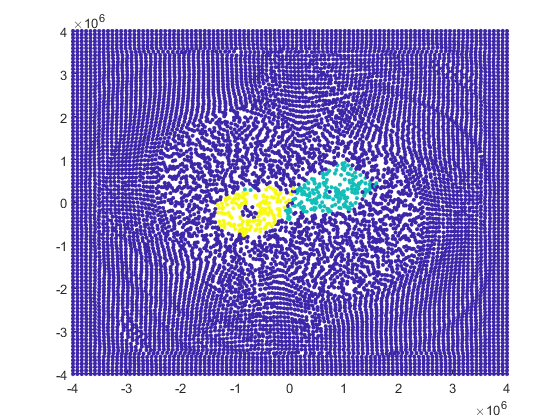}
	\end{minipage}
	\begin{minipage}[t]{0.22\textwidth}
		\includegraphics[width=\textwidth]{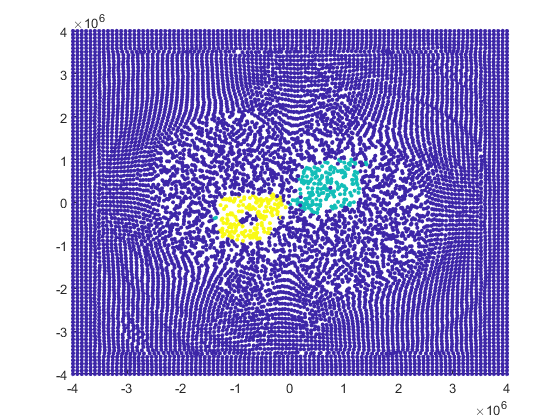}
	\end{minipage}
	\begin{minipage}[t]{0.22\textwidth}
		\includegraphics[width=\textwidth]{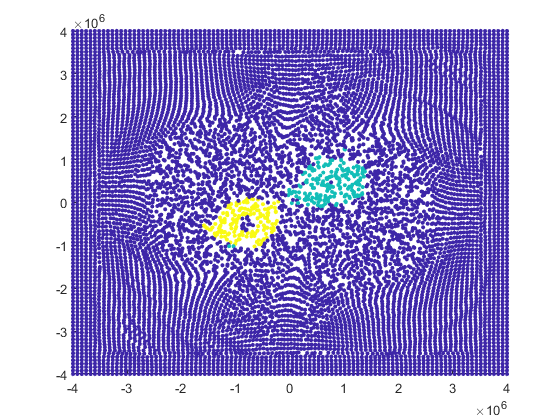}
	\end{minipage}
	\begin{minipage}[t]{0.22\textwidth}
		\includegraphics[width=\textwidth]{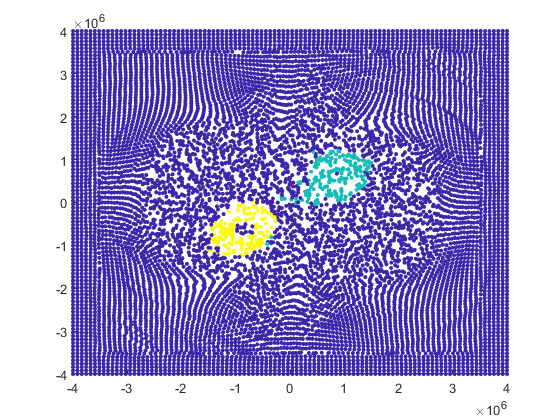}
	\end{minipage}
	
	\caption{
		Material coherent sets in the data set of the two synthetical cyclones (2TC case) (see figure~\ref{fig:synth_cyclones} and \ref{fig:partition_cyclones} for dynamical spectral clustering) with $10000$ trajectories for diffusion maps in a domain which covers a domain of \qty{8000}{\kilo\meter} in each horizontal direction and \qty{10}{\kilo\meter} in the vertical.
		The evolution under the flow is shown for eight time points.}
	\label{fig:DM2TC}
\end{figure}

In comparison to the previous analyses, figure~\ref{fig:DM2TC} demonstrates how Lagrangian trajectories can be used to detect materially coherent structures 
{for the 2TC case} (see section~\ref{subsec:material_coherence} for a description of the method).
Both individual cyclones are tracked as isolated coherent features (particles colored in yellow and cyan) that do not mix with the background (blue particles) over the full integration time of~$\qty{24}{\day}$.

\subsection{Detecting and tracking a Tropical Cyclone: Hurricane Florence}
\label{subsec:Florence_cyclones}

\begin{figure}
	\centering 
	\includegraphics[width=0.9\linewidth]{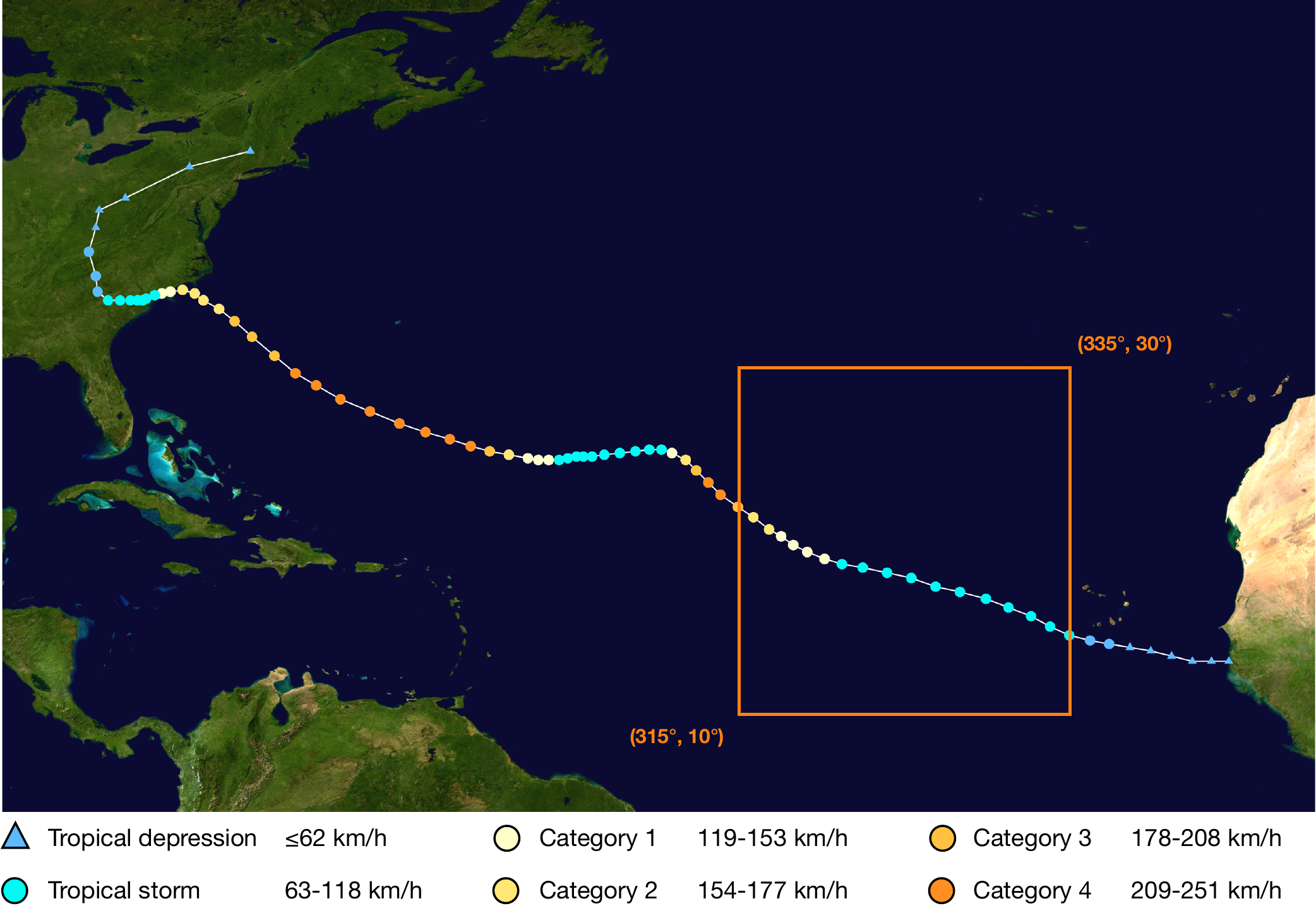}
	\caption{Track map of Hurricane Florence 2018 {\citep[basemap and track taken from][]{wiki:Florence2018}}. The points show the location of the storm at 6-hour intervals. The color represents the storm's maximum sustained wind speeds according to the Saffir–-Simpson scale (see legend). The orange box defines the area of interest which corresponds to the rapid intensification phase of the Hurricane Florence.}
\label{fig:florence_track}
\end{figure}

Here we investigate the intensification phase of hurricane Florence 2018 which corresponds to 2-3 September based on the ERA5 data set. The track of the hurricane and the region of interest are shown in figure~\ref{fig:florence_track}. The meteorological inputs to the algorithm described in \ref{sssec:trop_cyclones} are the pressure and 3D velocity vector fields, $p$ and $\mathbf{v}$, respectively, given for 48 hourly-distributed time steps.

\begin{figure}
	\centering 
    (a) \includegraphics[width=0.9\linewidth]{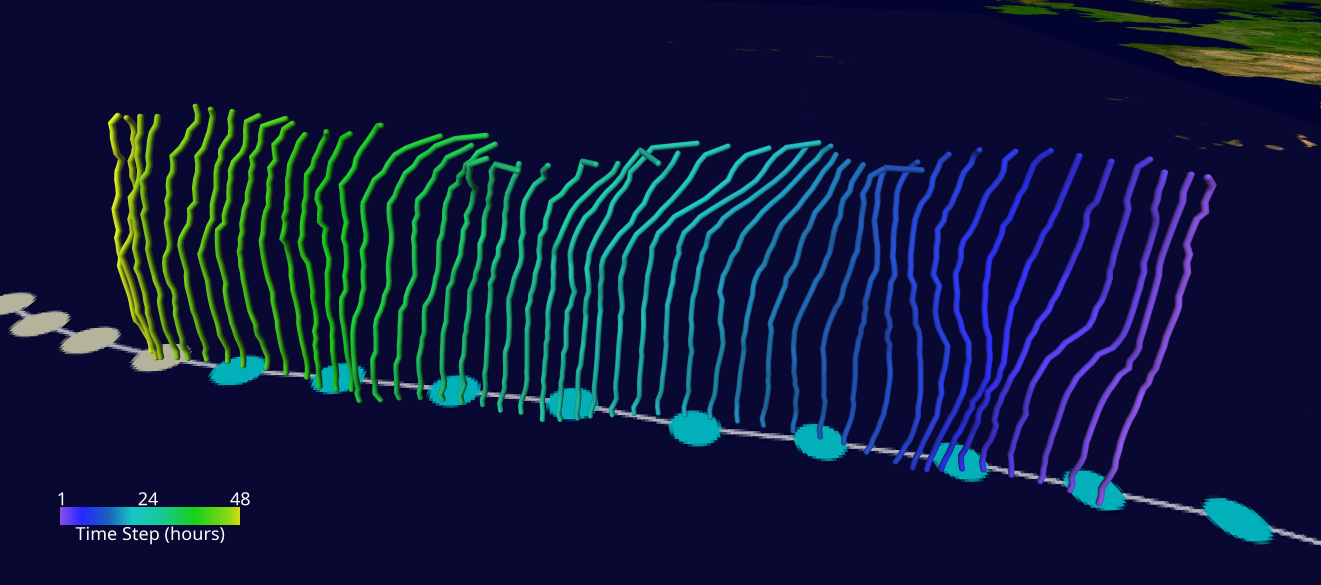} \\[0.2cm]
	(b)\includegraphics[width=0.45\linewidth]{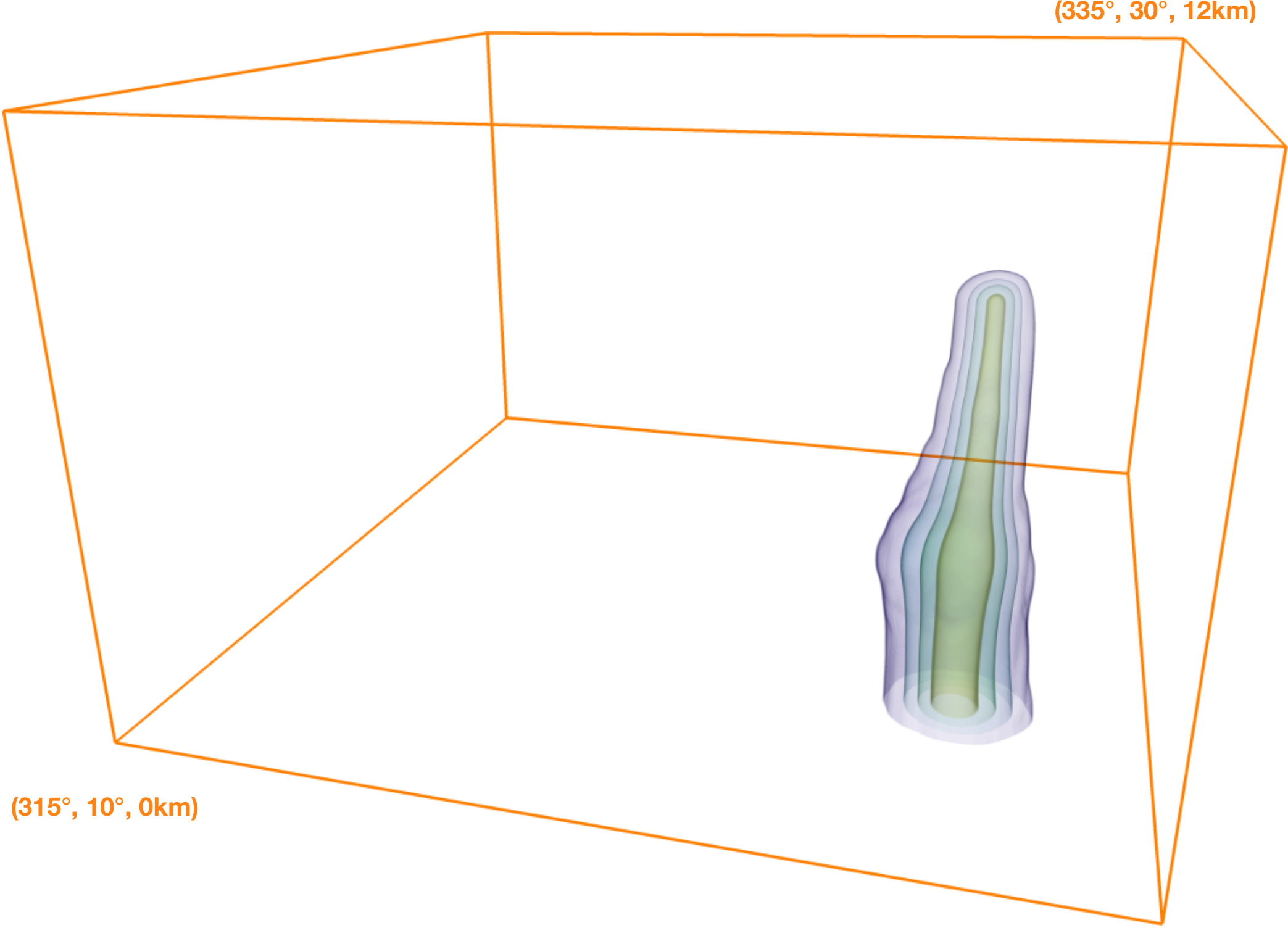} 
	(c) \includegraphics[width=0.45\linewidth]{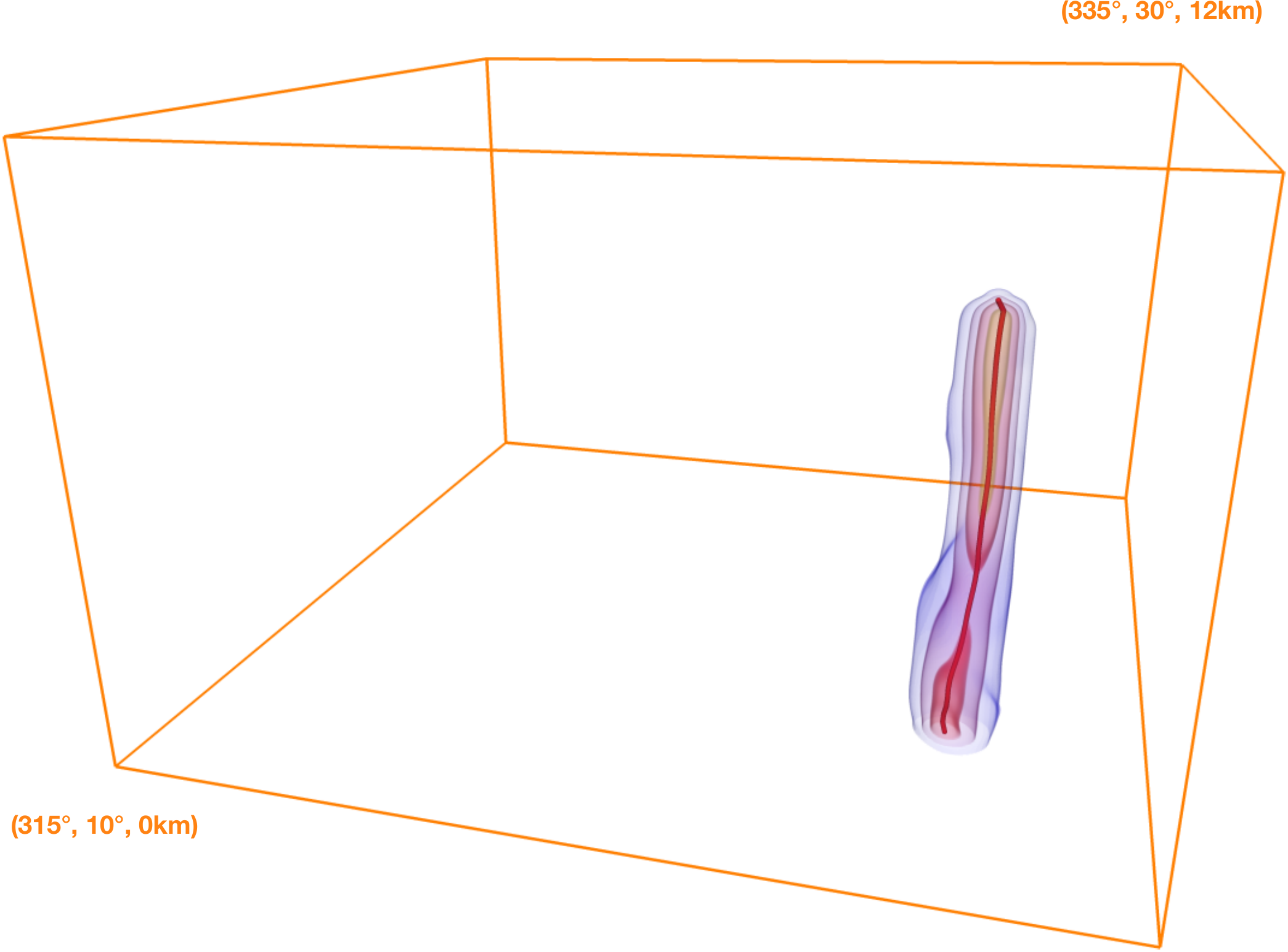}
	\caption{(a) Time evolution of centerlines computed from $Q$-value in regions where $p^*\leq 0.1$ based on ERA5 data of hurricane Florence 02-03 September 2018. (b) Nested iso-surfaces of the normalized pressure field perturbation for $p^*\in\{0.04, 0.05, 0.075, 0.1\}$ at the first time step (2 September 2018, 00:00). (c) Nested iso-surfaces of the $Q$-value at $Q\in\{0.02,0.03,0.04,0.05\}$ localized by $p^*\leq 0.1$. Red line represents the centerline computed using eq.~\eqref{eq:centerline}. }
	\label{fig:tc_pres_q}
\end{figure}

First we compute a normalized pressure perturbation $p^*$ as defined in eq.~\eqref{eq:pres_perturbation} and determine low-pressure areas for each time step by extracting regions where $p^*\leq 0.1$ (see figure~\ref{fig:tc_pres_q}b for an exemplary representation at initial time). Next, $Q$-values are computed, and the vortex core region is determined by finding regions with $Q>0$ inside the low-pressure areas (shown in figure~\ref{fig:tc_pres_q}c for initial time). Those regions are tracked over the presented time steps. The size of the extracted structure and appropriate time resolution make the simple volume overlap technique discussed in \ref{subsubsec:simpleOverlap} with $\tau_{\rm overlap} = 0.2$ applicable to this case. The centerlines are computed using \eqref{eq:centerline}, where $\omega$ is chosen to be $Q$-values. Extracted centerlines over the full time range of \qty{48}{\hour} are presented in figure~\ref{fig:tc_pres_q}a.

As discussed in section~\ref{sssec:trop_cyclones}, this --- somewhat involved --- procedure allows to robustly extract the centerline as scaffold of the surrounding core region throughout the vertical extent of the TC. The two-field approach of combined criteria to the two feature-defining fields is based on physical considerations and efficiently reduced spurious contribution to the vortex core that would enter due to small scale vortical structures that are not directly associated to the primary circulation. Adding the normalized pressure criterion \eqref{eq:pres_perturbation} restricts the detection to regions where the flow is in approximate gradient wind or cyclostrophic balance. This restriction also automatically provides an estimate of the vertical extent of the core structure, as can be seen by figure~\ref{fig:tc_pres_q}, where the top height of the vortex changes over the timesteps. Furthermore, we see that the vortex centerline, as a three-dimensional line feature, evolved non-trivially due to both transport by the external wind and self-induced motions \citep[cf.][for a thorough description of the underlying mechanism]{PaeschkeEtAl2012,DoerffelEtAl2021}.

\begin{figure}
	\centering
	
	\begin{minipage}[t]{0.3\textwidth}
		\includegraphics[width=\textwidth]{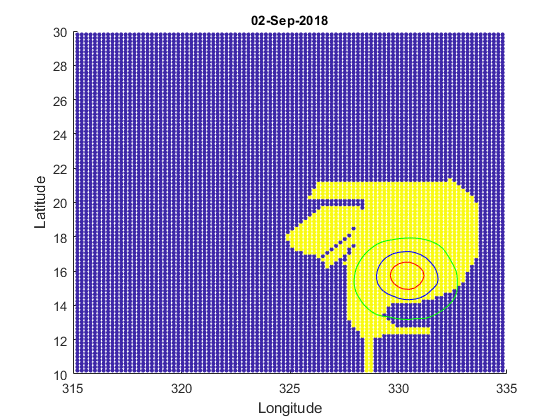}
	\end{minipage}
	\begin{minipage}[t]{0.3\textwidth}
		\includegraphics[width=\textwidth]{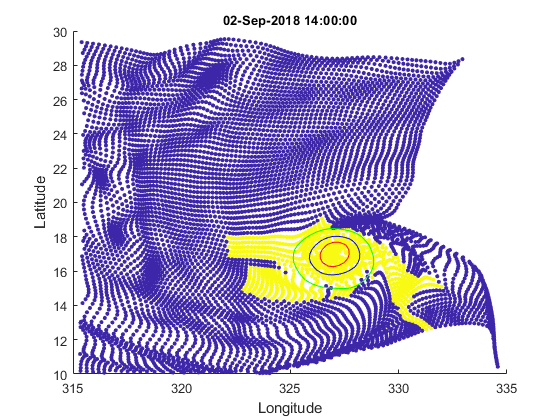}
	\end{minipage}
	\begin{minipage}[t]{0.3\textwidth}
		\includegraphics[width=\textwidth]{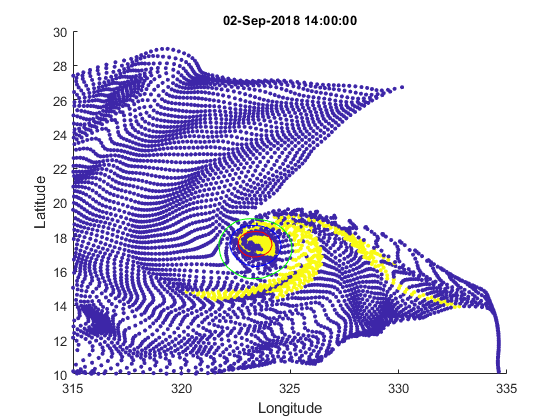}
	\end{minipage}

	\begin{minipage}[t]{0.3\textwidth}
	\includegraphics[width=\textwidth]{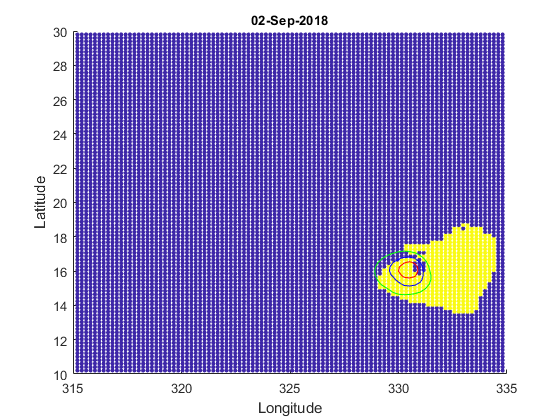}
\end{minipage}
\begin{minipage}[t]{0.3\textwidth}
	\includegraphics[width=\textwidth]{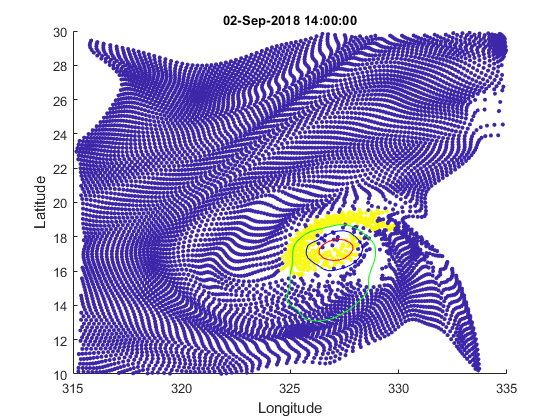}
\end{minipage}
\begin{minipage}[t]{0.3\textwidth}
	\includegraphics[width=\textwidth]{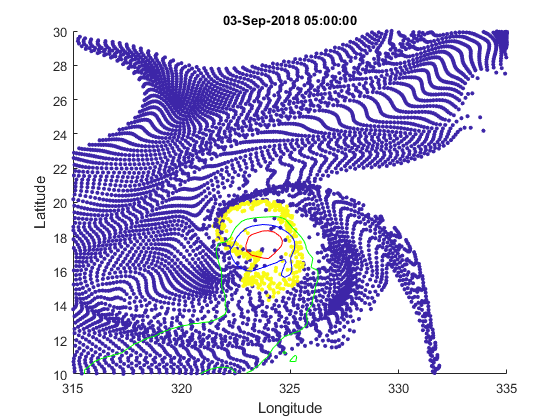}
\end{minipage}

	\begin{minipage}[t]{0.3\textwidth}
	\includegraphics[width=\textwidth]{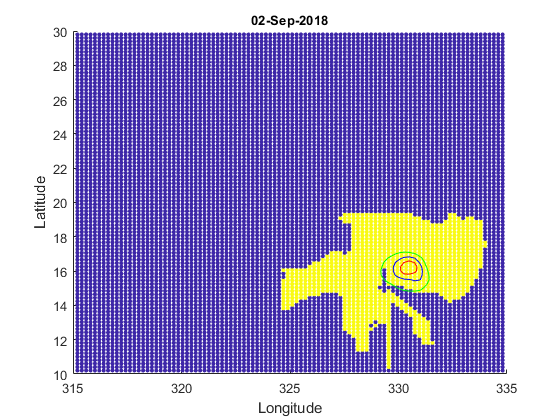}
\end{minipage}
\begin{minipage}[t]{0.3\textwidth}
	\includegraphics[width=\textwidth]{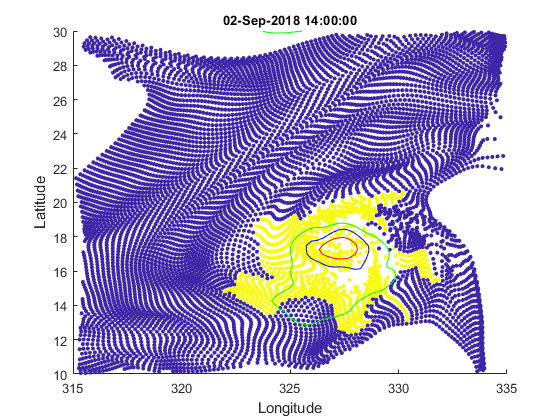}
\end{minipage}
\begin{minipage}[t]{0.3\textwidth}
	\includegraphics[width=\textwidth]{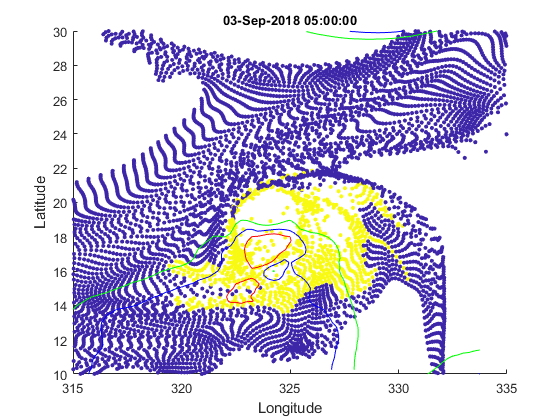}
\end{minipage}

	\begin{minipage}[t]{0.3\textwidth}
	\includegraphics[width=\textwidth]{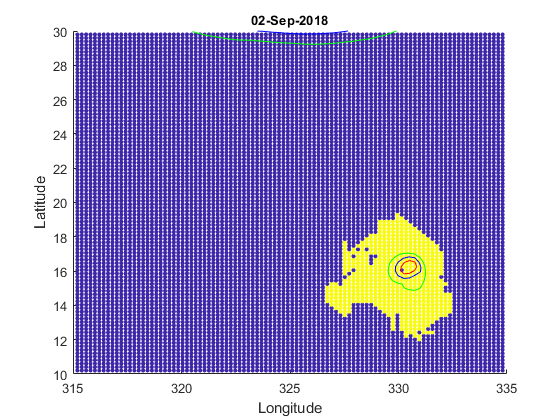}
\end{minipage}
\begin{minipage}[t]{0.3\textwidth}
	\includegraphics[width=\textwidth]{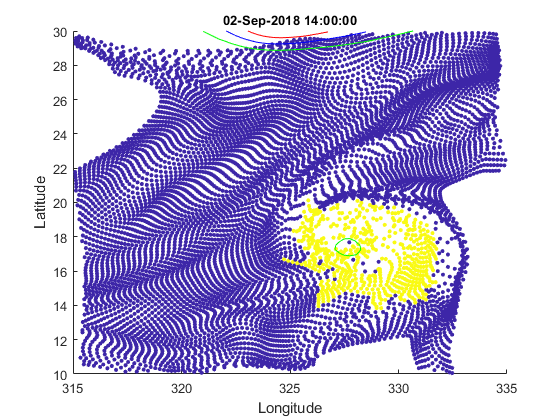}
\end{minipage}
\begin{minipage}[t]{0.3\textwidth}
	\includegraphics[width=\textwidth]{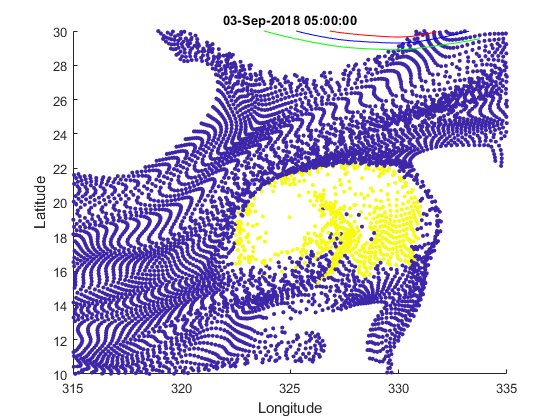}
\end{minipage}

	\caption{
		Material coherent sets in the data set of hurricane Florence with $10000$ trajectories for a domain which is bounded by $10^{\circ}\text{N -- }30^{\circ}$N latitude and $315^{\circ}\text{E -- }335^{\circ}$E longitude. The evolution under the flow is shown for three time points: 02-Sep-2018 00:00 (initial), 02-Sep-2018 14:00 and 03-Sep-2018 05:00 (final time) for height levels 0.2 km (top), 6.2 km, 7.2 km and 8.2 km (bottom). Time horizon for the method is time interval from  02-Sep-2018 00:00 to 02-Sep-2018 14:00. The pressure isocontours are shown (0.1 in red, 0.2 in blue and 0.3 in green). }
	\label{fig:DM}
\end{figure}

Next, we analyze constructed Lagrangian trajectories based on the velocity fields of storm Florence for material coherent sets, as described in section~\ref{subsec:material_coherence}. The evolution under the flow is shown in figure~\ref{fig:DM} for three time points: 02-Sep-2018 00:00 (initial), 02-Sep-2018 14:00 and 03-Sep-2018 05:00 (final time) for height levels 0.2 km (top), 6.2 km, 7.2 km and 8.2 km (bottom). The time horizon for the computational method is the time interval from 02-Sep-2018 00:00 to 02-Sep-2018~14:00.

Sets are called materially coherent for small dynamic isoperimetry ratio which is defined in \eqref{eq:cohset}. While the detected yellow set in figure~\ref{fig:DM} on height level \qty{6.2}{\kilo\meter} retains a comparably low boundary-to-volume ratio, all the detected sets at different height levels (\qty{0.2}{\kilo\meter}, \qty{7.2}{\kilo\meter}, \qty{6.2}{\kilo\meter}, and \qty{8.2}{\kilo\meter}, top to bottom panels of figure~\ref{fig:DM}) do not capture the structure as detected by the previous method based on pressure and $Q$-criterion (isocontours of the $Q$-criterion are plotted in each panel of figure~\ref{fig:DM}). Under the flow, all sets either reduce or increase their volume dramatically over time. Furthermore, most of the initial sets (except for \qty{6.2}{\kilo\meter}) appear spurious exhibiting partially grid-aligned and irregular boundaries. In the course of the time evolution, branches of the sets are formed. We therefore do not interpret the detected sets as being physically meaningful.

The behavior of the detected coherent sets is caused by the very nature of TCs: Near the ocean surface, turbulent friction causes orbiting air parcels to slow down, hence the radial force may not suffice to balance the pressure gradient, finally leading to radially converging inflow. Due to mass conservation, the inflow is balanced by vertical outflow to higher layers, eventually triggering convection. In terms of coherent sets, this means that, considering two-dimensional horizontal projections of the flow field, as done here, the volume of an arbitrary subset is not preserved. The 
top right of figure~\ref{fig:DM} nicely demonstrates the convergence towards a single point. For higher altitudes, the situation reverses and radial outflow causes particles to be expelled from a center region. The result are voids as can be seen for all the right panels of figure~\ref{fig:DM}, except the top one.

\begin{figure}
	\centering
	(a) \includegraphics[width=0.45\linewidth]{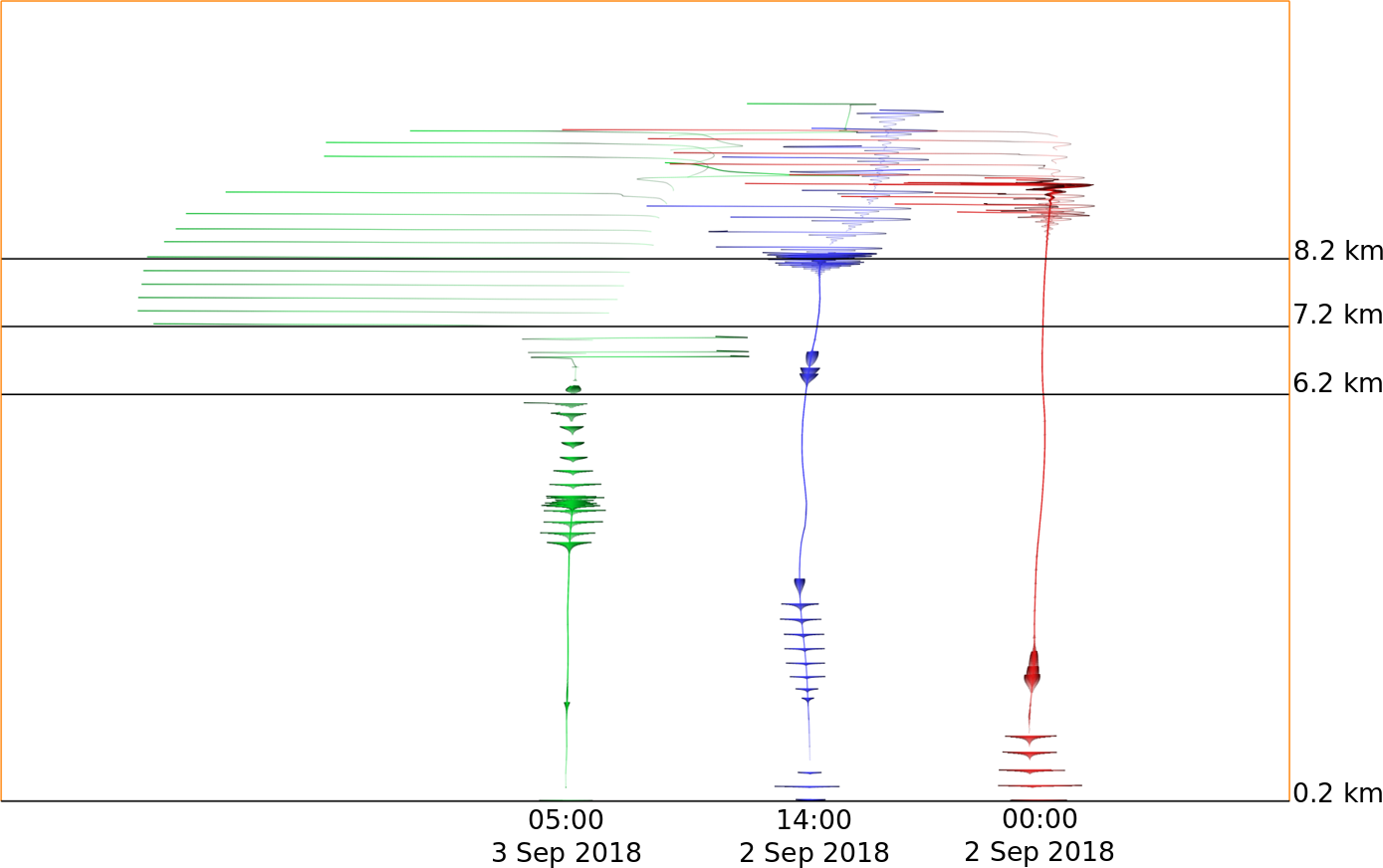}
	(b) \includegraphics[width=0.45\linewidth]{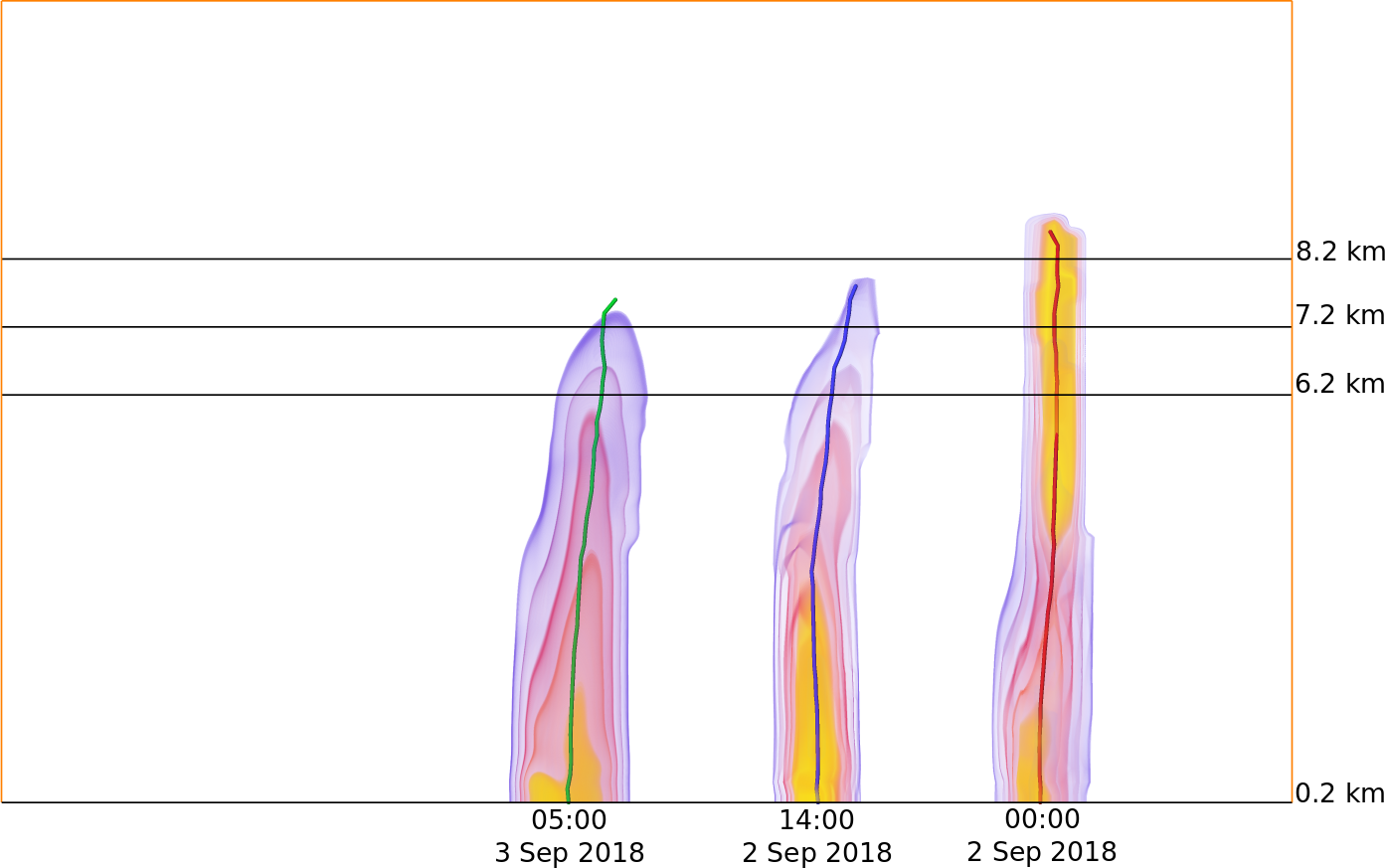} 
	\caption{ Height variation of the structure for 3 time frames represented by (a) streamlines initiated in the center of circular motion for each level and (b) nested iso-surfaces of the $Q$-value at $Q\in\{0.01,0.02,0.03,0.04,0.05\}$ localized by $p^*\leq 0.1$. } 
	\label{fig:tc}
\end{figure}

Figure~\ref{fig:tc} gives further insight into the difficulty of detecting coherent sets in the flow field of TCs: Over three time steps, height variation of the structure detected by the pressure-$Q$-criterion method are represented by (a) streamlines seeded at the extracted centerline and (b) nested iso-surfaces of the $Q$-value. It becomes apparent, that the TC shrinks during its evolution. By Sep 2, 2018, 14:00h, the top of the TC has fallen below \qty{8.2}{\kilo\meter}, the uppermost level of the present analysis. By means of streamlines, the basis for the analysis of diffusion maps, we see that the pressure-$Q$-criterion structure is topped by substantially diverging outflow layer, which, as previously discussed poses a major challenge to the present method.

While the present approach based on diffusion maps delivered reasonable results for the 2TC test case (cf. section~\ref{subsec:results_segm_cyclones}), we conclude that persistent structures in the sense of coherent sets may not be found when, as far as analyzed here, we restrict to two-dimensional horizontal flow fields and do not take into account the converging-diverging behavior as induced by the secondary circulation. Further improvement of the present method would require lifting the method to three-dimensions and accounting for a modified volume measure based on the three-dimensional anelastic constraint, \ie, a divergence constraint on the momentum field instead of the velocity.


\section{Discussion}
\label{sec:discussion}

All previous considerations were targeting the characterization, identification, or the tracking of persistent atmospheric structures. Next, we will adapt a level of abstraction that allows us to view all of these analyses within one unified framework. 


\subsection{Persistent structures - a common framework}
\label{ssec:PersistentStructureFramework}

Let us refer to a \emph{spatial flow structure} if we can identify a time dependent subset $A_t$ of the flow domain -- the spatial support of the structure --  inside of which one or more observables or \emph{features} $F_t = (F_i(\vxi_t,x))_{i\in I}$ of the system's \emph{state} $\vxi_t$ and position $x\in\Omega$ exceed prescribed significance thresholds $F^c = (F_i^c)_{i\in I}$. Here $I$ is an appropriate, possibly uncountable, index set that allows us to label individual features. For any $t$, the structure described by $(A_t, F_t, F^c)$ is called \emph{spatially coherent} if $A_t$ exhibits some logical properties $G = (G_1, G_2, ...)$. These could describe certain geometrical or topological attributes, or small temporal changes of these. The structure is called \emph{persistent} if it maintains these logical properties over long times relative to a suitable intrinsic flow time scale,~$T_f$.

Thus, to describe what in our individual applications is considered a ``persistent structure'', we will need three ingredients:
\begin{itemize}

\item \textbf{State.} The ``state'' $\vxi_t$ of the system at time $t$ describes a set of variables that is sufficient to completely describe the system at time~$t$.

\item \textbf{Feature and its spatial support.} A time-dependent physically defined collection of one or more properties that define the sets of interest through the ``feature indicating map'' $F_t = F(\vxi_t)$. Based on the user-defined significance thresholds $F^c$ for its components, this map yields the spatial support 
\begin{equation}\label{eq:SpatialSupport}
A : [0,T] \to P(\Omega); \quad 
t \mapsto A_t = \setdef{x \in \Omega}{F_i(\vxi_t,x) \geq F^c_i \ \ (i = 1, 2, ...)}
\end{equation}
of the structure, where $P(\Omega)$ denotes the power set of~$\Omega$.

\item \textbf{Spatial coherence and persistence.} A scalar quantity
\begin{equation}\label{eq:PersistenceMeasure}
M(A, G) = \frac{T_L(A, G)}{T_f} \gg 1  
\qquad \text{with} \qquad
T_L(A, G) = \left|\setdef{t \in \reals^+}{G(A,t)}\right|
\end{equation}
that measures the time scale over which the set valued function $A$ maintain the spatial coherence property (or properties) $G$, relative to a characteristic time scale $T_f$ of the flow inside the set.

\end{itemize}

Notice that a given combination of features, thresholds and of the persistence measure may identify a multiply connected set within a flow.
In this case we may consider each singly connected subset as one persistent structure and speak of an ensemble of persistent structures of the same kind. 


\subsection{Volume Overlap indicates persistence of 3D objects in scalar fields}
\label{ssec:VolumeOverlapPersistence}

In this section, we map the tracking technique of ``volume overlap with image registration'' from section~\ref{sssec:Persistence_overlap} to the common framework introduced in the previous section. This helps us understand what a "persistent structure" is from the point of view of volume overlap and how it is used to track  weather fronts, tropical cyclones and extratropical cyclones. The three necessary components are,

\medskip
\noindent
\textbf{State.} The physical set of variables which completely describe the tropical and extratropical cyclones in our examples at time $t$ are $(v, p, T)$, where $v$ is the velocity, $p$ is the pressure and $T$ is the temperature. In the case of weather fronts, the fields are $(v, p, T, q)$ where $q$ is the tuple of  mixing ratios of the moisture constituents, such as water vapor, cloud water, rain water, and several types of frozen water particles. 

\medskip
\noindent
\textbf{Feature and its spatial support.} We rely on two feature indicators namely the $Q$-criterion and the $\DSI$ to identify the objects for tracking. Therefore, $F_t = Q = \frac{1}{2} \Vert A \Vert^2 - \Vert S \Vert^2$ or $F_t = |\DSI|$. From equation \eqref{eq:SpatialSupport}, it follows that,
\begin{equation}
	A_t = \{ x\in \Omega \, | \, F_{t}(\vxi_t, x) \geq F^c\}
\end{equation}
where $F^c = \tau$ is the user-defined threshold that identifies the relevant level sets of $F$. For weather fronts, the feature indicator is described in section~\ref{subsec:wassersteinDistancePersistence}.

\medskip
\noindent
\textbf{Spatial coherence and persistence.} To test the persistence of the structure described by $(A_{t}, F_{t}, \tau)$, we compute a similarity measure between $t_{n}$ and $t_{n+1}$ as described in equation \eqref{eq:DSC}. Therefore, 
\begin{equation} \label{eq:DSCOverlap}
	G = \text{DSC} > \tau_{\rm overlap}
\end{equation}
where $\tau_{\rm overlap}$ is a second user-defined threshold measuring the quality of overlap between two timesteps. If equation \eqref{eq:DSCOverlap} is not satisfied, then image registration is used to find the appropriate structure at time $t_{n+1}$. More precisely, we apply the transformation parameters $\hat{\mu}$ described in equation \eqref{eq:ImageRegistrationEq1} on the structure defined by $(A_{t}, F_{t}, \tau)$ at time $t_{n}$ and deform the points to a region in~$t_{n+1}$. Since transformation between timesteps is not usually known, \ie, the structure at $t_{n}$ may have rotated, translated, changed its shape/size or a combination of these, the B-spline transform only guesses where the ``correct'' structure will be at~$t_{n+1}$. Therefore, we overlap the deformed structure with all structures at $t_{n+1}$ to identify the ``correct'' best-fitting structure. Then, equation \eqref{eq:DSCOverlap} becomes,
\begin{equation} 
	G = \{\hat{\mu}, \DSC > \tau_{\rm overlap}\}, \quad M(A, G) = \frac{T_L(A, G)}{T_f} \gg 1, 
\end{equation}
where a proper choice of the characteristic intrinsic time scale $T_{f}$ for cyclones is the turnover time of the primary circulation, \ie, the time it takes an air parcel located near the radius of maximum wind to once surround the center of the storm. This choice is consistent with the asymptotic analysis of tropical storms in \citep{PaeschkeEtAl2012,DoerffelEtAl2021}, which explicitly utilizes a time scale separation between the primary circulation and the overall motion of the storm. 


\subsection{Wasserstein distance indicates persistence of weather fronts}
\label{subsec:wassersteinDistancePersistence}

Here we describe how weather fronts, captured by the approach from sections~\ref{george_method}, \ref{subsec:class_synop_loc} fit into the common framework for persistent atmospheric structures from section~\ref{ssec:PersistentStructureFramework}.

\medskip
\noindent
\textbf{State.} The state $\vxi_t$ comprises all physical variables characterizing the atmospheric system at hand, \ie, the fields $(v, p, T, \boldmath{q})$ of velocity, pressure, temperature, and the mixing ratios of the moisture constituents (\eg, $\boldmath{q} = (q_v, q_c, q_r)$ for water vapor, cloud water, and rain water), respectively.
	
\medskip
\noindent
\textbf{Feature and its spatial support.} The (only) feature to be accounted for is the magnitude of the gradient of the smoothed temperature field, $F_t = \left|\nabla (\sigma(T_t))\right|:\Omega\to \R$, where $\Omega\subset \R^2$ is a subset of a horizontal slice through the atmosphere and constitutes the domain of interest, and $\sigma\colon L_2(\Omega) \to L_2(\Omega)$ is a spatial averaging filter, see section~\ref{george_method}. Frontal structures are then identified by 
\begin{equation}
	A_t = \{ x\in \Omega\colon \Vert \nabla (\sigma(T_t))(x)\Vert_2 \geq \tau \}
\end{equation}
for some user-defined gradient magnitude threshold~$F^c = \tau>0$.
	
\medskip
\noindent
\textbf{Spatial coherence and persistence.} Fronts can be more or less coherent in space, and this is measured by the mean or median Wasserstein distance indicator $\iota_t = \OT_\varepsilon(\tilde \mu_t, \tilde \mu_{t+\dt})$ from (\refeq{eq:WassersteinCoherenceFronts}) (see also section~\ref{subsec:dyn_spec_cluster}). It takes as input the appropriately normalized uniform probability measures $\tilde \mu_t, \tilde \mu_{t+\dt}\in \mathcal P(\Omega)$ supported on the front boundaries $\partial A_t, \partial A_{t+\dt}$. That is, we choose
\begin{equation}
\widetilde G(A,t) = \iota_t 
\qquad\text{and}\qquad
G(A,t) = \left(\widetilde G(A,t) < G^c\right)
\end{equation}
as the spatial coherence measure and the associated geometrical property, respectively. 

Based on these definitions, one may now assess the degree of spatial coherence maintained by a frontal structure over the simulation time window $[0,T]$ by 
\begin{equation}
	\mathrm{median}_{t\in [0,T]}(\iota_t)\,,
\end{equation}
and 
\begin{equation}
M(A,G) = \frac{T_L(A,G)}{T_f}
\end{equation}
as its persistence measure (see \eq{eq:PersistenceMeasure}). An appropriate intrinsic flow time scale in this case should be a characteristic advection time scale across the frontal structure, \ie, $T_f = \mathrm{median}_{t\in[0,T]}\diam(A_t)/v_{\mathrm{ref}}$, with $v_\mathrm{ref} \sim 10\, \mathrm{m/s}$ a typical wind velocity.


\subsection{Material coherence indicates persistence of storms}

Here we restrict ourselves to sets that are ``advectively'' or ``materially'' coherent. This also means that we can restrict the state of the system to the information that describes advection.

The effect of an incompressible flow is completely described by the deformation which the fluid mass has undergone in the course of time. Therefore, one option is to associate the time dependent state $\vxil_t$ of such a system with the forward flow map, $\phi_0^t$ (the flow from time $0$ forward in time by $t$ time units), \ie, 
\begin{equation}
\vxil_t : \left\{
\begin{array}{c@{\ }c@{\ }c@{\ }l} 
& \Omega_0
& \to 
& \Omega_t := \phi_0^t \Omega_0
\\[5pt]
& \vx 
& \mapsto 
& \vxil_t(\vx) = \phi_0^t(\vx)
\end{array}
\right.
\qquad (t \in [0,T])\,,
\end{equation}
where $\Omega_0 \subset \reals^d$ is the flow domain at time $0$ and $t$ denotes time. In  particular, we have $\vxil_0 = \mathrm{id}$, and we see that
\begin{equation}
\vxil_{t+\tau} = \phi_t^{\tau} \vxil_t \,.
\end{equation}
This first option represents the Lagrangian perspective: The argument $x \in \Omega_0$ of $\vxil_t$ labels particles according to their initial position and $\vxil_t(x) \in \Omega_t$ describes to which positions the particles have moved at time~$t$. 

A second option to define the state of an incompressible flow that is closer to the usual viewpoint in meteorology corresponds to the Eulerian perspective and is given by

\medskip
\noindent
{\bfseries State.} Original positions of all particles
\begin{equation}\label{eq:FlowMapEuler}
\vxi_t : \left\{
\begin{array}{c@{\ }c@{\ }c@{\ }l} 
& \Omega_t
& \to 
& \Omega_0
\\[5pt]
& \vx 
& \mapsto 
& \vxi_t(\vx) = \phi^{-t}_t(\vx)
\end{array}
\right.
\qquad (t \in [0,T])\,.
\end{equation}
That is, the state is given by the inverse flow map. With \eqref{eq:FlowMapEuler} the flow state associates a state variable (the Lagrange coordinate) with each location $x$ in the (current) flow domain, and this is akin to the association of pressure, temperature, wind field, etc.\ to position in the Eulerian representation of the state of the atmosphere. We follow this Eulerian perspective in the rest of this section. 

The time dependent subsets $A_t$, the (material) coherence of which we intend to characterize, have the common \emph{feature} that they always contain the same fluid parcels. That is, for material coherence we have the

\medskip

\noindent
{\bfseries Feature and its spatial support.} A ``material set'' always contains the same particles, \ie, 
\begin{equation}
\qquad 
A : 
\left\{
\begin{array}{rcl}
\dss [0,T] 
  & \to 
    & \dss P(\Omega) 
      \\
 t
   & \mapsto
     & A_t
\end{array}
\right.
\ \ \text{is ``material''}
\quad \Leftrightarrow \quad
\forall \ t \in [0,T] : \ \vxi_t (A_t) = A_0\,,
\end{equation}
where $\Omega = \cup_{t\in[0,T]} \Omega_t$, $P(\Omega)$ is the power set of $\Omega$, and for given $A_0 \subset \Omega_0 \subset \Omega$ the \emph{feature indicating map} is
\begin{equation}\label{eq:FeatureMapMaterial}
F(\vxi_t) = \chi_{A_0}(\vxi_t)\,,
\end{equation}
where $\chi_D$ is the indicator function of the set $D$, and a significance threshold would be simply~$F^c=1$. In the sequel we will focus on initial sets $A_0$ with smooth boundary~$\partial A_0$. 

\medskip
\noindent
{\bfseries Spatial coherence and persistence.} The coherence of an identified material set can be assessed on the basis of its surface-to-volume ratio
\begin{equation}\label{eq:mat_coh_svratio}
\widetilde{G}(D,t) := S(D, \Omega_t) = \frac{|\partial D|}{\min \{|D|,|\Omega_t\setminus D|\}}\,,
\end{equation}
by considering the indicator $G(D,t) = \big( \widetilde{G}(D,t) < \lambda\, h(\Omega)\big)$, where $\lambda \geq 1$ is a scaling factor and $h(\Omega)$ a reference value, discussed below.
The set $A_t$ for some $t\in [0, T]$ is then called spatially coherent if $G(A_t, t) = 1$, where, $1$ corresponds to the Boolean ``true''. Further, the time dependent family of sets $A$ is called materially coherent \footnote{The word ``material'' emphasizes that the coherence property is measured for a material (\ie, purely advected) feature.} if $G(A_t,t) = 1$ for all $t$ in some time-interval of interest, and is called (materially) persistent if
\begin{equation}
\label{eq:coh_crit}
M(A,G) = \frac{1}{T_f} \int_0^T G(A_t,t)\, dt \gg 1,
\end{equation}
where~$T_f \ll T$.
Thus, \eqref{eq:coh_crit} requires a coherent set to be ``unfilamented'' on average for long times in comparison with a suitable reference value~$h(\Omega)$. A meaningful reference in this context could be the \emph{Cheeger isoperimetric constant} of the domain~$\Omega$,
\begin{equation}\label{eq:mat_coh_ref_val}
h(\Omega) := \inf_{t\in [0, T]} \inf_{E \subset \Omega_t} S(E,\Omega_t)\,,
\end{equation}
with the surface-to-volume ratio $S$ from \eq{eq:mat_coh_svratio}. The infimum here is taken over all sets $E$ with smooth boundary, and thus $h(\Omega)$ describes the smallest perimeter-to-volume ratio (``filamentation'') of sets forming a two-partition of the reference domain~$\Omega$. The criterion~\eqref{eq:coh_crit} allows for a coherent set a reasonable but not too large deviation from this optimum. Hence, a scaling factor $\lambda \ge 1$, with $\lambda \approx 1$ is reasonable.

Naturally, it is in the eye of the beholder what she still considers to be coherent. Thus, $\lambda$ and \eqref{eq:coh_crit} give a quantity to order sets according to their material coherence. Since the perimeter increases more slowly than the volume under upscaling, larger sets (unless their volume exceeds $\frac12 |\Omega|$) tend to appear higher up in the coherence hierarchy given by~\eqref{eq:coh_crit}, assuming an analogous but scaled evolution.

An important problem that has attracted a lot of interest recently is how to compute sets $A$ that are maximally materially coherent, given $\Omega, \phi_t^\tau$, and $T$, and this is, in particular, also pursued in settings where it is important that the flow domain itself is time dependent.

Note that colleagues working on material coherence may be used to considering the advected sets $A_t$ themselves to be the relevant ``feature''. For them, the introduction of the indicator function $\chi_{A_0}$ in \eq{eq:FeatureMapMaterial} may come across as equivalent, but somewhat artificial. We have nevertheless chosen to use this definition of the feature indicating map as it reveals that the concept of material coherence fits the general framework given in section~\ref{ssec:PersistentStructureFramework} above. 


\subsection{Dynamical Spectral Clustering indicates persistence of moving feature clusters}

Coherent sets with respect to dynamical spectral clustering as in the example of two tropical cyclones given in section~\ref{subsec:results_segm_cyclones} adhere to the proposed framework as follows.

\medskip
\noindent
\textbf{State.} The state $\vxi_t$ comprises all physical variables to characterize the atmospheric system at hand, as, \eg, the two tropical cyclones (see sections~\ref{ssec:VolumeOverlapPersistence} and \ref{subsec:wassersteinDistancePersistence} above).

\medskip
\noindent
\textbf{Feature and its spatial support.} The feature for dynamical spectral clustering is a time-dependent non-negative scalar field $F_t\colon \Omega\to \R$, where $\Omega\subset \R^d$ is the domain of interest. In the example of the two tropical cyclones, we used the vorticity magnitude $F_t = \Vert \nabla \times (u(t),v(t),w(t))^T \Vert_2$, but this could be replaced by other observables, like the absolute dynamic state index $\vert \mathrm{DSI}_t\vert$, a thresholded $Q$-criterion $Q_t\cdot 1_{Q_t \geq 0}$ etc. The choice of a significance threshold $F^c$ generally yields multiply connected spatial supports
\begin{equation}
A_t = \setdef{x \in \Omega}{F(\vxi_t,x) \geq F^c}.
\end{equation}
Tracking its connected components (by OT in the time-discretized setting) yields multiple simply connected spatial supports
\begin{equation}
A^1, \dots, A^k\colon [0, T]\to P(\Omega), \quad t\mapsto A_t^j\subset A_t,
\end{equation}
where $A_t^j$, $j=1,\dots,k$ are the simply connected components of $A_t$. Note that for some features, multiple such connected components might correspond to the same structure. We interpret the union $\smash{ A_t(k) \coloneqq \cup_k A_t^{j_k} }$ of all components of the $k$-th structure as a ``feature cluster''. One example of multiply connected feature clusters is $\vert \mathrm{DSI}_t\vert$: The structures indicated by the DSI, \eg, for a moving cyclone, exhibit a characteristic dipole structure as shown in section~\ref{subsubsec:dsi}. In this case, any choice $F^c>0$ will yield pairs of subsets as the natural constituents of a cluster. This is typically not a problem when using the dynamical spectral clustering algorithm, since it will first partition structures that are far apart as indicated by the singular value hierarchy of the concatenated transfer operator, see section~\ref{subsec:dyn_spec_cluster}.

\medskip
\noindent
\textbf{Spacial coherence and persistence.} The spacial coherence and persistence of one feature cluster $A_t(k)$ is assessed in the same way as in the case of material coherence by the equations \eqref{eq:mat_coh_svratio}--\eqref{eq:mat_coh_ref_val}.
Note that the dynamical spectral clustering algorithm does not necessarily find the most persistent sets according to this definition, but instead a more complicated criterion in terms of transfer operators and probability mass leakage, which will not be stated here.
The latter can, however, be seen as an approximation of the former.


\section{Conclusion}
\label{sec:conclusion}

In this work, we have detected persistent structures in four scale dependent meteorological phenomena and tracked them in time with multiple methods. Although, at first, these methods may appear to be largely unrelated, our abstraction framework reveals the existence of three basic components at the heart of each method namely, the state, a set of features and their spatial support, and measures of spatial coherence and persistence. Taken together, these components describe the definition and temporal evolution of a persistent structure.

\begin{table}[h]
\begin{tabularx}{\textwidth}{l l c c}
\makecell[l]{Meteorological phenomena} 
& \makecell[l]{Case} 
 & \makecell{Feature} 
  & \makecell{Persistence} \\
\hline
\makecell[l]{Extratropical cyclones}
& \makecell[l]{Lothar and Martin \\ (December 1999)} 
 & \makecell[l]{$Q$-criterion, $|DSI|$, \\ set indicator functions}
  & \makecell{VOIR, CS}\\
\hline
\makecell[l]{Synoptic and local fronts}
& \makecell[l]{Germany \\ (6th August 2013)} 
 & \makecell[l]{Smoothed temperature \\ gradient magnitude} 
  & \makecell{Classification - $\iota$ \\ Tracking - SVO} \\
\hline
\makecell[l]{Idealized tropical cyclones}
& \makecell[l]{Simulation} 
 & \makecell[l]{Vorticity magnitude, \\ $Q$-criterion, $|DSI|$}
  & \makecell{DC, SVO}\\
\hline
\makecell[l]{Hurricane}
& \makecell[l]{Florence \\ (August - September 2018)} 
 & \makecell[l]{$Q-$criterion with \\ pressure minima, \\ set indicator functions}
  & \makecell{SVO, CS}\\
\hline
\end{tabularx}
\caption{A summary of the meteorological phenomena, the indicators used for identification and the applied tracking methodologies are shown here. Different persistent structures associated with the extratropical cyclones Lothar and Martin are tracked using the Coherent Sets (CS) approach and by Volume Overlap with Image Registration (VOIR) based on vorticity criteria in section~\ref{subsec:results_lothar}. Weather fronts are classified as local and synoptic with the Wasserstein distance indicator $\iota$ and tracked with Simple Volume Overlap (SVO) in section~\ref{subsec:class_synop_loc}. Idealized tropical cyclones are tracked with both Dynamical Spectral Clustering (DC) and SVO in section~\ref{subsec:results_segm_cyclones}. In the case of hurricane Florence, SVO and CS were used for tracking related persistent structures in section~\ref{subsec:Florence_cyclones}.
}
\label{tab:summary}
\end{table}

All investigated meteorological phenomena, the indicators used to identify relevant structures, and the tracking approaches are summarized in Table~\ref{tab:summary}. The boundaries of weather fronts are identified with user-defined threshold on the smoothed scalar magnitude of a temperature gradient field at a chosen height level. Synoptic fronts are differentiated from local fronts by a high degree of temporal persistence, which is measured using an indicator based on the Wasserstein distance. Our results show that the method is capable of correctly separating the frontal structures by assigning higher values to local fronts, \ie, low persistence, and vice versa for synoptic fronts. Due to the erratic nature of local fronts, sometimes they exhibit branching behavior which is adequately captured with simple volume overlap. This technique is also applicable to synoptic fronts which due to their high temporal persistence have large volume overlaps.

Application of Eulerian and Lagrangian tracking and identification methods to extratropical cyclone and hurricane datasets provides interesting complementary results. 
Analyzing the synthetic TC datasets, the presented methods performed equally in that they identified the single TC and the two TCs as isolated and persistent features.
For hurricane Florence, however, both, coherent sets, applied in horizontal slices, and simple volume overlap, identify regions around the vortex core as coherent and persistent structures, respectively (see figure~\ref{fig:DM}). The persistence of the coherent sets found by the Lagrangian approach is not quantifiable as clearly within the horizontal slices, due to the more complex flow structure, and extended analyses of the full three-dimensional fields may provide a remedy. Within the time slice featuring the extratropical cyclones Lothar and Martin, however, volume overlap with image registration finds both storms whereas the Lagrangian coherent set approach identifies large-scale air masses in the vicinity the storms but does not reveal the storm centers themselves. The detected coherent sets are indicative of the larger \textit{flow system} involving rapidly advected air masses that feed and are fed by the storms for some period of time and later disperse when they have left the storm behind. The initial positions of these material coherent sets reveal where the air masses participating in the storm have originally come from, and these regions may lie far away from where the storm itself first reached noticeable strength.

Often, there are scenarios where the relevant persistent structure cannot be tracked by the standard overlap technique: this can be either due to poor time resolution or a small-scale feature being advected fast in the domain. We have successfully demonstrated two techniques that deal with this scenario namely, dynamical spectral clustering and volume overlap with image registration. The former is applied to a simulation of two idealized tropical cyclones represented by merely eight available snapshots far enough apart in time to not allow for any overlap of the pertinent structures between them. The latter technique was used to track the storm Lothar with $Q$-criterion scalar fields separated $\qty{6}{\hour}$ in time. 

This paper has introduced the concept of ``persistent structures'' in the context of meteorological applications. The concept is surely more widely applicable in that it could straightforwardly be used to identify solitons in nonlinear dispersive systems, shock and sound waves in compressible media, reaction-diffusion fronts in biology or combustion, or materially coherent sets in many different contexts. Central to the definition of a particular type of persistent structure is the feature-indicating map which encodes the physical properties of the structure. An important example of how different feature selections yield complementary information in one and the same application was given on the basis of mid-latitude storms: Vorticity-based features enable to localize the storms themselves, while material indicator functions provide information on how the storms exchange mass with their environment.
%
%
This highlights once more that in a given flow different types of persistent structures can coexist. These will differ by the physical properties that distinguishing them from each other and that can be encoded through different feature-indicating maps. In conclusion, we hope that the present work may stimulate advanced multi-criteria data analyses for enhanced physical insights into complex fluid flows.


\begin{acknowledgments}
This research has been funded by Deutsche Forschungsgemeinschaft (DFG) through grant CRC 1114 'Scaling Cascades in Complex Systems, Project Number 235221301, Projects A01,  'Coupling a multiscale stochastic precipitation model to large scale atmospheric flow dynamics', B07 'Selfsimilar structures in turbulent flows and the construction of LES closures', C06 'Multiscale structure of atmospheric vortices'. P.K., A.M., and P.N.~thank Henning Rust for discussions on material coherence.
\end{acknowledgments}

\medskip



\bibliography{bibliography}


\appendix



\end{document}